\documentclass[3p,times]{elsarticle}

\usepackage{mathrsfs}

\usepackage{amssymb}

 \usepackage{graphicx}
 \usepackage{epstopdf}
 \usepackage{epsfig}
 \usepackage{booktabs}
 \usepackage{lscape} 

\usepackage[figuresright]{rotating}
\usepackage[pdftex, colorlinks]{hyperref}

\begin{document}

\begin{frontmatter}


\title{Sensitivity analysis and determination of free relaxation parameters for the weakly-compressible MRT-LBM schemes}
 \author{Hui Xu\corref{cor1}}%
 \ead{xuhuixj@gmail.com or xu@lmm.jussieu.fr}
\cortext[cor1]{Corresponding author. }
\author{Orestis Malaspinas}
 \author{Pierre Sagaut}
 \ead{sagaut@lmm.jussieu.fr}
\address{Institut Jean le Rond d'Alembert, UMR CNRS 7190, Universit\'e Pierre et Marie Curie - Paris 6, 4 Place Jussieu case 162 Tour 55-65, 75252 Paris
Cedex 05, France}





\begin{abstract}
It ha been proved that the lattice Boltzmann Methods (LBMs) are very efficient for computational fluid dynamics purposes, and for capturing the dynamics of weak acoustic fluctuations. Especially,  the multi-relaxation-time lattice Boltzmann method (MRT-LBM) appears as a very robust scheme with  high precision. It is well-known that there exist several free relaxation parameters in the MRT-LBM. Although these parameters have been tuned via linear analysis, the sensitivity analysis of these parameters and other related parameters are still not sufficient for detecting the behaviors of the dispersion and dissipation relations of the MRT-LBM. Previous researches have shown that the bulk dissipation in the MRT-LBM induces a significant over-damping  of acoustic disturbances. This indicates that   MRT-LBM  cannot be used to obtain the correct behavior of pressure fluctuations because of  the fixed bulk relaxation parameter. In order to cure this problem, an effective algorithm  has been proposed for recovering the linearized Navier-Stokes equations (L-NSE ) from the linearized MRT-LBM (L-MRT-LBM).  The recovered L-NSE appear as in matrix form with arbitrary order of the truncation errors with respect to $\delta t$.  Then,  in wave-number space,  the first/second-order sensitivity analyses of matrix eigenvalues  are  used  to address the sensitivity of the wavenumber magnitudes to  the dispersion-dissipation relations.  By the first-order sensitivity analysis, the numerical behaviors of the group velocity of the  MRT-LBM are first obtained. Afterwards, the distribution sensitivities of the matrix eigenvalues corresponding to the linearized form of the MRT-LBM are investigated in the complex plane.  Based on the sensitivity analysis and the recovered L-NSE, we propose some simplified optimization strategies to determine the free relaxation parameters in the MRT-LBM. Meanwhile, the dispersion and dissipation relations of the optimal MRT-LBM are quantitatively compared with the exact dispersion and dissipation relations. At last, some numerical validations on classical  acoustic benchmark problems are shown to assess the new optimal MRT-LBM.
\end{abstract}


\end{frontmatter}


\section{Introduction}\label{Intro}
The lattice Boltzmann method (LBM) has emerged as an innovative mescoscopic numerical method for the computational modelling of a wide variety of complex fluid flows \cite{chendoolen}. The early researches
about the dispersion and dissipation relations of the classical LBM (BGK-LBM) show that the LBM are well suited to capture the weak acoustic
pressure fluctuations
\cite{buickgreatedcampbell,simondenispierre,ricotmariesagautbailly,buickbuckleygreated}. 
But, the BGK-LBM always suffers from an unstability drawback. In order to improve the stability of the BGK-LBM, the MRT-LBM was derived in moment space based on the generalized lattice Boltzmann equation \cite{dhumieres2,lallemandluo}.   
Thanks to its adjustable bulk viscosity,  the MRT-LBM can guarantee a better stability by a high value of the bulk viscosity \cite{simondenispierre,lallemandluo}. 
In most researches and simulations, the relaxation parameters of the MRT-LBM were chosen according to the results of the linear analysis \cite{dhumieres2,lallemandluo}.  
Results of the linear  von Neumann analysis of the BGK-LBM and the MRT-LBM yielded quantitative results about the dispersion and dissipation relations. 
Von Neumann analysis shows that the LBM have  lower numerical dissipation than many aeroacoustic-optimized high-order schemes for the  Navier-Stokes equations \cite{simondenispierre}. 
The 2nd-order accurate LBM has the better dispersion capabilities than the classical Navier-Stokes schemes with the 2nd-order accuracy in space combined with 3-step Runge-Kutta time integration \cite{simondenispierre}.   
The dispersion errors in  the LBM are larger than those induced by  3rd-order accurate in space the finite difference method combined with  4th-order accurate time-integration methods, and also higher than those of  6th-order accurate dispersion relation preserving (DRP) schemes. 
But it is important to emphasize that thanks their compactness,  for a given dispersion error, the LBM are faster than  high-order  Navier-Stokes equations \cite{simondenispierre} and are therefore more efficient for practical computations.  
Compared with the BGK-LBM, it is observed that there exists a higher dissipation of the acoustic modes in the MRT-LBM \cite{simondenispierre, xusagaut}.  
The free-parameter sensitivity  of the MRT-LBM is missing in open literature. 
Therefore, it is difficult to guarantee that the the results of the linear analysis are always acceptable.  
In order to clarify the sensitivity issue, we will use a novel methodology to reflect the relation between the L-NSE and the L-MRT-LBM. 
This methodology offers us a way to determine the free relaxation parameters in the MRT-LBM. This  entirely new work is different from the linear analysis.  
Using the  linear analysis,  it is impossible  to obtain the deterministic values of free relaxation parameters   \cite{dhumieres2,lallemandluo}. 
This difficulty is attributed to a drawback of the linear analysis of L-MRT-LBM:  the linear analysis \cite{lallemandluo} cannot reflect the influence of the linearized high-order truncated terms corresponding to the L-NSE on the dispersion and dissipation relations. 
But, using our new methodology, the influence of linearized high-order terms coupled with free relaxation parameters on dispersion and dissipation relations is revealed under a certain accuracy with respect to the higher-order truncation errors. Meanwhile, this methodology offers us a way to deal with the acoustic problems for any values of the shear viscosity and the bulk viscosity.  
The important fact is that the optimal MRT-LBM can be used in the weakly-compressible acoustic problems,  for which the shear viscosity is not always equal to the bulk viscosity. 
It is well-known that the classical BGK-LBM can only be used to solve acoustic problems with  equal shear viscosity and bulk viscosity, while the original MRT-LBM uses  fixed bulk relaxation parameters. As mentioned above, if the compressible effects are considered, the bulk viscosity becomes important for the sound wave propagation \cite{Kinsler, Lord, Morse}. So, it is necessary for acoustic problems  to specify both the shear viscosity and the bulk viscosity in the LBM. 

By our novel methodology, the recovered L-NSE and the exact L-NSE can be expressed in  matrix form in the wave-number space as follows
\begin{equation}
\partial_t W=B^{(n)}\cdot W+O(\delta t^n)\  {\rm (recovered\ L-NSE)},\  \partial_t W=B\cdot W\  {\rm (exact\ L-NSE)},
\end{equation}

where the square matrices $B^{(n)}$ and $B$ are functions of wave-number $\rm \bold{k}$ and base-flow uniform velocity $\rm \bold u$, and  $B^{(n)}$  can be regarded as a perturbation of $B$ \cite{xusagaut}. 
The vector $W$ denotes the perturbed macroscopic fluid flow conservative quantities (such as, density $\rho$, momentum $\rm \bold j$, $\cdots$).  Denoting $\mathcal{M}_{\epsilon}^{(n)}=B^{(n)}-B$ the perturbation matrix, the matrix $\mathcal{M}_{\epsilon}^{(n)}$ will be the function of all relaxation parameters of the MRT-LBM for $n>2$. 
With the aid of  the matrix $B^{(n)}$, we only consider the influence of the free parameters in the L-MRT-LBM on the hydrodynamic modes. 
This is of great benefit to determine the free relaxation parameters by our proposed optimization strategies. 
In order to establish the optimization strategies,  the parameter sensitivities are investigated in the wave-number space.  Then,  the sensitivity properties  of the free parameters on the hydrodynamic modes and the non-hydrodynamic modes are obtained.  
According to the sensitivity analysis, we propose a class of the simplified optimization strategies to minimize the dispersion and dissipation errors for the MRT-LBM in the wave-number space, based on the matrix perturbation theory and the theory of the modified equations. 
The proposed methodology leads to  a relation between the recovered L-NSE and the exact L-NSE, and  free parameters can be determined by this process.  Then, the optimal dispersion and dissipation relations are investigated by von Neumann analysis. It is noted that this famous and reliable analysis method has been revisited and extended \cite{senguptadipankarsagaut}.

It is  necessary to point out that our proposed original  methodology  is more general than the usual method for recovering the higher-order lattice Boltzmann schemes \cite{duboislallemand,dubois}, although we also use the  Taylor expansion method. 
Our method appears as an easy-to-hand recursive algorithm in the wave-number space, which is completely different from the  method proposed in \cite{duboislallemand}.

In next section, the fundamentals of the LBM are surveyed, and the algorithm used to establish the transformation relation from the L-MRT-LBM to the L-NSE is given.  In the third section, with the aid of von Neumann method, the sensitivity properties of the free parameters to the dispersion and dissipation relations are analyzed in the wave-number space. Then,  the optimization strategies are studied in section 4. In the last section, the optimal MRT-LBM
schemes are validated considering some benchmark problems.

\section{Fundamentals of the LBM and the methodology from the L-MRT-LBM to the L-NSE }\label{Method}

In this section, the fundamental theory of the LBM is briefly recalled. Then, we give the linearized LBM  and show 
the method to establish the relation between the linearized LBM and the
L- NSE.

\subsection{Lattice Boltzmann schemes}\label{subMRTLBM}

The governing equations of  the lattice Boltzmann schemes are described as follows \cite{dhumieres2,lallemandluo,xusagaut}
\begin{equation}\label{blbs}
f_i(x+v_i\delta t,t+\delta
t)=f_i(x,t)+\Lambda_{ij}\left(f_j^{\rm(eq)}(x,t)-f_j(x,t)\right),\quad
0\leq i,j\leq N,
\end{equation}

where $v_i$ belongs to the discrete velocity set $\mathcal{V} $,
$f_i(x,t)$ is the discrete single particle distribution function
corresponding to $v_i$ and $f_i^{\rm(eq)}$ denotes the discrete
single particle equilibrium distribution function. $\delta t$
denotes the time step and $N+1$ is the number of discrete
velocities. $\Lambda_{ij}$ is the generalized relation matrix.  
From here on, the repeated index  indicates the Einstein summation is used except for some special explanations. Let
$\mathcal{L}\in \mathbb{R}^d $ ($d$ denotes the spatial dimension)
denotes the lattice system, and the following condition is required
\cite{duboislallemand}
\begin{equation}
x+v_j\delta t\in \mathcal{L},
\end{equation}
that is to say, if $x$ is a node of the lattice, $x+v_j\delta t$ is
necessarily another node of the lattice.

Generally, for BGK-LBM, the relaxation matrix is given by
\begin{equation}\label{eq_BGK_LBM_R}
\Lambda_{ij}=s\delta_{ij},
\end{equation}
where $s$ denotes the relaxation frequency of BGK-LBM.

The standard isothermal MRT-LBM has the following form
\cite{lallemandluo,duboislallemand}
\begin{equation}\label{eq_conservative}
m_i=W_i=m_i^{\rm(eq)},0\leq i\leq d,
\end{equation}
and
\begin{equation}\label{eq_noconservative}
m_i(x+\delta t v_j,t+\delta
t)=m_i(x,t)+s_i\left(m_i^{\rm(eq)}(x,t)-m_i(x,t)\right),d+1\leq
i\leq N.
\end{equation}
It is necessary to point out that for isothermal flow, the number of the conservative quantities is equal to $d+1$. According to the work of Lallemand and Luo \cite{lallemandluo}, the
relaxation parameters in Eq. (\ref{eq_noconservative}), should
satisfy the following stability constraints
\begin{equation}
s_i\in (0,2),d+1\leq i\leq N.
\end{equation}
The relaxation matrix $\Lambda$ corresponding to Eqs.
(\ref{eq_conservative}) and (\ref{eq_noconservative}) is defined by
\begin{equation}
\Lambda=M^{-1}SM,
\end{equation}
where $S$ is a diagonal matrix which denotes the relaxation
parameters of MRT-LBM. $M=\left(M_{ij}\right)_{0\leq i\leq N,0\leq
i\leq N}$ is the transformation matrix, which satisfies the
following basic conditions (refer to \ref{app:1} for the detail definitions) \cite{lallemandluo} 
\begin{equation}
M_{0j}=1,M_{\alpha j}=v_j^{\alpha},(1\leq\alpha\leq d).
\end{equation}{
The macroscopic quantities are defined by
\cite{lallemandluo,duboislallemand}
\begin{equation}\label{q-macro}
m_i=M_{ij}f_j,\quad m_i^{\rm (eq)}=M_{ij}f_j^{\rm(eq)}.
\end{equation}

\subsection{Theoretical dispersion and dissipation relations for the L-MRT-LBM and the L-NSE}\label{t:dis2}

In order to implement von Neumann analysis for the L-MRT-LBM, the expression of the L-MRT-LBM in the
frequency-wave number space is given in this part. Considering a uniform mean part
$f_i^0$ and a fluctuating part $f_i{'}$, the equilibrium
distribution function can be linearized as
\cite{simondenispierre,ricotmariesagautbailly}
\begin{equation}\label{perturb-f}
f^{\rm(eq)}_i(\{f_j^0+f_j{'}\}_{0\leq j\leq
N})=f_i^{\rm(eq),0}+\left.\frac{\partial f^{\rm(eq)}_i}{\partial
f_j}\right|_{f_j=f_j^0}\cdot f_j{'}+O\left(f_j'^2\right).
\end{equation}
Then, considering a plane wave solution of the linearized equation
\begin{equation}\label{perturb-w-f}
f_j'=A_j{\rm exp}[\mathrm{i}({\bf k}\cdot {\bf x}-\omega t)],
\end{equation}
by Eqs. (\ref{perturb-f}), (\ref{perturb-w-f}) and
(\ref{q-macro}), we get the following eigenvalue problem for the
L-MRT-LBM in the frequency-wave number space
\cite{simondenispierre,ricotmariesagautbailly,lallemandluo}
\begin{equation}\label{sec:l-mrt}
e^{-\mathrm{i}\omega}{\bf f}'=M^{\rm mrt}{\bf f}',
\end{equation}
where the matrix $M^{\rm mrt}=A^{-1}\left[I-M^{-1}SMN^{\rm
bgk}\right]$, and $N^{\rm bgk}$ is defined by
\begin{equation}
N^{\rm bgk}_{ij}=\delta_{ij}-\left.\frac{\partial
f_i^{\rm(eq)}}{f_j}\right|_{f_j=f_j^0}.
\end{equation}
For the L-NSE, the analytical acoustic modes $\omega^{\pm}(\bf k)$ and shear
modes $\omega^s(\bf k)$ are given by \cite{landaulifshitz}
\begin{equation}\label{th:d}
\left\{\begin{array}{l} Re[\omega^{\pm}({\bf k})]=|{\bf k}|(\pm
c_s+|{\bf u}|{\bf
cos}(\widehat{\bf k\cdot u})),\\
Im[\omega^{\pm}({\bf k})]=-|{\bf
k}|^2\frac{1}{2}\left(\frac{2d-2}{d}\nu+\eta\right),\\
Re[\omega^s(\bf k)]=|{\bf k}||{\bf u}|{\rm cos}(\widehat{\bf k\cdot u}),\\
Im[\omega^s(\bf k)]=-|{\bf k}|^2\nu,
\end{array}\right.
\end{equation}
where $\nu$ is the shear viscosity and $\eta $ is the bulk
viscosity.

\subsection{New form of the L- MRT-LBM and  corresponding the higher-order L- NSE}\label{subL-MRT-LBM}

 The equilibrium function $m_i^{\rm(eq)}$ ($d+1\leq i\leq
N$) is expressed as a function of the  conservative quantities $W_i$
($W_i=m_i^{\rm(eq)}, 0\leq i\leq d$) (we  use the same notations
as Dubois and Lallemand \cite{duboislallemand})
\begin{equation}\label{mrteq} m_i^{\rm(eq)}=G_i\left(\{W_j\}_{0\leq j\leq d}\right),d+1\leq
i \leq N.
\end{equation}
or
\begin{equation}\label{l-mrt-eq} m_i^{\rm(eq)}=G_i\left(\{W_j\}_{0\leq j\leq
d}\right)=G_{ij}W_j=G_{ij}m_j
\end{equation}

In order to implement the linear stability analysis and recover
linearized macroscopic equations, we introduce the linearized form
of Eq. (\ref{mrteq}) around the reference state
\cite{simondenispierre,lallemandluo,duboislallemand}.
Using Eq. (\ref{l-mrt-eq}), the linearized description of Eqs.
(\ref{eq_conservative}) and (\ref{eq_noconservative}) can be
written  as follows

\begin{equation}\label{l-mrt-lbs}
m_i(x,t+\delta t)=M_{il}M_{lp}^{-1}\Psi_{pr}m_r(x-v_l\delta t,t),
\end{equation}
where the matrix $\Psi $ has the following form
\begin{equation}\label{mrt-psi}
\Psi=\left(\begin{array}{cc}\left(\mathrm{I}_{ij}\right)_{0\leq i\leq d,0\leq j\leq d}&0\\
\Theta&\left(\mathrm{I}_{ij}-S_{ij}\right)_{d+1\leq i\leq N,d+1\leq
j\leq N}
\end{array}\right)
\end{equation}
where $\Theta_{ij}=s_iG_{ij} $ (A matrix with the size
$(N-d-1)\times (d+1) $), $\mathrm{I}$ is an identity matrix and the
diagonal matrix $S$ is defined by
\begin{equation}\label{S_relaxation}
S={\rm
Diag}(\underbrace{0,\ldots,0}_{d+1},\underbrace{s_{d+1},\ldots,s_{N}}_{N-d-1}).
\end{equation}

In order to derive the linearized high-order equations, one assumes that
the discrete single particle distribution $f_i $
belongs to $C^{\infty}(T\times\mathcal{L}) $ (a functional set, in
which the element possesses a sufficiently smooth property with
respect to the time domain $T$ and the spatial domain $\mathcal{L} $ ).
This assumption is also used by Junk {\it et. al.} for asymptotic
analysis of LBM \cite{junklarluo}.
This regularity hypothesis indicates that macroscopic
quantities $m_i$ are  smooth ones and that  the linearized system (\ref{l-mrt-lbs}) is well defined.

The next step consists of performing Taylor series expansion of the right
hand side of Eq. (\ref{l-mrt-lbs}),  yielding
\begin{equation}\label{expansion_1}
m_i(x,t+\delta t)=\sum_{n=0}^\infty\frac{\delta t^n}{n!}
M_{il}(-v_l^{\alpha}\partial_\alpha)^nM^{-1}_{lp}\Psi_{pr}m_r,
\end{equation}
where  $\alpha$ indicates the summation from $1$ to $d$.

 Now, we define the matrix $A^{({n}),*}=\left(A^{({n}),*}_{ij}\right)_{0\leq i\leq
N,0\leq j\leq N} $ as follows
\begin{equation}
A^{({n}),*}_{ir}=\frac{1}{n!}M_{il}\left(-v_l^\alpha\partial_\alpha\right)^nM_{lp}^{-1}\Psi_{pr}.
\end{equation}
When we need to derive the equivalent equations or the modified equations,
it is difficult to use the matrix $A^{({n}),*} $ to carry out the
calculations. In order to overcome this difficulty, we use the
differential operators in spectral space. Let us note
$\partial_\alpha=\mathrm{i}k_\alpha $, with $k_\alpha $ the
wave-number in the $\alpha$ -direction and  $\mathrm{i}^2 = -1$.
Then, in spectral
space, the matrix $A^{({n}),*} $ has the following form
$\left(A^{({n})}=\left(A^{({n})}_{ij}\right)_{0\leq i\leq N,0\leq j\leq N} \right)$
\begin{equation}
A^{({n})}_{ir}=\frac{1}{n!}M_{il}\left(- \mathrm{i} v_l^\alpha
k_\alpha\right)^nM_{lp}^{-1}\Psi_{pr}.
\end{equation}
Therefore, Eq. (\ref{expansion_1}) can be rewritten as follows

\begin{equation}\label{expansion_2}
m_i(x,t+\delta t)=\sum_{n=0}^{J-1}\delta t^n A^{({n})}_{ir}m_r+O(\delta
t^J).
\end{equation}

In order to derive L-NSE corresponding to L-MRT-LBM defined by Eq. (\ref{expansion_2}),
we introduce an original   recursive algorithm.
Given $m_i=W_i (0\leq i\leq d) $ ($W_i$ denotes the macroscopic
conservative quantities, $d$ indicates the number of the macroscopic
conservative quantities), the algorithm is given as follows

\begin{itemize}

\item {\bf Initial step}. The initial $\Phi^{(1)} $ and $B^{(1)} $ are given as follows
\begin{equation}
\Phi^{(1)}_{ij}=\delta_{ij} (0\leq i\leq
d),\Phi^{(1)}_{ij}=\frac{1}{s_i}\Psi_{ij} (d+1\leq i\leq N),
\end{equation}
\begin{equation}
B^{(1)}_{ij}=A^{(1)}_{ir}\Phi^{(1)}_{rj}.
\end{equation}
Let $W=\{W_i\}_{0\leq i\leq d} $ and $m=\{m_i\}_{0\leq i\leq N} $
denote the vector of the conservative quantities and the vector of
all macroscopic quantities, respectively.

At the first order of $\delta t$, for all macroscopic quantities, we
have
\begin{equation}
m_i=\Phi^{(1)}_{ij}W_j+O(\delta t),0\leq i\leq N.
\end{equation}
In the matrix form, we obtain
\begin{equation}\label{R-IA}
m=\Phi^{(1)}W+O(\delta t) .
\end{equation}
At the first-order of $\delta t$, for conservative quantities, we
have
\begin{equation}
\partial_t W_i= A^{(1)}_{ir}\Phi^{(1)}_{rj}W_j+O(\delta t).
\end{equation}
The matrix form is
\begin{equation}\label{R-IB}
\partial_t W= A^{(1)}\Phi^{(1)}W_j+O(\delta t).
\end{equation}

\item {\bf Recursive formula for all macroscopic quantities}.
$\Phi^{(n)}_{ij}$ can be given as follows
\begin{equation}\label{R-A}
\Phi^{(n)}_{ij}=\frac{1}{s_i}\left(\Psi_{ij}-\sum_{l=1}^{n-1}\frac{\delta
t^l}{l!} \Phi^{(n-1)}_{ir}\left(B^{(n-1)}\right)_{rj}^l+\sum_{l=1}^{n-1}\delta t^l
A_{ir}^{(l)}\Phi_{rj}^{(n-1)}\right),d+1\leq i\leq N
\end{equation}
and
\begin{equation}
\Phi_{ij}^{(n)}=\delta_{ij}, (0\leq i\leq d).
\end{equation}
Eliminating the higher-order term of $\delta t^{n-1}$, we have
\begin{equation}\label{R-B}
\Phi_{ij}^{(n)}=\sum_{l=0}^{n-1}\delta t^l{\rm Coeff}\left(\Phi_{ij}^{(n)},\delta
t,l\right),
\end{equation}
where ${\rm Coeff(\cdot,\cdot,\cdot)} $ is a function which extracts
the coefficients of the polynomials, for example,
$f(x)=\sum_{i=0}^na_ix^i $
\begin{equation}
{\rm Coeff}\left(f(x),x,i\right)=a_i.
\end{equation}
According to Eqs.(\ref{R-A}) and (\ref{R-B}), we have
\begin{equation}\label{R-C}
m=\Phi^{(n)}W+O(\delta t^n).
\end{equation}

\item {\bf Recursive formula for conservative quantities}. $B_{ij}^{(n)}$ is
presented as follows
\begin{equation}\label{R-D}
B_{ij}^{(n)}=-\sum_{l=1}^{n-1}\frac{\delta
t^l}{(l+1)!}\left(B^{(n-1)}\right)_{ij}^{l+1}+\sum_{l=1}^n\delta
t^{l-1}A_{ir}^{(l)}\Phi_{rj}^{(n+1-l)}.
\end{equation}
Eliminating the higher-order term of $\delta t^{n-1}$, we have
\begin{equation}\label{R-E}
B_{ij}^{(n)}=\sum_{l=0}^{n-1}\delta t^l{\rm Coeff}\left(B_{ij}^{(n)},\delta t,l\right).
\end{equation}
Now, for the conservative quantities, we have the following equation
system
\begin{equation}\label{R-F}
\partial_tW=B^{(n)}\cdot W+O(\delta t^n).
\end{equation}
Using  Eqs. (\ref{R-B}),~(\ref{R-C}),~(\ref{R-E}) and (\ref{R-F}),
we can get the coefficient matrix of the conservative quantities at
any order of $\delta t$ . Details and validations are displayed in \cite{xusagaut}.
\end{itemize}

\subsection{Illustrative examples }\label{L-NSE-WNS}

In this section, we will use the above proposed algorithm to recover the L-NSE.  Here, we will show two examples corresponding to 2D and 3D MRT-LBM that are similar to the incompressible NSE \cite{dhumieres2,lallemandluo}.  

\subsubsection{Application to the 2D MRT-LBM}\label{re:2d}

In this part, we  illustrate the algorithm presented
in Sec. \ref{subL-MRT-LBM} considering the 2D MRT-LBM. For the standard 2D
 MRT-LBM, the equilibrium distribution functions are described as
 follow (similary to the incompressible LBS)\cite{lallemandluo}
 \begin{equation}
 m^{\rm(eq)}=\left\{\rho,\ j_x,j_y,\ -2\rho+3\left(j_x^2+j_y^2\right),\ \rho-3\left(j_x^2+j_y^2\right),\ -j_x,\ -j_y,\left(j_x^2-j_y^2\right),\ j_xj_y\right\},
 \end{equation}
 where $j_x$ and $j_y$ denote the x-momentum and y-momentum
 respectively, and $\rho$ represents the density 
 (the reference density $\rho_0=1$, $W_0=m_0=\rho,W_1=m_1=j_x=u,W_2=m_2=j_y=v$).  The corresponding matrix $\Psi $ is given in \ref{App:d2q9}. The diagonal elements of the corresponding diagonal matrix $S$ are
set as follows
\begin{equation}
s_0=s_1=s_2=0,s_3=s_e,s_4=s_\epsilon,s_5=s_6=s_q,s_7=s_8=s_\nu.
\end{equation}
For the original MRT-LBM \cite{lallemandluo}, only $s_\nu$ is a free
parameter, and $s_e=1.64$, $s_\epsilon=1.54$, $s_q=1.9$. The
analogous form of $\Psi $ can be found in the existent literature
\cite{lallemandluo,duboislallemand}. The derivation
of $\Psi$ can be achieved by means of the first-order Taylor series expansion with
respect to $\rho$, $j_x$ and $j_y$ at reference states \cite{simondenispierre,lallemandluo,xusagaut}. In the expression
of $\Psi$, $(U,V)$ to denote the mean flow velocity components. In order to express the L-NSE by intuitive parameters, we introduce the following relation
\begin{equation}\label{sigma}
\sigma_\eta=\frac{1}{s_\eta}-\frac{1}{2},
\end{equation}
where $\eta$ stands for any notations in the set
$\Xi=\{e,\epsilon,q,\nu\} $. Now, when the truncated error term is equal to $O(\delta t^2)$ , the coefficient matrix $B^{(2)}$ with the mean flow can be described
by the summation of two matrices (the coefficient matrices of $\delta^0$ and $\delta t$) \cite{xusagaut}. These two matrices  describe the specific terms in the linearized Navier-Stokes equations. The coefficient matrix associated with $\delta t^0$ is
\begin{equation}
 \left[ \begin {array}{rrr} 
0&-{\rm i}{ k_x}&-{\rm i}{ k_y}
\\\noalign{\medskip}-\frac{1}{3}\,{\rm i}{ k_x}&-2\,{
\rm i}{ k_x}\,U-{\rm i}{ k_y}\,V&-{\rm i}{ k_y}\,U\\\noalign{\medskip}-\frac{1}{3}\,{
\rm i}{ k_y}&-{\rm i}{ k_x}\,V&-2\,{\rm i}{ k_y}\,V-{\rm i}{ k_x}\,U
\end {array} \right] 
\end{equation}
The coefficient matrix associated with $\delta t$ is given in \ref{App:firstorder2d}.
The higher-order truncated error matrices of $\delta t$ are omited for the sake of brevity (see \cite{xusagaut} for more detailed descriptions). 

\subsubsection{Application to the 3D MRT-LBM}\label{re:3d}

In this part, the algorithm presented in Sec.\ref{subL-MRT-LBM} will  be used to recover the 3D L-NSE. For the sake of convenience, we consider the standard 3D MRT-LBM and the equilibrium distribution functions with 15 discrete velocities are given  by \cite{dhumieres2,lallemandluo}
\begin{equation}
     m^{\rm(eq)}=\left\{\rho,j_x, j_y, j_z,  -\rho+\left(j_x^2+j_y^2+j_z^2\right), -\rho, -\frac{7}{3} j_x, -\frac{7}{3} j_y,  -\frac{7}{3} j_z, 2 j_x^2-j_y^2-j_z^2,  j_y^2-j_z^2,  j_xj_y,  j_yj_z,   j_xj_z,  0, \right\}.
\end{equation}
where $j_x$, $j_y$ and $j_z$ denote the x-momentum and y-momentum
 respectively, and $\rho$ represents the density 
 (the reference density $\rho_0=1$, $W_0=m_0=\rho,W_1=m_1=j_x=u,W_2=m_2=j_y=v, W_3=m_3=j_z=w$). 
The relaxation parameters corresponding to the matrix (\ref{S_relaxation}) is defined by 
\begin{equation}
\left\{0,0,0,0,s_4,s_5,s_6,s_7,s_8,s_9,s_{10},s_{11},s_{12},s_{13},s_{14}\right\},
\end{equation}
where
\begin{equation}
s_4=s_e, s_5=s_\epsilon, s_6 = s_7 = s_8=s_q, s_9 = s_{10} = s_{11} = s_{12} = s_{13}=s_\nu, s_{14}=s_t.
\end{equation}
For the original MRT-LBM \cite{dhumieres2,lallemandluo}, $s_\nu$ is a free paramter and $s_e=1.6$, $s_\epsilon=1.2$, $s_q=1.6$ , $s_t=1.2$ \cite{dhumieres2}. 
For convenience, we redefined an enlarged set $\Xi=\{e, \epsilon, q, \nu,  t\}$. The cooresponding defination (\ref{sigma}) can also be used for 3D MRT-LBM. When the truncated error term is equal to O($\delta t^2$), the coefficient matrix $B^{(2)}$ with the mean flow can be expressed by the summation of two matrices. The coefficient matrix corresponding to $\delta t^0$ is
\begin{equation}
\left[ \begin {array}{rrrr} 0&-{\rm i}{ k_x}&-{\rm i}{ k_y}&-{\rm i}{ k_z}
\\\noalign{\medskip}-1/3\,{\rm i}{ k_x}&-2\,{\rm i}{ k_x}\,U-{\rm i}{ k_z}\,W-{\rm i}{
 k_y}\,V&-{\rm i}{ k_y}\,U&-{\rm i}{ k_z}\,U\\\noalign{\medskip}-1/3\,{\rm i}{ 
k_y}&-{\rm i}{ k_x}\,V&-2\,{\rm i}{ k_y}\,V-{\rm i}{ k_x}\,U-{\rm i}{ k_z}\,W&-{\rm i}{ k_z
}\,V\\\noalign{\medskip}-1/3\,{\rm i}{ k_z}&-{\rm i}{ k_x}\,W&-{\rm i}{ k_y}\,W&-i
{ k_x}\,U-{\rm i}{ k_y}\,V-2\,{\rm i}{ k_z}\,W\end {array} \right] 
\end{equation}
The coefficient matrix  corresponding to $\delta t$ is given in \ref{App:d3q19} .

In Sec. \ref{re:2d} and Sec \ref{re:3d}, the recovered L-NSE are given and the corresponding truncated error  is up to $\delta t^2$. In order to save space, the matrices $B^{(n)}$ with the higher-order truncated terms of  $\delta t^n$ ($n>2$) are not given in this paper.  For readers, by the algorithm given in Sec. \ref{subL-MRT-LBM} and Maple symbolic calculations, it is easy to obtain the matrix $B^{(n)}$ with any value of $n$.

From the matrices corresponding to the dissipation, it is clear that the dissipation action of the recovered L-NSE is dependent on the mean flow.  This dependency always leads  to the breakdown of Galilean invariance. This point remains an open issue to the knowledge of the authors. 

\section{Sensitivity and stability analysis of free parameters in MRT-LBM}\label{sec:sensitivity}

In this section, the sensitivity  of the free parameters and the  stability for the MRT-LBM will be addressed by means of the first/second-order sensitivity of the eigenvalues  \cite{stewartsun} and the distribution of the matrix eigenvalues in the complex plane, respectively.  
The free relaxation parameters in $\Xi$ determined by the linear analysis in the wave-number spaces  are ``sub-optimal"  \cite{dhumieres2,lallemandluo}.  However, the original sub-optimal free relaxation parameters are still used for  the simulations  of fluid flows in open literature. Because of the linear character of the analysis, the influence of the free relaxation parameters on the hydrodynamic modes is still not completely clarified. 
In this part, all of these will be analyzed. The analysis offers us a new methodology  to propose our simplified optimization strategies for determining the free relaxation parameters.

\subsection{Methodology for sensitivity and stability analysis}

In this part, the methods for investigating the sensitivity and stability are presented. 

\subsubsection{The first/second-order derivatives of matrix engenvalues with respect to free parameters}\label{fisrtsecondder}

For the convenience of analysis, let  $\Upsilon$ denotes the free parameter set for both 2D and 3D problems. 
From Eq. (\ref{sec:l-mrt}), the linear behavior of MRT-LBM is fully determined by the eigvalues of the matrix $M^{\rm mrt}$.  
Let $\rm X_{\it M}=\{x_i\}_{0\leq i\leq N}$ and $\Lambda_{M}=\{\lambda_i\}_{0\leq i\leq N}$  denote the eigenvector set  and the eigenvector set of the matrix $M^{\rm mrt}$, respectively.  It is clear that $\rm X_{\it M}$ and $\Lambda_{M}$ are function of $\Upsilon$. Let $\varepsilon\in \Upsilon$ denote the the concerned parameter which exists in the matrix $M^{\rm mrt}$. 
Because the eigenvalues of the matrix $M^{\rm mrt}$ are distinct from each other ($|\bf k|>0$), the  first-order sensitivity method can be used to address the linear response behaviors or the group velocity behaviors.  
According to the definition of the eigenvalue problem, the following equations are satisfied

\begin{equation}
M^{\rm mrt}(\varepsilon){\rm x}_i(\varepsilon)=\lambda_i(\varepsilon){ \rm x}_i(\varepsilon).
\end{equation}
Differentiating with repect to $\varepsilon$ yields
\begin{equation}
M^{\rm mrt}_{\varepsilon}(\varepsilon){\rm x}_i(\varepsilon)+M^{\rm mrt}(\varepsilon){\rm x}_i^{\prime}(\varepsilon)=\lambda_i^{\prime}(\varepsilon){\rm x}_i(\varepsilon)+\lambda_i(\varepsilon){\rm x}_i^{\prime}(\varepsilon),
\end{equation}
where $M^{\rm mrt}_{\varepsilon}(\varepsilon)=\partial_{\varepsilon}M^{\rm mrt}(\varepsilon)$.
If we define ${\rm Y}_{M}^H={\rm X}_{M}^{-1}=\{{\rm y}_i^H\}_{0\leq i\leq d+1}$ ($H$ denotes the conjugate transpose) as the left eigenvector set corresponding to $\Lambda_M$,  we obtain

\begin{equation}\label{matrixder}
{\rm Y}_{M}^H(\varepsilon)M^{\rm mrt}_{\varepsilon}(\varepsilon){\rm X}_{M}(\varepsilon)+{\rm Y}_{M}^H(\varepsilon)M^{\rm mrt}(\varepsilon){\rm X}_{M}^{\prime}(\varepsilon)={\Lambda}_{M}^{\prime}(\varepsilon)+{\rm Y}_{M}^H(\varepsilon){\rm X}^{\prime}(\varepsilon)\Lambda_{M}(\varepsilon).
\end{equation}
Considering the diagonal elements in Eq. (\ref{matrixder}), we have
\begin{equation}\label{firstder}
{\rm y}_i^H(\varepsilon)M^{\rm mrt}_{\varepsilon}(\varepsilon){\rm x}_i(\varepsilon)+{\rm y}_i^H(\varepsilon)M^{\rm mrt}(\varepsilon){\rm x}_i^{\prime}(\varepsilon)=\lambda_i^{\prime}(\varepsilon){\rm y}_i^H(\varepsilon){\rm x}_i(\varepsilon)+\lambda_i(\varepsilon){\rm y}_i^H(\varepsilon){\rm x}_i^{\prime}(\varepsilon).
\end{equation}
Since ${\rm y}_i^H{\rm x}_i=1$, we get
\begin{equation}\label{diag}
\lambda_i^{\prime}(\varepsilon)={\rm y}_i^H(\varepsilon)M^{\rm mrt}_{\varepsilon}(\varepsilon){\rm x}_i(\varepsilon).
\end{equation}
It is clear that the derivatives of eigenvalues with respect to the free parameters can be expressed by the right and left eigenvectors  and the derivative of the matrix $M^{\rm mrt}$. The first-order derivative of eigenvectors can be described by \cite{AndrewTan,aamorsche}
\begin{equation}
{\rm x}_i^{\prime}(\varepsilon)=C_{ij}{\rm x}_j(\varepsilon),  {\rm y}_i^{H {\prime}}(\varepsilon)=-C_{ji}{\rm y}_j(\varepsilon)\end{equation}
where the matix $C$ is defined by
\begin{equation}
C_{ij}=\left\{\begin{array}{ll}
{\rm y}_j^H(\varepsilon)M^{\rm mrt}_{\varepsilon}(\varepsilon){\rm x}_i(\varepsilon)/(\lambda_i(\varepsilon)-\lambda_j(\varepsilon)),& i\neq j\\
0,& i=j
\end{array}\right.
\end{equation}
 From Eq. (\ref{diag}),  the second-order derivative can be described by \cite{AndrewTan,aamorsche}
\begin{equation}
\lambda_i^{\prime\prime}(\varepsilon)={\rm y}_i^H(\varepsilon)M^{\rm mrt}_{\varepsilon^2}(\varepsilon){\rm x}_i(\varepsilon)+{\rm y}_i^{H{\prime}}(\varepsilon)M^{\rm mrt}_{\varepsilon}(\varepsilon){\rm x}_i(\varepsilon)+{\rm y}_i^H(\varepsilon)M^{\rm mrt}_{\varepsilon}(\varepsilon){\rm x}_i^{\prime}(\varepsilon).
\end{equation}
From Eq. (\ref{sec:l-mrt}), the following relation between $\omega=\{\omega_i\}_{0\leq i\leq N}$ and $\Lambda_{M}$ holds
\begin{equation}
{\rm exp}(-{\rm i}\omega_i(\varepsilon))=\lambda_{i}(\varepsilon).
\end{equation}
So, the first and second order derivatives of $\omega_i(\varepsilon))$ can be expressed by
\begin{equation}
\omega_i^{\prime}(\varepsilon)={\rm i}\lambda_{i}^{\prime}(\varepsilon)/\lambda_{i}(\varepsilon),\ \omega_i^{\prime\prime}(\varepsilon)={\rm i}\lambda_{i}^{\prime\prime}(\varepsilon)/\lambda_{i}(\varepsilon)-{\rm i}\lambda_{i}^{\prime}(\varepsilon)^2/\lambda_{i}(\varepsilon)^2.
\end{equation}
Now, the first/second-order sensitivity of free parameters to $\omega_i(\cdot)$ can be investigated easily. The behavior of the numerical group velocity of the MRT-LBM can be addressed.
\subsubsection{Hydrodynamic stability in the complex planes}\label{pseudo}
 
 It is known that the changes of any free parameters always result in the variation of the matrix $M^{\rm mrt}$. 
 By investigating the eigenvalue distribution of the matrix $M^{\rm mrt}$ in the complex planes, the basic sensitivity behaviors of the L-MRT-LBM can be revealed.  
 According to Eq. (\ref{sec:l-mrt}),  the frequency $\omega$ and the eigenvalues of the matrix  $M^{\rm mrt}$ satisfy the following relation
\begin{equation}\label{stab:fre}
\omega_{i}={\rm i}\cdot{\rm Log}(\lambda_{i}).
\end{equation}
The stability of MRT-LBM requires that the imaginary part of $\omega_i$ should be smaller than or equal to 0.  It is well-known that $\lambda_i$ could be rewritten as follows
\begin{equation}
\lambda_{i}=|\lambda_i|({\rm cos}(\theta)+{\rm i\cdot sin}(\theta))=|\lambda_i|{\rm exp}({\rm i}\cdot \theta),
\end{equation}
where $|\lambda_i|$ is the modulus of $\lambda_{i}$ and $\theta={\rm Arg}(\lambda_i)$ is the argument of $\lambda_i$. 
From Eq. (\ref{stab:fre}), the following relation can be obtained
\begin{equation}\label{stability:c}
\omega_{i}={\rm i}\cdot{\rm Log}(|\lambda_i|)-{\rm Arg}(\lambda_i).
\end{equation}
The stability requires that ${\rm Log}(|\lambda_i|)\leq 0$. That means all eigenvalues of $M^{\rm mrt}$ should lie within the unit circle the origin of which is located at the point $(0,0)$. The eigenvalue behavior of the matrix $M^{\rm mrt}$ with respect to the $\varepsilon$-parameter  reflects the sensitivity of free parameters. 

\subsection{Analysis of the sensitivities}\label{ch:sensitivities}
In this section, we consider the first- and second-order sensitivities of the eigenvalues with respect to the wave-number $\kappa$ and observe the real behaviors corresponding to $\nu$, $\eta$ and $c_s$ in the L-MRT-LBM. Finally, in the complex plane, according to the distribution of the eigenvalues of the matrix $M^{\rm mrt}$ and the stability condition, the influence of free parameters is investigated.

\subsubsection{The first- and second-order sensitivities with respect to  the wave-number magnitude $|{\bf k}|$}\label{ana:firstsec}
Although there exist several papers that are focused on the analysis of dispersion and dissipation relations, as far as authors know, the investigations about  dispersion and dissipation behaviors with respect to  $|{\bf k}|$ still haven't been  completely perfomed \cite{simondenispierre,ricotmariesagautbailly,lallemandluo} for the MRT-LBM. 

From Eq. (\ref{th:d}),  the first-order derivatives of acoustic modes (i.e. group velocity) $\omega^{\pm}({\rm k})$ and shear modes $\omega^{s}({\bf k})$ with respect to $\kappa=|{\bf k}|$ are given as follows ($\widehat{\bf k\cdot u}$ is the  angle  between $\bf k$ and $\bf u$)
\begin{equation}\label{th:d-der1}
\left\{\begin{array}{ll} Re[\omega^{\pm}_\kappa({\bf k})]&=\pm
c_s+|{\bf u}|{\rm
cos}(\widehat{\bf k\cdot u}),\\
Im[\omega^{\pm}_\kappa({\bf k})]&=-|{\bf
k}|\left(\frac{2d-2}{d}\nu+\eta\right),\\
Re[\omega^s_\kappa({\bf k})]&=|{\bf u}|{\rm cos}(\widehat{\bf k\cdot u}),\\
Im[\omega^s_\kappa({\bf k})]&=-2|{\bf k}|\nu.
\end{array}\right.
\end{equation}

 The second-order derivatives are given by
\begin{equation}\label{th:d-der2}
\left\{\begin{array}{ll} Re[\omega^{\pm}_{\kappa\kappa}({\bf k})]&=0,\\
Im[\omega^{\pm}_{\kappa\kappa}({\bf k})]&=-\left(\frac{2d-2}{d}\nu+\eta\right),\\
Re[\omega^s_{\kappa\kappa}({\bf k})]&=0,\\
Im[\omega^s_{\kappa\kappa}({\bf k})]&=-2\nu.
\end{array}\right.
\end{equation}
From Eqs. (\ref{th:d-der1}) and (\ref{th:d-der2}), we can detect the lattice sound speed $c_s$ (for $|{\bf u}|=0$ or $\widehat{\bf k\cdot u}=\pi/2$) and the effective shear and bulk viscosities. Theoretically, the dispersion and dissipation relations are dependent on $|{\bf k}|$, $|{\bf u}|$ and $\widehat{\bf k\cdot u}$ for the fixed $\nu$ and $\eta$, and independent on the individual orientations of ${\bf k}$ and ${\bf u}$. 

\begin{table}[!htbp]
\caption{Illustrative parameters for the investigations of the original D2Q9 MRT-LBM ($\delta t=1$): ${\bf e_x}$ denotes the unit vector (1,0) co-directional with x-axis. The symbol ``$\diagdown$" indicates that the corresponding parameters are determined according to the cases.}\label{table:1}
\begin{tabular*}{\textwidth}{@{\extracolsep{\fill}}cllllllll}\toprule[1pt]
${\rm Index}$&$s_{\nu}$& $s_e$ & $\nu$ & $\eta$ & $|{\bf u}|$& $\widehat{\bf k\cdot u}$ & $\widehat{\bf u\cdot e_x}$& $\widehat{\bf k\cdot e_x}$\\ \midrule[1pt]
0 & 1.9 & 1.64 & 8.771929833e-3 & 0.03658536587 & 0.0 & 0 &0&0\\
1 & 1.99 & 1.64 & 8.375209333e-4 & 0.03658536587 & 0.0 & 0 &0&0\\
2 & 1.999 & 1.64 & 8.337503333e-5 & 0.03658536587 & 0.0 & 0 &0&0\\
3 & 1.9999 & 1.64 & 8.333733333e-6 & 0.03658536587 & 0.0 & 0 &0&0\\
4 & 1.99 & 1.64 & 8.375209333e-4 & 0.03658536587 & 0.1 & $\diagdown$ &$\diagdown$&$\diagdown$\\
5 & 1.9999 & 1.64 & 8.333733333e-6 & 0.03658536587 & 0.1 & $\diagdown$ &$\diagdown$&$\diagdown$
\\
\bottomrule[1pt]
\end{tabular*}
\end{table}

 \begin{figure}[!htbp]
\begin{center}
\scalebox{0.45}[0.45]{\includegraphics[angle=0]{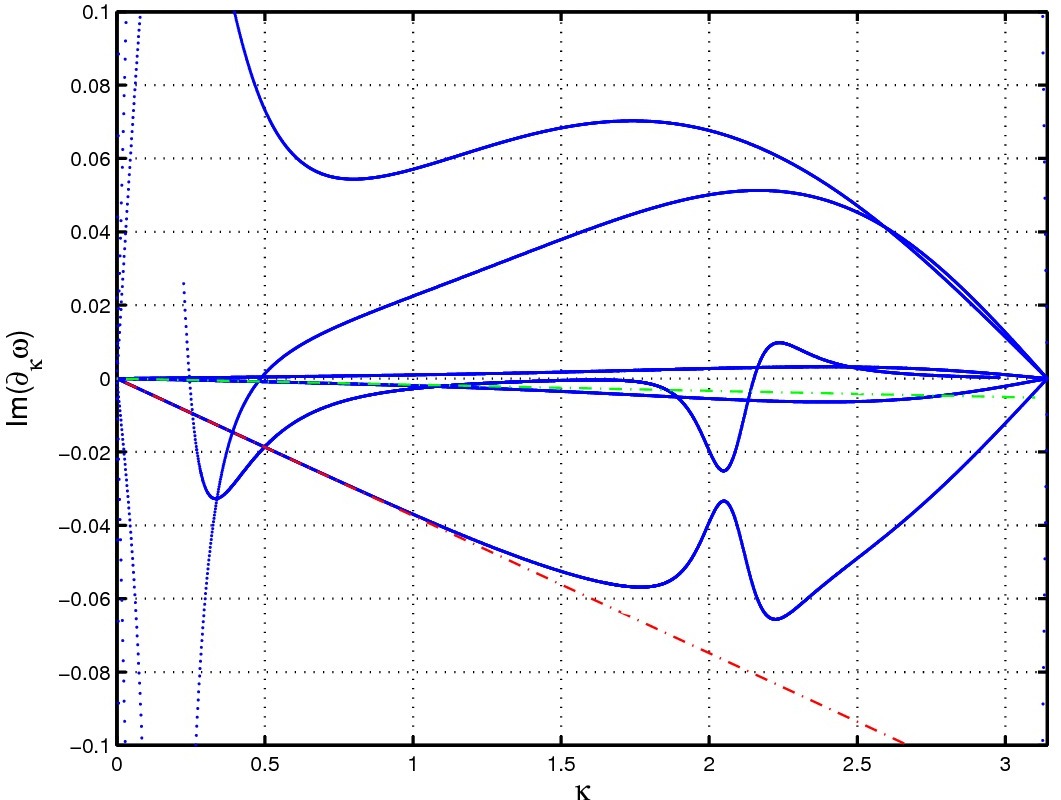}}
\scalebox{0.45}[0.45]{\includegraphics[angle=0]{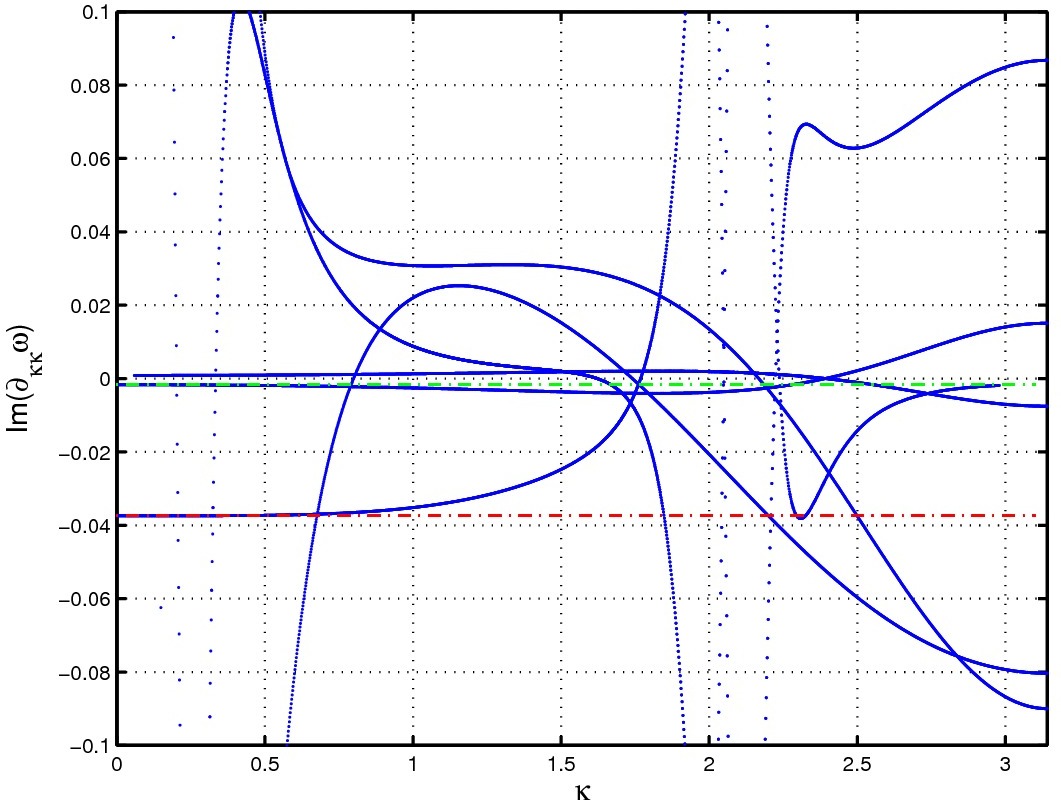}}\\
\caption{Local information for derivatives of $Im[\omega]$: (Left): $Im[\omega_{\kappa}]$ ; (Righ): $Im[\omega_{\kappa\kappa}]$.   In the left subfigure, Green line denotes $Im[\omega^{s}_{\kappa}]$; Red  line denotes $Im[\omega^{\pm}_{\kappa}]$.  In the right subfigure, Green  lines denote $Im[\omega^{\pm}_{\kappa\kappa}]$; Red  line denotes $Im[\omega^{s}_{\kappa\kappa}]$. The parameters are from the index 1 of Tabel \ref{table:1}.}\label{fig:3}
\end{center}
\end{figure}
 \begin{figure}[!htbp]
\begin{center}
\scalebox{0.45}[0.45]{\includegraphics[angle=0]{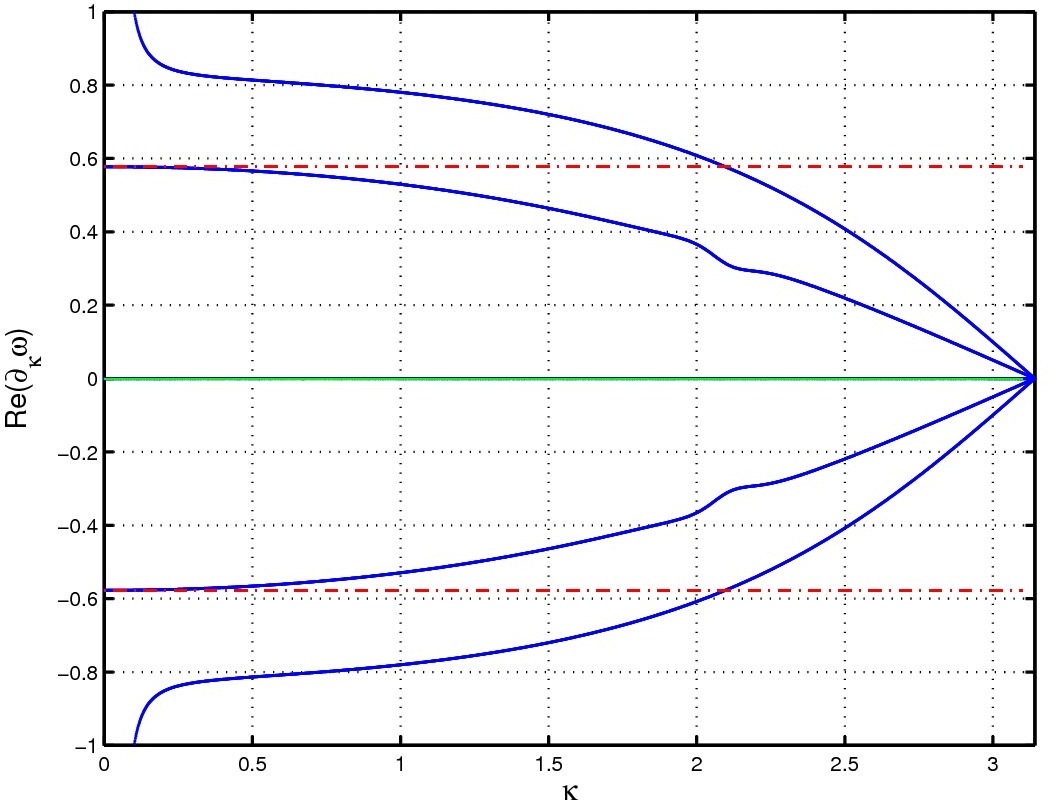}}
\scalebox{0.45}[0.45]{\includegraphics[angle=0]{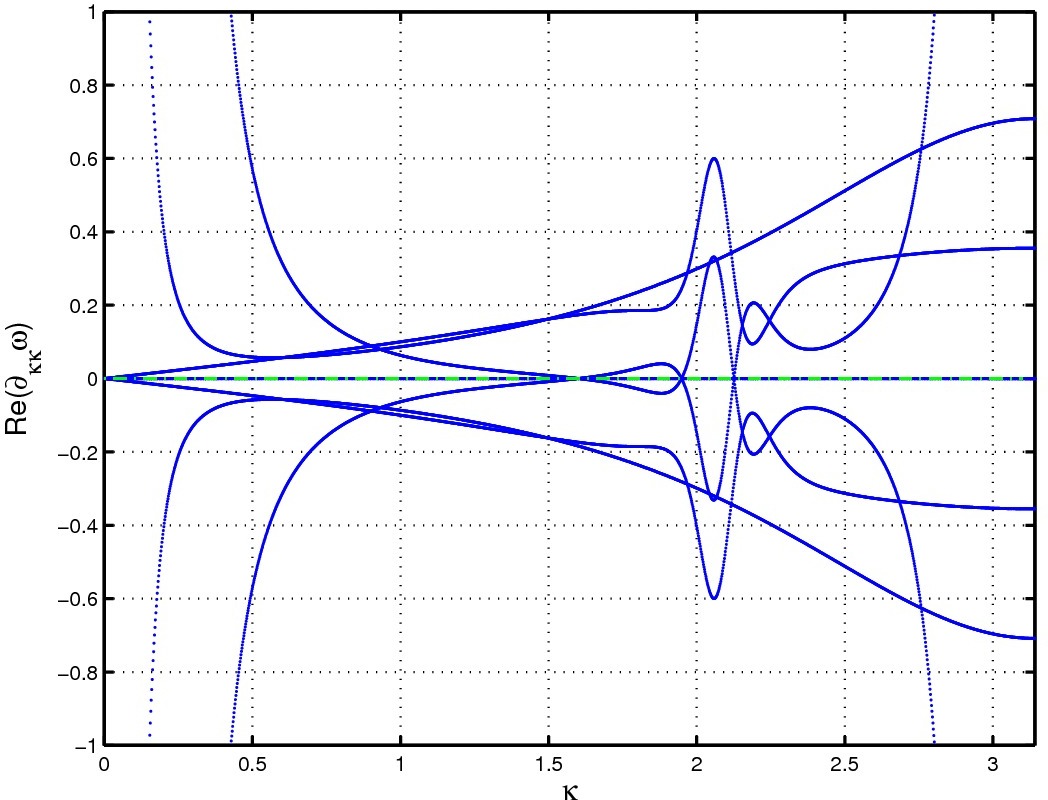}}\\
\caption{Local information for derivatives of $Re[\omega]$: (Left): $Re[\omega_{\kappa}]$; (Righ): $Re[\omega_{\kappa\kappa}]$. In the left subfigure, Green  lines denote $Re[\omega^{s}_{\kappa}]$; Red  line denotes $Re[\omega^{\pm}_{\kappa}]$.  In the right subfigure, Green  lines denote $Re[\omega^{\pm}_{\kappa\kappa}]$; Red  line denotes $Re[\omega^{s}_{\kappa\kappa}]$. The parameters are from the index 1 of Tabel \ref{table:1}.}\label{fig:4}
\end{center}
\end{figure}

 In order to observe the sensitivity of the hydrodynamic modes with respect to  the wave-number $\kappa$, the locally-magnified profiles are given in Figs. \ref{fig:3} and \ref{fig:4}.  
 In the left sub figure of Fig. \ref{fig:3}, the green profile  stands for  $Im[\omega_\kappa^{s}]$ and the red profile  denotes $Im[\omega_\kappa^{\pm}]$. 
 It is noted that the  acoustic dissipation has a significant change around $\kappa=2$. 
 When $\kappa$ is larger than about 2.25,  $Im[\omega_\kappa^{\pm}]$ grows very fastly.  
 From the right sub-figure of Fig. \ref{fig:3},  when $\kappa$ is smaller than about 2.4, the shear viscosity $\nu$ in the L-MRT-LBM will be smaller than the given $\nu$ and becomes negative.  
 This means that the real behavior of $\nu$ and $\eta$  depends on the wavenumber $\kappa$. 
 Meanwhile, the important fact is that the computed $Im[\omega^{\pm}_{\kappa\kappa}]  $ is varies from negative to positive values when $\kappa$ is larger that about 1.75. 
 This phenomenon means that  $\nu+\eta$ is changes from positive to negative values after $\kappa=1.75$ and the instability will appear for acoustic modes.   In Fig. \ref{fig:4}, the profiles of the left subfigure denote  the numerical $Re[\omega_{\kappa}]$. The red lines are the exact  $Re[\omega^{\pm}_{\kappa}]$. It is clear that when the second term of $Re[\omega^{\pm}_{\kappa}]$ in Eq. (\ref{th:d-der1}),  the red lines are the exact profiles of $\pm c_s$. From the subfigure, the computed sound speed $c_s$ is dependent on the wavenumber $\kappa$ under the condition of zero-mean flow. 
The green line indicates the exact $Re[\omega^{s}_{\kappa}]$. 
It is clear that the computed $Re[\omega^{s}_{\kappa}]$ coincides with the exact $Re[\omega^{s}_{\kappa}]$ very well. 
From the right sub-figure in Fig. \ref{fig:4}, we can see that computed $Re[\omega^{s}_{\kappa}]$ is exact for zero-mean flows. And the computed $Re[\omega^{\pm}_{\kappa}]$ is sensitive to the wave-number $\kappa$. Especially, when the wave-number $\kappa$ is around 2,  there exists some significant changes of  the computed $Re[\omega^{\pm}_{\kappa}]$.  These results means the the numerical group velocity become very sensitive around $\kappa=2$. 

 \begin{figure}[!htbp]
\begin{center}
\scalebox{0.45}[0.45]{\includegraphics[angle=0]{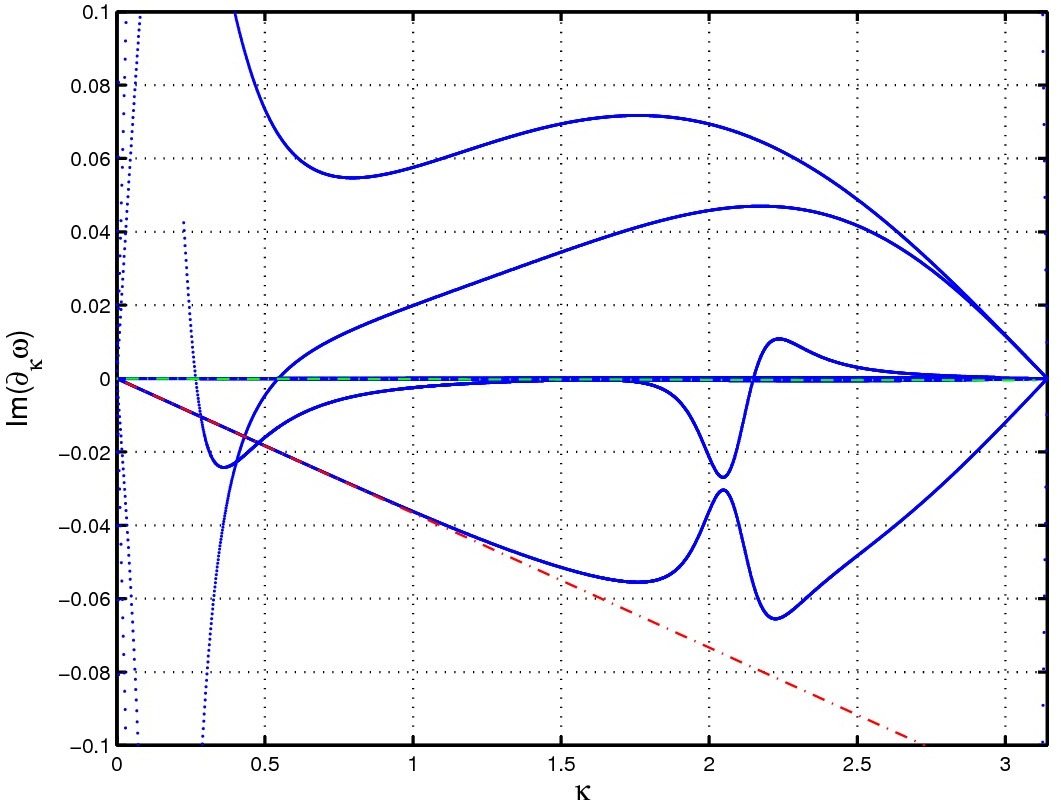}}
\scalebox{0.45}[0.45]{\includegraphics[angle=0]{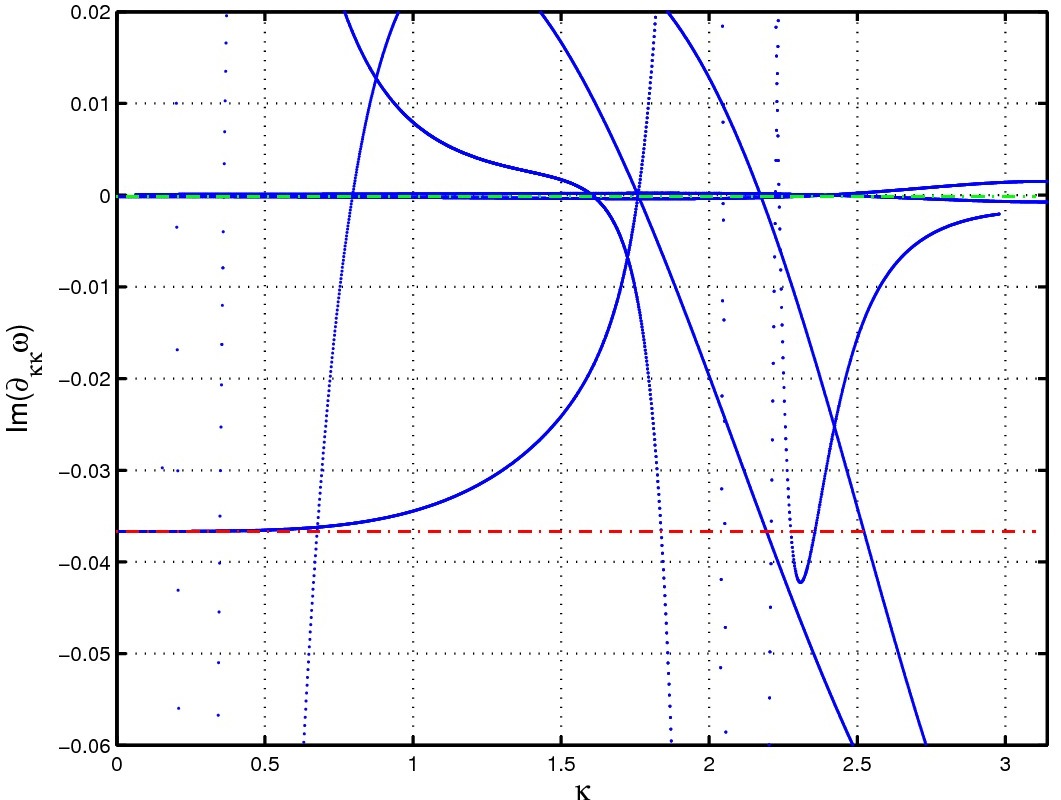}}\\
\caption{Local information for derivatives of $Im[\omega]$: (Left): $Im[\omega_{\kappa}]$ ; (Righ): $Im[\omega_{\kappa\kappa}]$.   In the left subfigure, Green line denotes $Im[\omega^{s}_{\kappa}]$; Red  line denotes $Im[\omega^{\pm}_{\kappa}]$.  In the right subfigure, Green  lines denote $Im[\omega^{\pm}_{\kappa\kappa}]$; Red  line denotes $Im[\omega^{s}_{\kappa\kappa}]$. The parameters are from the index 2 of Tabel \ref{table:1}.}\label{fig:5}
\end{center}
\end{figure}
 \begin{figure}[!htbp]
\begin{center}
\scalebox{0.45}[0.45]{\includegraphics[angle=0]{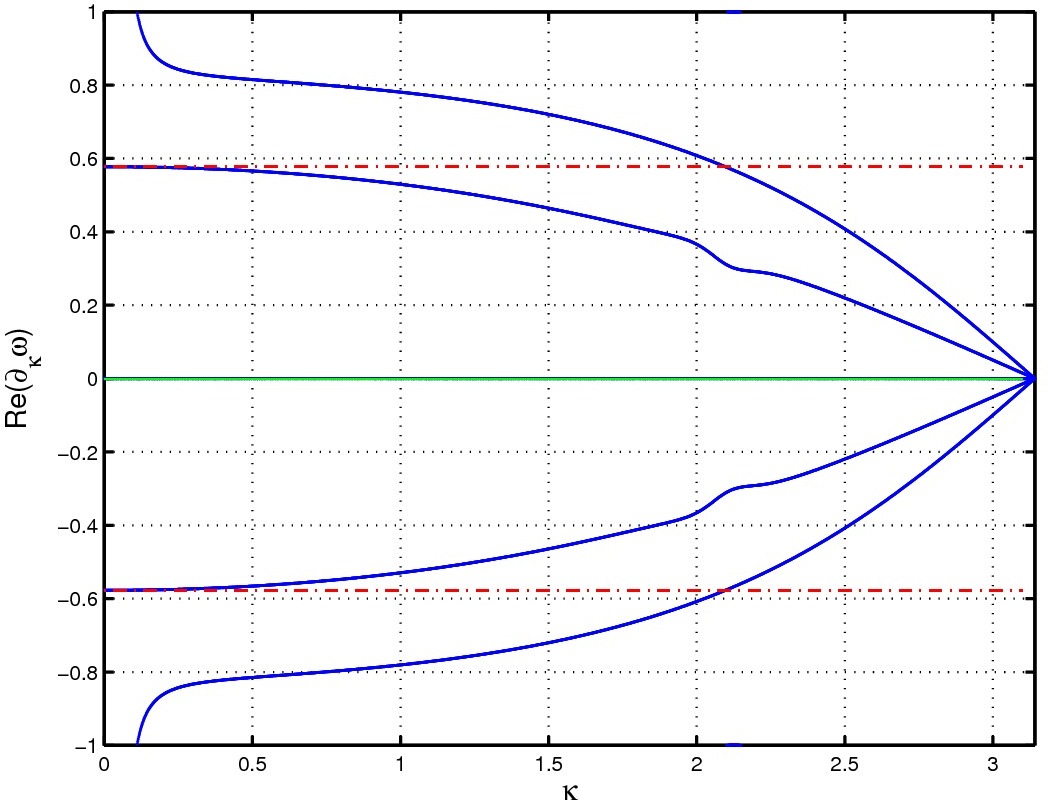}}
\scalebox{0.45}[0.45]{\includegraphics[angle=0]{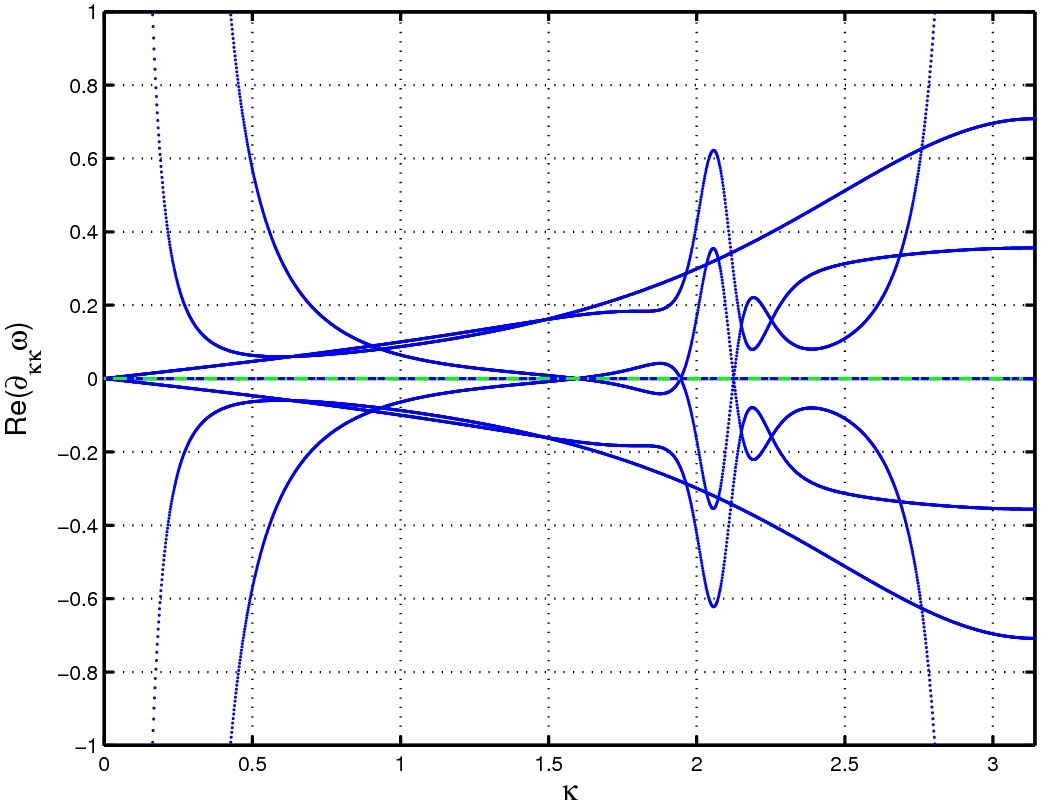}}\\
\caption{Local information for derivatives of $Re[\omega]$: (Left): $Re[\omega_{\kappa}]$; (Righ): $Re[\omega_{\kappa\kappa}]$. In the left subfigure, Green  lines denote $Re[\omega^{s}_{\kappa}]$; Red  line denotes $Re[\omega^{\pm}_{\kappa}]$.  In the right subfigure, Green  lines denote $Re[\omega^{\pm}_{\kappa\kappa}]$; Red  line denotes $Re[\omega^{s}_{\kappa\kappa}]$. The parameters are from the index 2 of Tabel \ref{table:1}.}\label{fig:6}
\end{center}
\end{figure}

In order to observe the impact of the influence of the small shear viscosity on the sensitivities, in Figs. \ref{fig:5}$\sim$\ref{fig:8}, the profiles of the numerically-computed $Im[\omega_{\kappa}]$, $Im[\omega_{\kappa\kappa}]$, $Re[\omega_{\kappa}]$ and $Re[\omega_{\kappa\kappa}]$ are displayed for $s_\nu=1.999$ (indicated by the index 2 in Table \ref{table:1}) and $s_\nu=1.9999$  (indicated by the index 3 in Table \ref{table:1}). Clearly, the similar results are obtained compared with the results ( $s_\nu=1.99$, in the index 1 in Table \ref{table:1}).  Now, we choose a smaller $s_\nu=1.9$ corresponding to the larger shear viscosity (indicated by the index 0 in Table \ref{table:1}). From  Figs. \ref{fig:9}$\sim$\ref{fig:10}, it is observed that when $\kappa\leq 2$, there does not exist any significant change for the sensitivities of the bulk viscosity $\nu+\eta$ and the lattice sound speed $c_s$. However, Fig. \ref{fig:9} shows that there exist some large deviations between the exact relations and the computed relations for $Im[\omega_{\kappa}^s]$ and $Im[\omega_{\kappa\kappa}^s]$. These deviations indicate that for  the large shear viscosity $\nu$, the real $\nu$ of the MRT-LBM deviates significantly from the exact $\nu$  in a large range of wave numbers.

\begin{table}[!htbp]
\caption{Illustrative parameters for investigations of the D3Q19 MRT-LBM ($\delta t=1$): ${\bf e_x}$ denotes the unit vector (1,0,0) codirectional with x-axis.  The symbol ``$\diagdown$ " indicates that the corresponding parameters are determined according to the cases. }\label{table:2}
\begin{tabular*}{\textwidth}{@{\extracolsep{\fill}}cllllllll}\toprule[1pt]
${\rm Index}$&$s_{\nu}$& $s_e$ & $\nu$ & $\eta$ & $|{\bf u}|$& $\widehat{\bf k\cdot u}$ & $\widehat{\bf u\cdot e_x}$& $\widehat{\bf k\cdot e_x}$\\ \midrule[1pt]
0 & 1.9 & 1.6 & 8.771929833e-3 & 0.02777777778 & 0.0 & 0 &0&0\\
1 & 1.99 & 1.6 & 8.375209333e-4 & 0.02777777778 & 0.0 & 0 &0&0\\
2 & 1.9999 & 1.6 & 8.333733333e-6 & 0.02777777778 & 0.0 & 0 &0&0\\
3 & 1.99 & 1.6 & 8.375209333e-4 & 0.02777777778 & 0.1 & $\diagdown$ &$\diagdown$&$\diagdown$\\
4 & 1.9999 & 1.6 &  8.333733333e-6 & 0.02777777778 & 0.1 & $\diagdown$ &$\diagdown$&$\diagdown$
\\
\bottomrule[1pt]
\end{tabular*}
\end{table}
 \begin{figure}[!htbp]
\begin{center}
\scalebox{0.45}[0.45]{\includegraphics[angle=0]{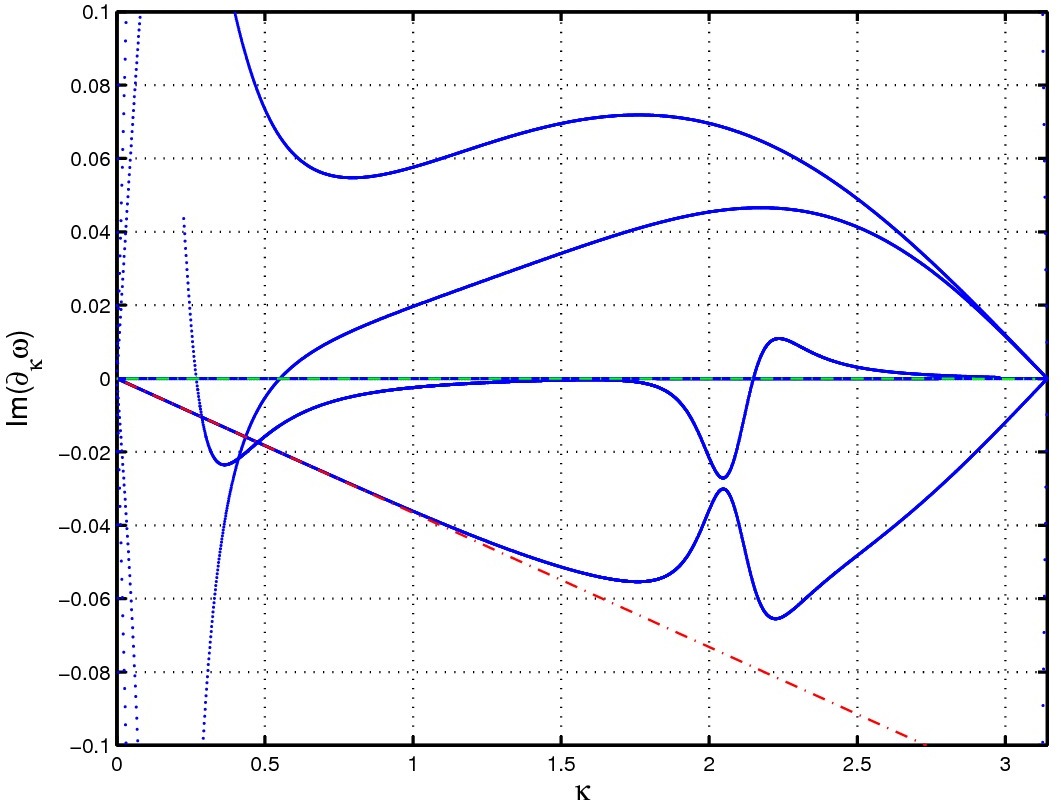}}
\scalebox{0.45}[0.45]{\includegraphics[angle=0]{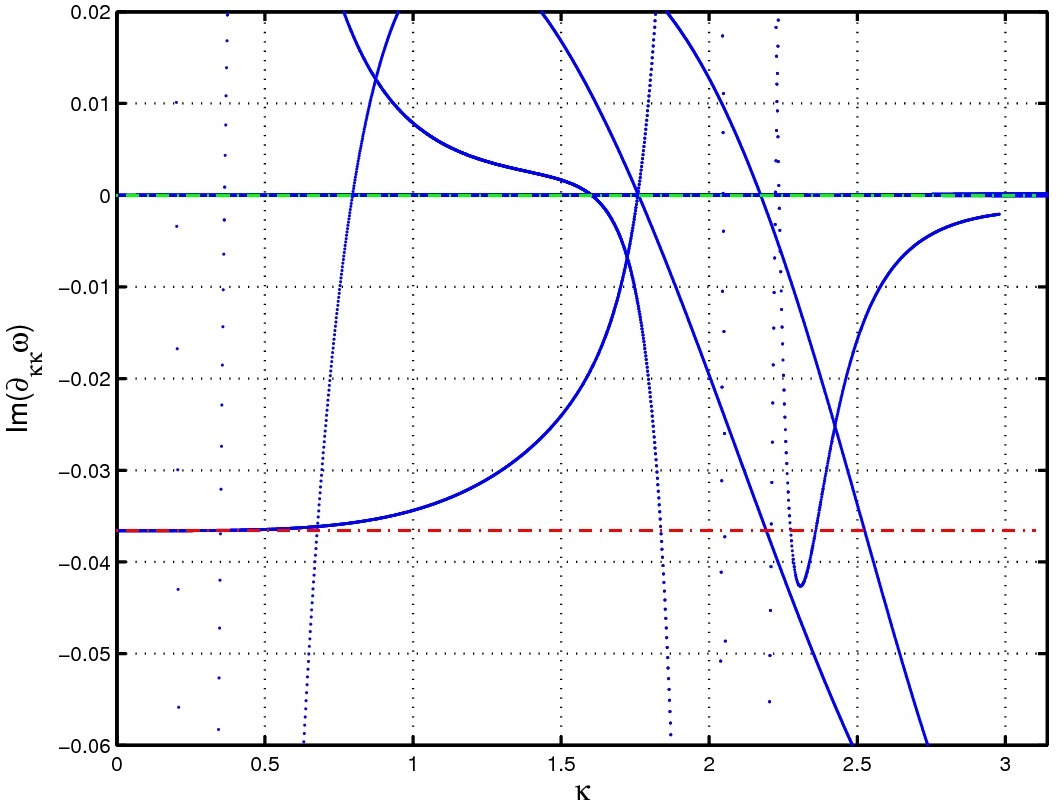}}\\
\caption{Local information for derivatives of $Im[\omega]$: (Left): $Im[\omega_{\kappa}]$ ; (Righ): $Im[\omega_{\kappa\kappa}]$.   In the left subfigure, Green line denotes $Im[\omega^{s}_{\kappa}]$; Red  line denotes $Im[\omega^{\pm}_{\kappa}]$.  In the right subfigure, Green  lines denote $Im[\omega^{\pm}_{\kappa\kappa}]$; Red  line denotes $Im[\omega^{s}_{\kappa\kappa}]$. The parameters are from the index 3 of Tabel \ref{table:1}.}\label{fig:7}
\end{center}
\end{figure}
 \begin{figure}[!htbp]
\begin{center}
\scalebox{0.45}[0.45]{\includegraphics[angle=0]{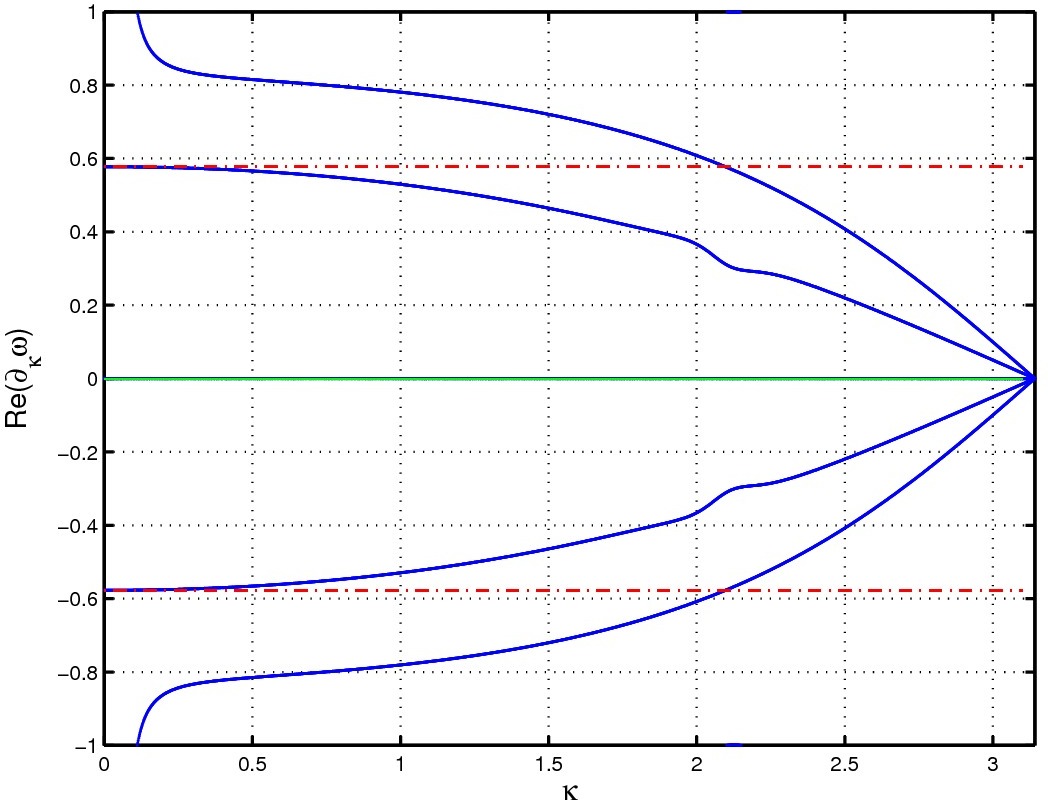}}
\scalebox{0.45}[0.45]{\includegraphics[angle=0]{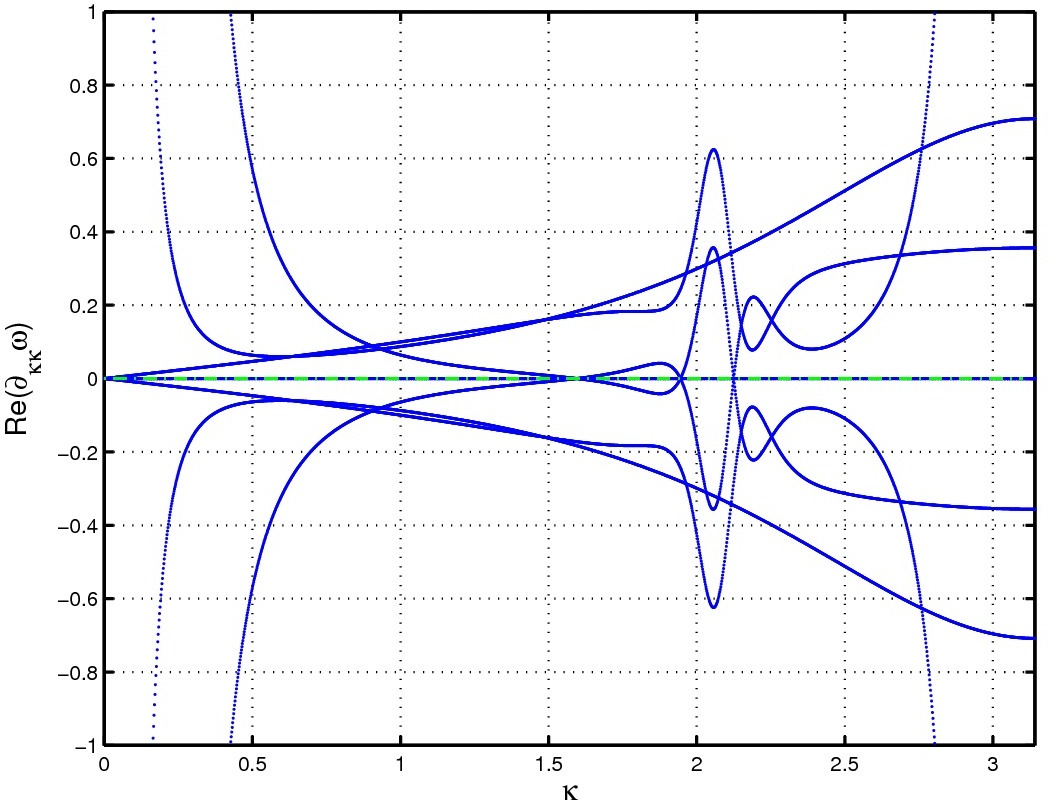}}\\
\caption{Local information for derivatives of $Re[\omega]$: (Left): $Re[\omega_{\kappa}]$; (Righ): $Re[\omega_{\kappa\kappa}]$. In the left subfigure, Green  lines denote $Re[\omega^{s}_{\kappa}]$; Red  line denotes $Re[\omega^{\pm}_{\kappa}]$.  In the right subfigure, Green  lines denote $Re[\omega^{\pm}_{\kappa\kappa}]$; Red  line denotes $Re[\omega^{s}_{\kappa\kappa}]$. The parameters are from the index 3 of Tabel \ref{table:1}.}\label{fig:8}
\end{center}
\end{figure}

 \begin{figure}[!htbp]
\begin{center}
\scalebox{0.45}[0.45]{\includegraphics[angle=0]{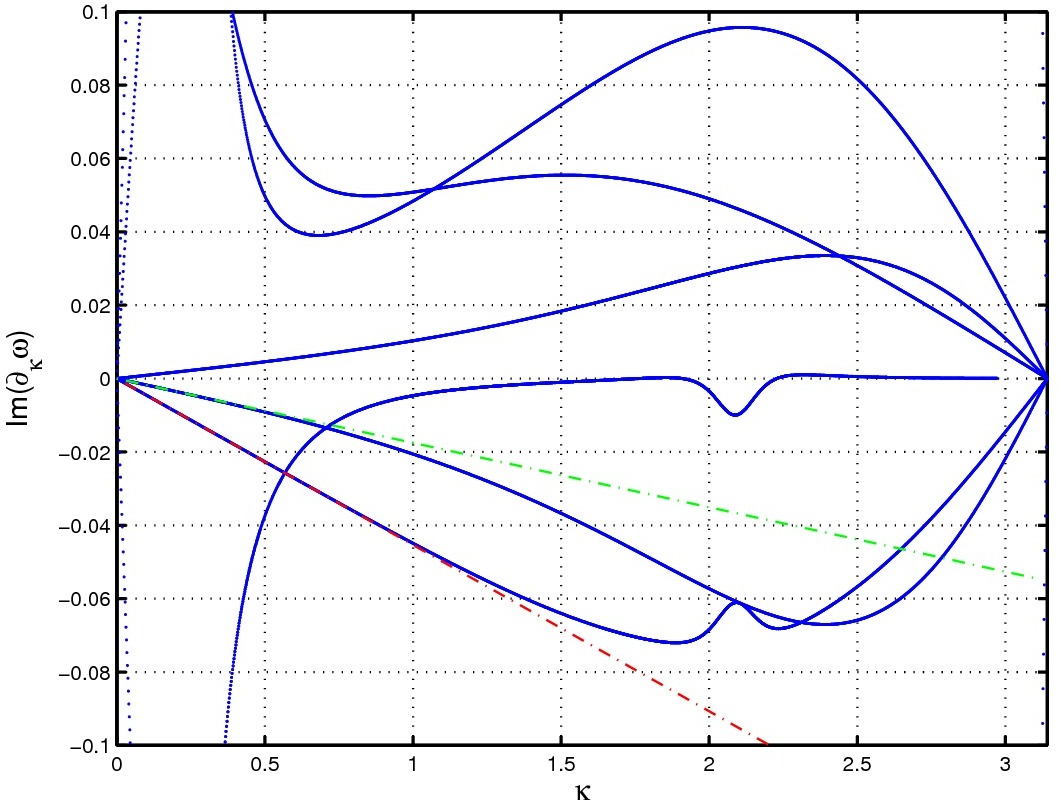}}
\scalebox{0.45}[0.45]{\includegraphics[angle=0]{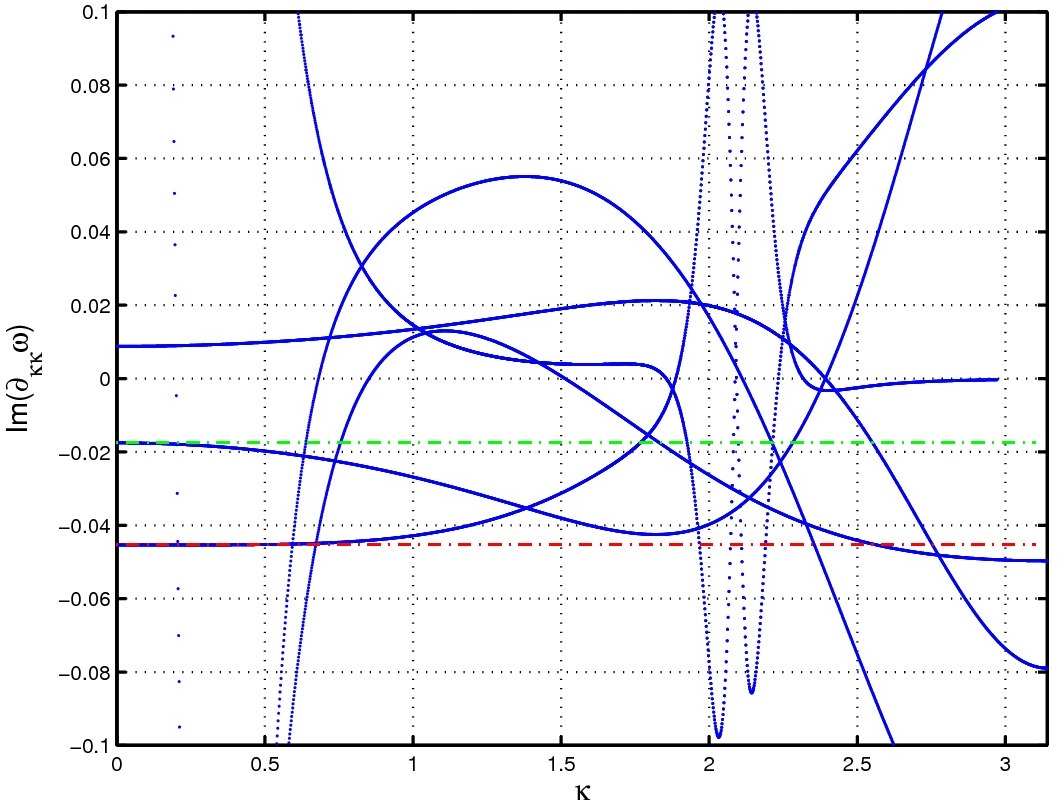}}\\
\caption{Local information for derivatives of $Im[\omega]$: (Left): $Im[\omega_{\kappa}]$ ; (Righ): $Im[\omega_{\kappa\kappa}]$.   In the left subfigure, Green line denotes $Im[\omega^{s}_{\kappa}]$; Red  line denotes $Im[\omega^{\pm}_{\kappa}]$.  In the right subfigure, Green  lines denote $Im[\omega^{\pm}_{\kappa\kappa}]$; Red  line denotes $Im[\omega^{s}_{\kappa\kappa}]$. The parameters are from the index 0 of Tabel \ref{table:1}.}\label{fig:9}
\end{center}
\end{figure}
 \begin{figure}[!htbp]
\begin{center}
\scalebox{0.45}[0.45]{\includegraphics[angle=0]{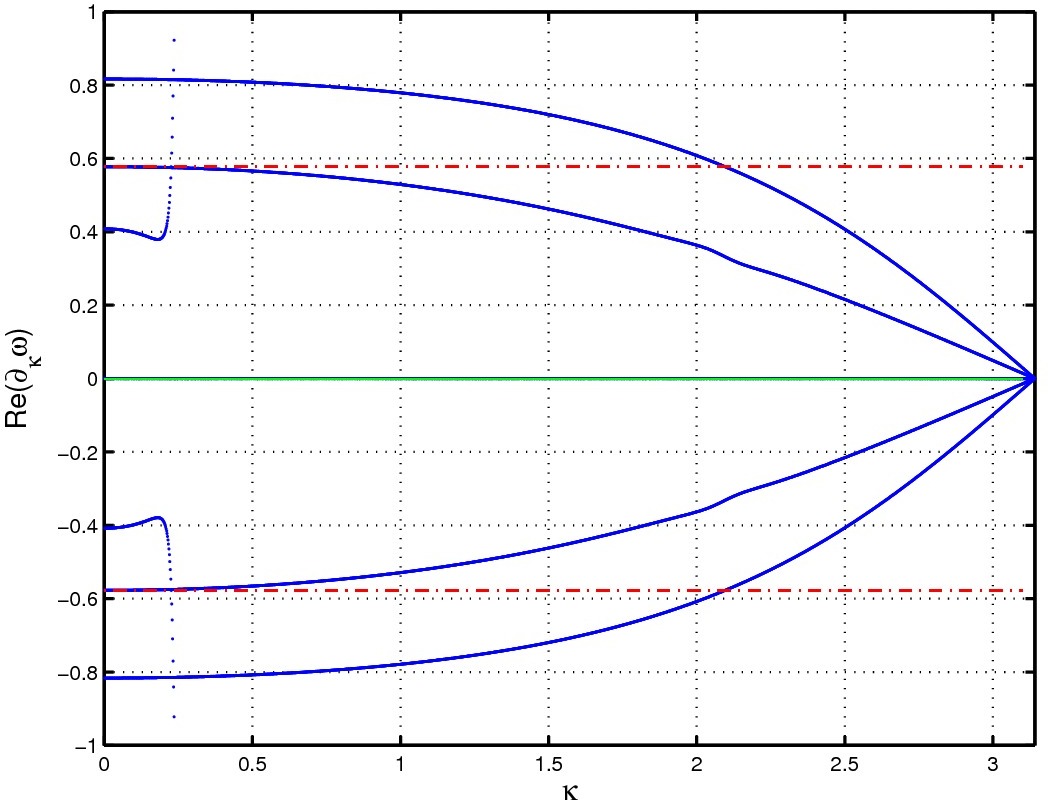}}
\scalebox{0.45}[0.45]{\includegraphics[angle=0]{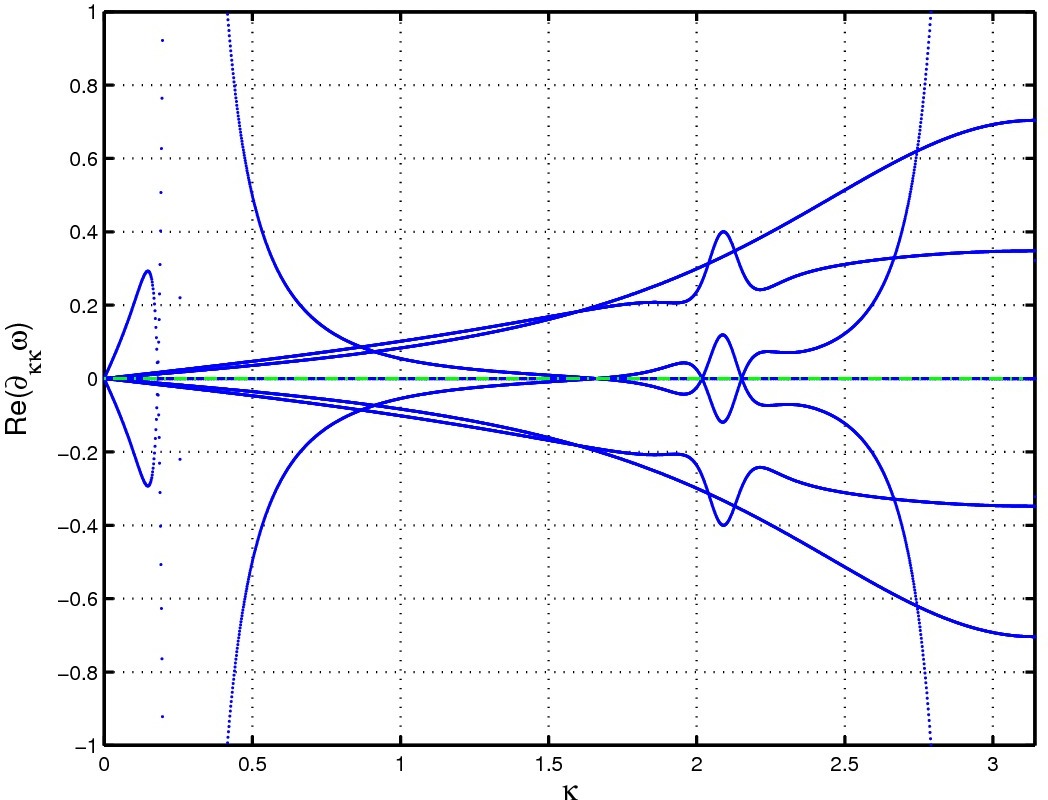}}\\
\caption{Local information for derivatives of $Re[\omega]$: (Left): $Re[\omega_{\kappa}]$; (Righ): $Re[\omega_{\kappa\kappa}]$. In the left subfigure, Green  lines denote $Re[\omega^{s}_{\kappa}]$; Red  line denotes $Re[\omega^{\pm}_{\kappa}]$.  In the right subfigure, Green  lines denote $Re[\omega^{\pm}_{\kappa\kappa}]$; Red  line denotes $Re[\omega^{s}_{\kappa\kappa}]$. The parameters are from the index 0 of Tabel \ref{table:1}.}\label{fig:10}
\end{center}
\end{figure}

\begin{figure}[!htbp]
\begin{center}
\scalebox{0.45}[0.45]{\includegraphics[angle=0]{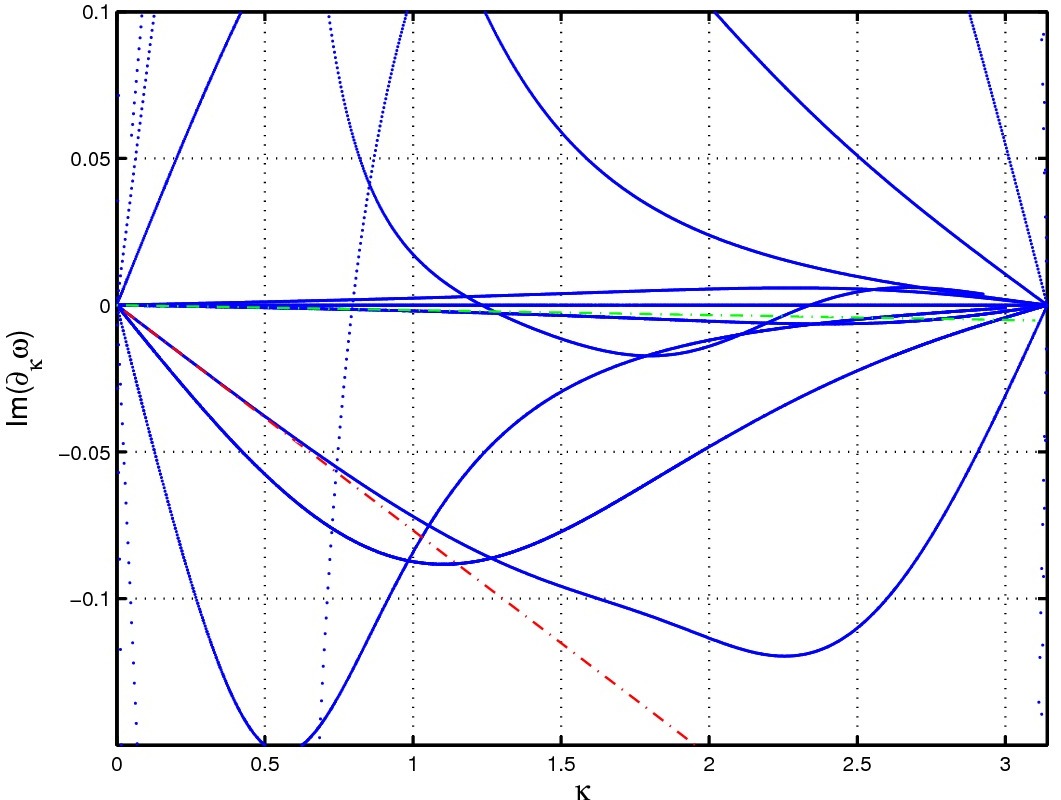}}
\scalebox{0.45}[0.45]{\includegraphics[angle=0]{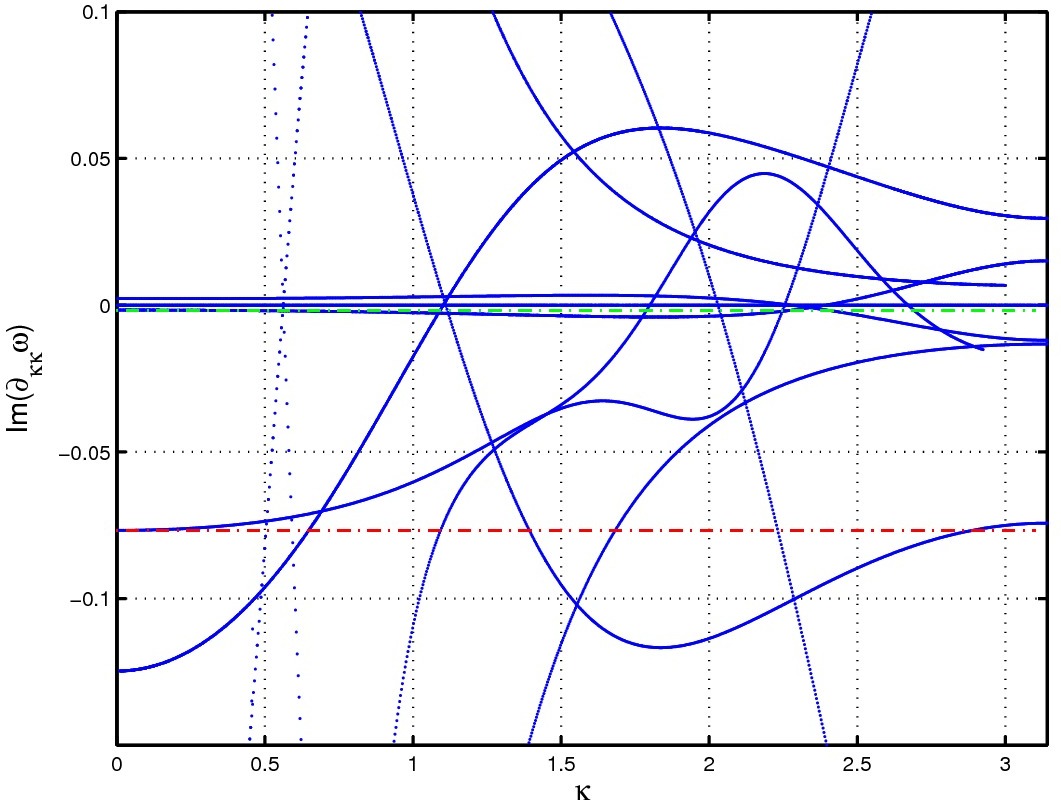}}\\
\caption{Local information for derivatives of $Im[\omega]$: (Left): $Im[\omega_{\kappa}]$ ; (Righ): $Im[\omega_{\kappa\kappa}]$.   In the left subfigure, Green line denotes $Im[\omega^{s}_{\kappa}]$; Red  line denotes $Im[\omega^{\pm}_{\kappa}]$.  In the right subfigure, Green  lines denote $Im[\omega^{\pm}_{\kappa\kappa}]$; Red  line denotes $Im[\omega^{s}_{\kappa\kappa}]$. The parameters are from the index 1 of Tabel \ref{table:2}.}\label{fig:11}
\end{center}
\end{figure}
 \begin{figure}[!htbp]
\begin{center}
\scalebox{0.45}[0.45]{\includegraphics[angle=0]{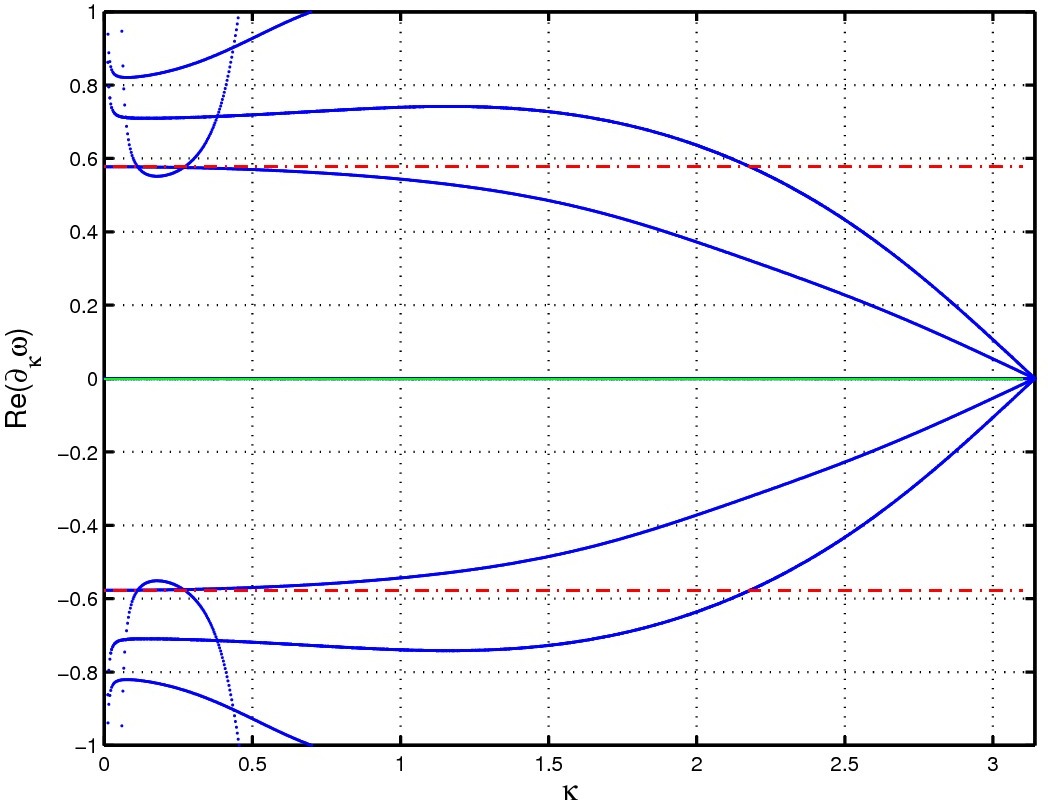}}
\scalebox{0.45}[0.45]{\includegraphics[angle=0]{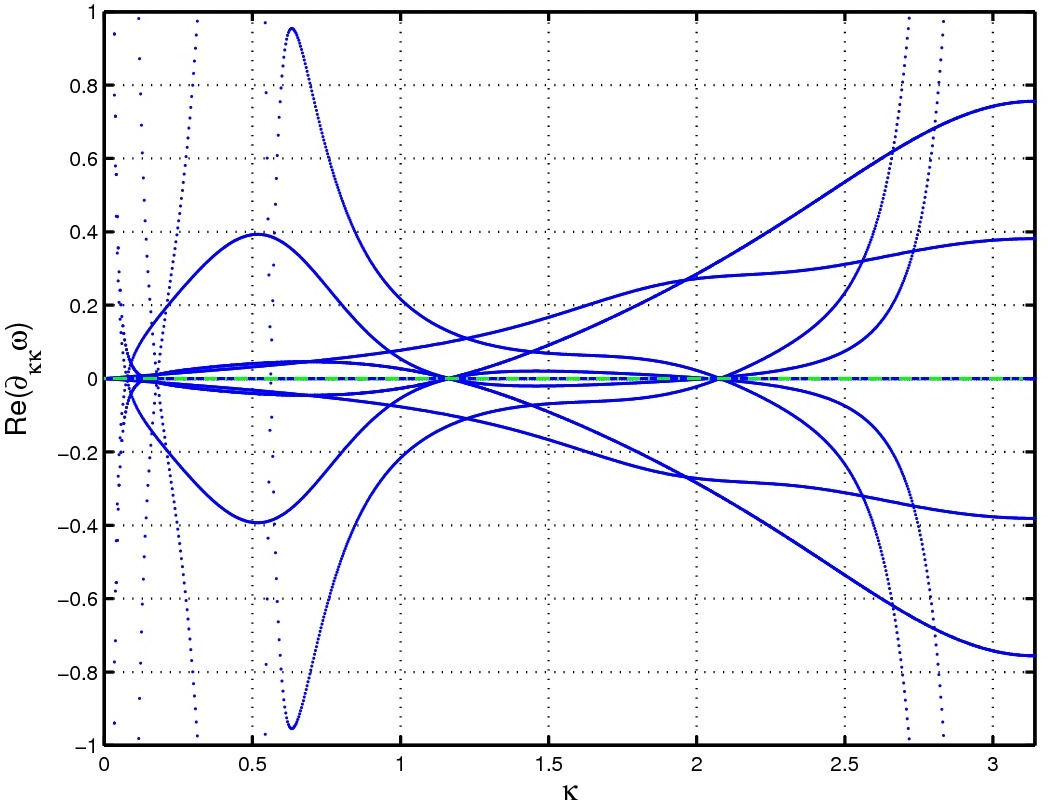}}\\
\caption{Local information for derivatives of $Re[\omega]$: (Left): $Re[\omega_{\kappa}]$; (Righ): $Re[\omega_{\kappa\kappa}]$. In the left subfigure, Green  lines denote $Re[\omega^{s}_{\kappa}]$; Red  line denotes $Re[\omega^{\pm}_{\kappa}]$.  In the right subfigure, Green  lines denote $Re[\omega^{\pm}_{\kappa\kappa}]$; Red  line denotes $Re[\omega^{s}_{\kappa\kappa}]$. The parameters are from the index 1 of Tabel \ref{table:2}.}\label{fig:12}
\end{center}
\end{figure}

\begin{figure}[!htbp]
\begin{center}
\scalebox{0.45}[0.45]{\includegraphics[angle=0]{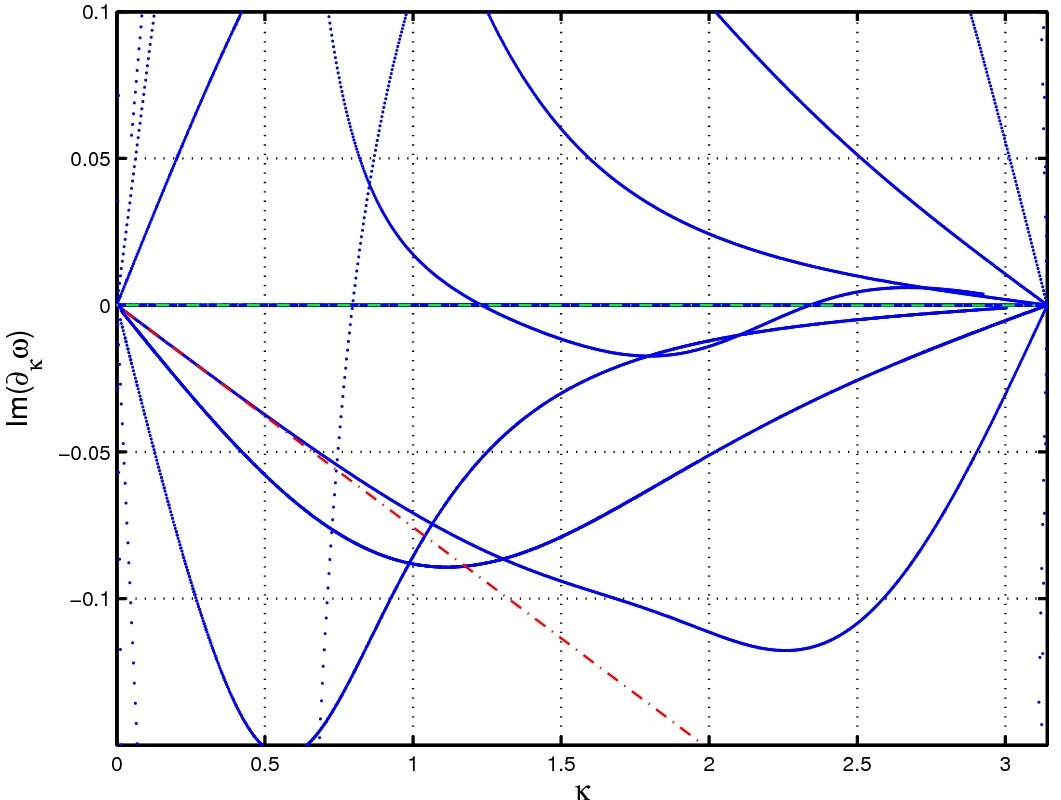}}
\scalebox{0.45}[0.45]{\includegraphics[angle=0]{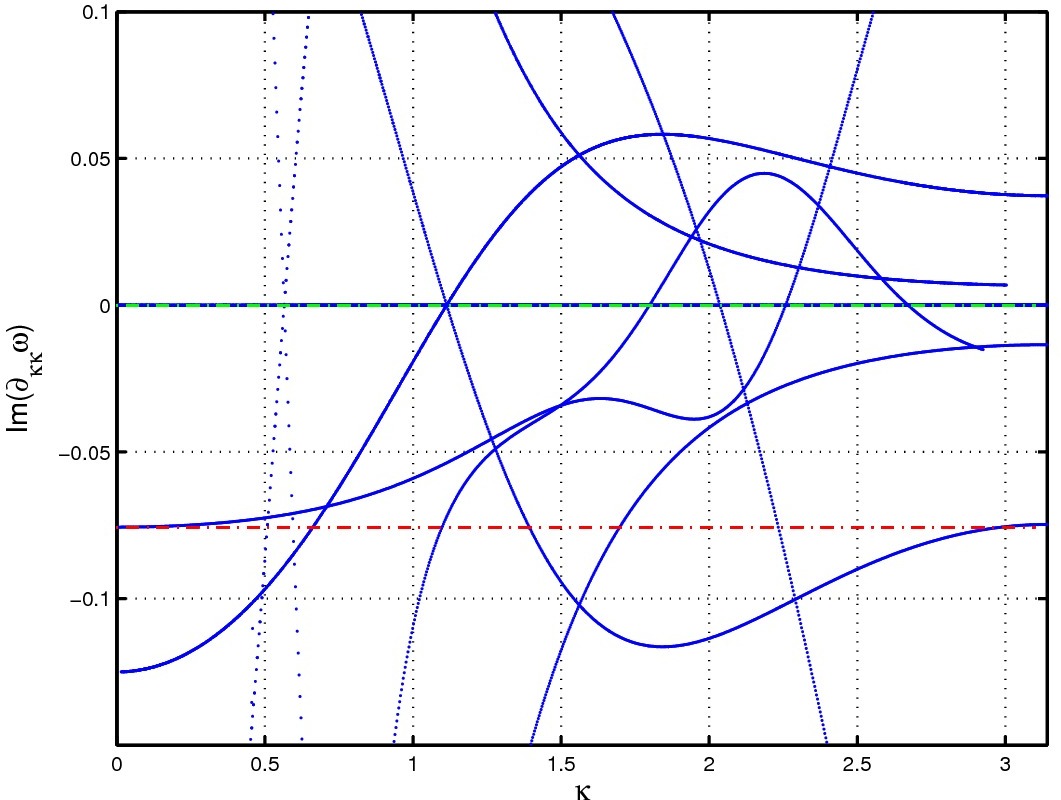}}\\
\caption{Local information for derivatives of $Im[\omega]$: (Left): $Im[\omega_{\kappa}]$ ; (Righ): $Im[\omega_{\kappa\kappa}]$.   In the left subfigure, Green line denotes $Im[\omega^{s}_{\kappa}]$; Red  line denotes $Im[\omega^{\pm}_{\kappa}]$.  In the right subfigure, Green  lines denote $Im[\omega^{\pm}_{\kappa\kappa}]$; Red  line denotes $Im[\omega^{s}_{\kappa\kappa}]$. The parameters are from the index 2 of Tabel \ref{table:2}.}\label{fig:13}
\end{center}
\end{figure}
 \begin{figure}[!htbp]
\begin{center}
\scalebox{0.45}[0.45]{\includegraphics[angle=0]{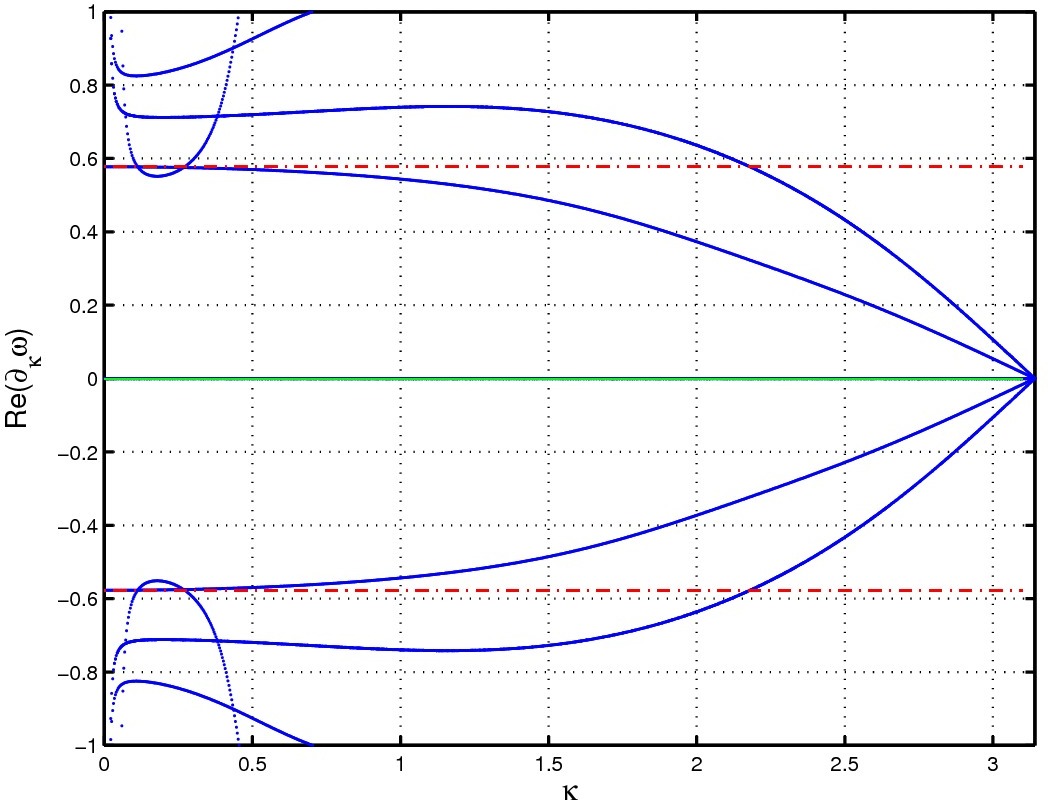}}
\scalebox{0.45}[0.45]{\includegraphics[angle=0]{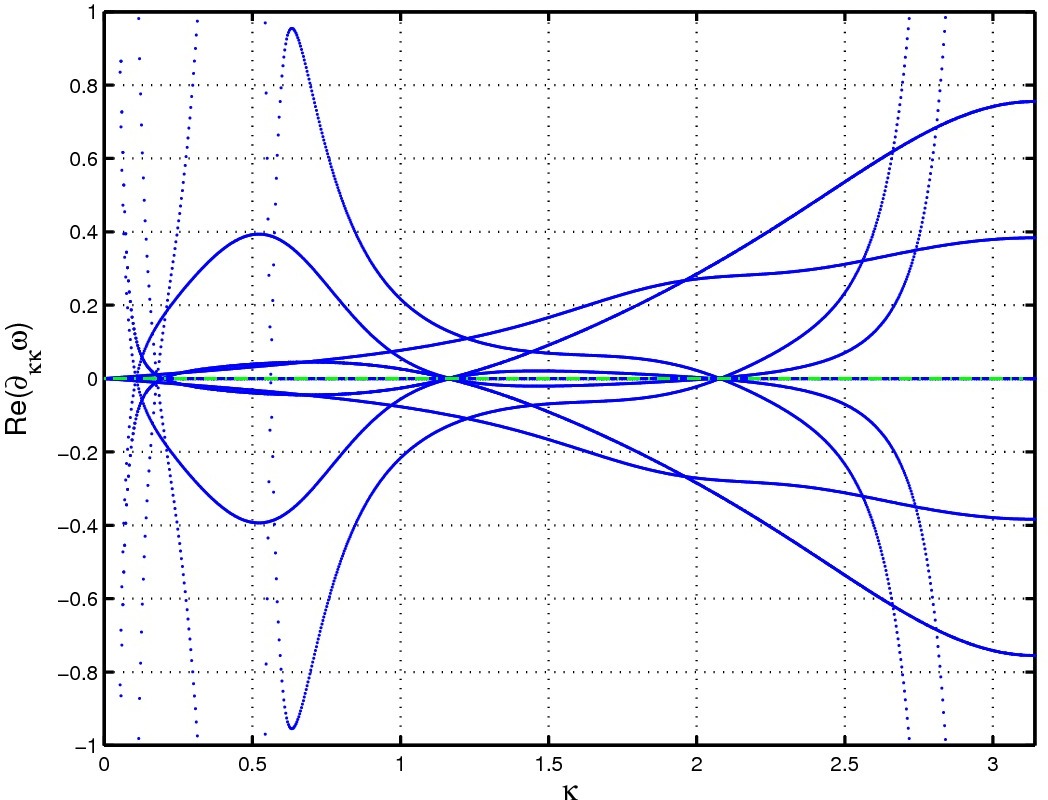}}\\
\caption{Local information for derivatives of $Re[\omega]$: (Left): $Re[\omega_{\kappa}]$; (Righ): $Re[\omega_{\kappa\kappa}]$. In the left subfigure, Green  lines denote $Re[\omega^{s}_{\kappa}]$; Red  line denotes $Re[\omega^{\pm}_{\kappa}]$.  In the right subfigure, Green  lines denote $Re[\omega^{\pm}_{\kappa\kappa}]$; Red  line denotes $Re[\omega^{s}_{\kappa\kappa}]$. The parameters are from the index 2 of Tabel \ref{table:2}.}\label{fig:14}
\end{center}
\end{figure}

\begin{figure}[!htbp]
\begin{center}
\scalebox{0.45}[0.45]{\includegraphics[angle=0]{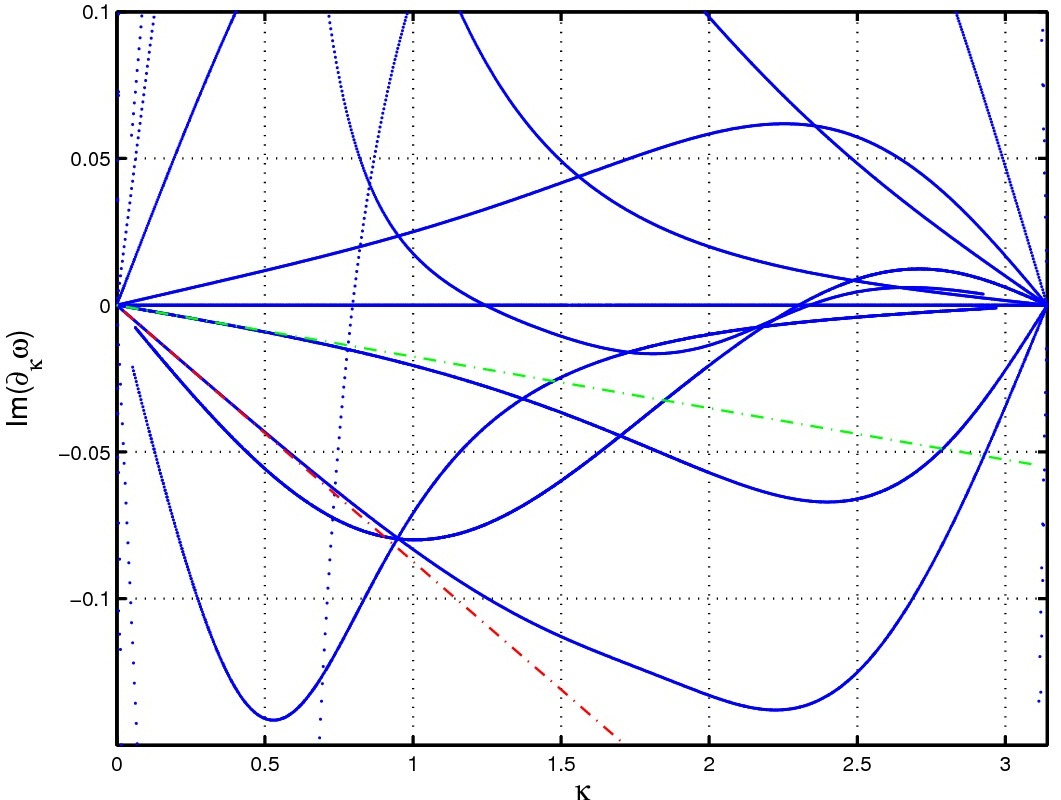}}
\scalebox{0.45}[0.45]{\includegraphics[angle=0]{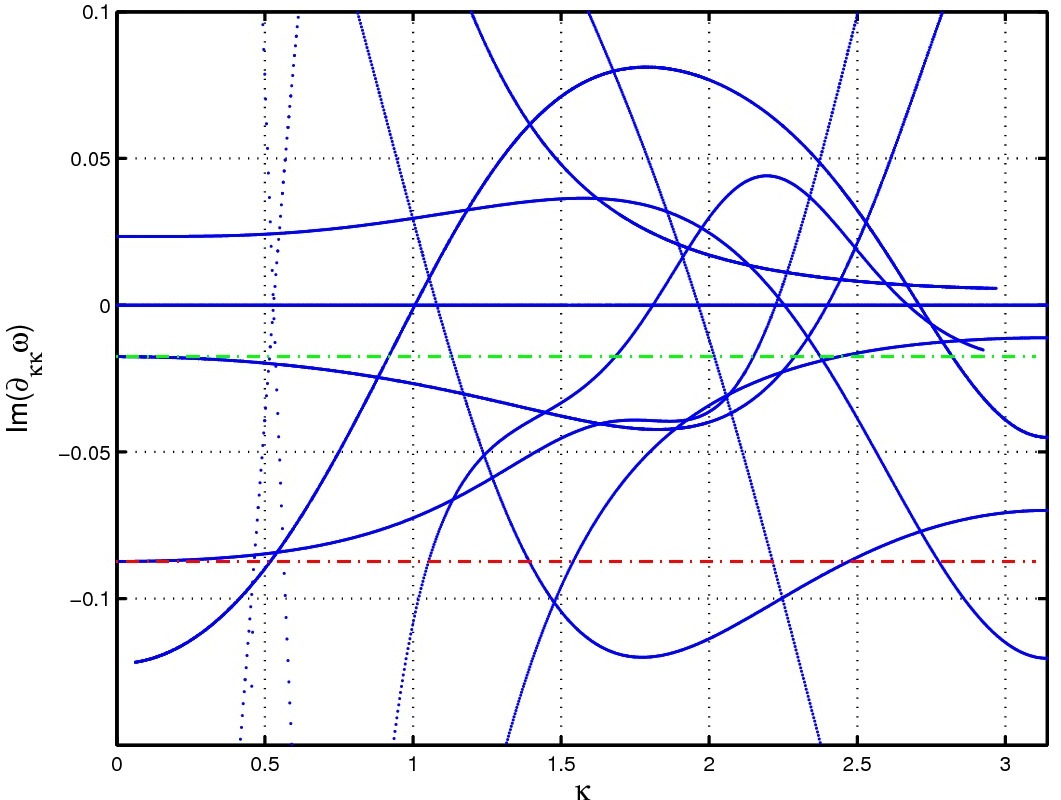}}\\
\caption{Local information for derivatives of $Im[\omega]$: (Left): $Im[\omega_{\kappa}]$ ; (Righ): $Im[\omega_{\kappa\kappa}]$.   In the left subfigure, Green line denotes $Im[\omega^{s}_{\kappa}]$; Red  line denotes $Im[\omega^{\pm}_{\kappa}]$.  In the right subfigure, Green  lines denote $Im[\omega^{\pm}_{\kappa\kappa}]$; Red  line denotes $Im[\omega^{s}_{\kappa\kappa}]$. The parameters are from the index 0 of Tabel \ref{table:2}.}\label{fig:15}
\end{center}
\end{figure}
 \begin{figure}[!htbp]
\begin{center}
\scalebox{0.45}[0.45]{\includegraphics[angle=0]{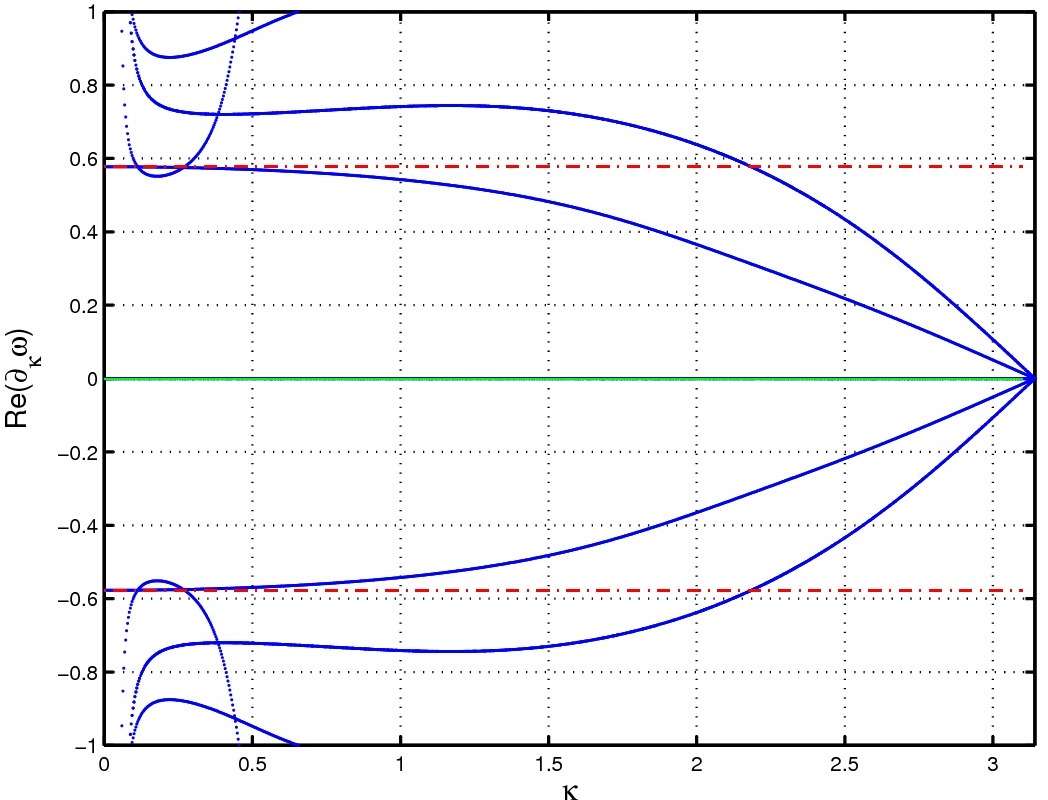}}
\scalebox{0.45}[0.45]{\includegraphics[angle=0]{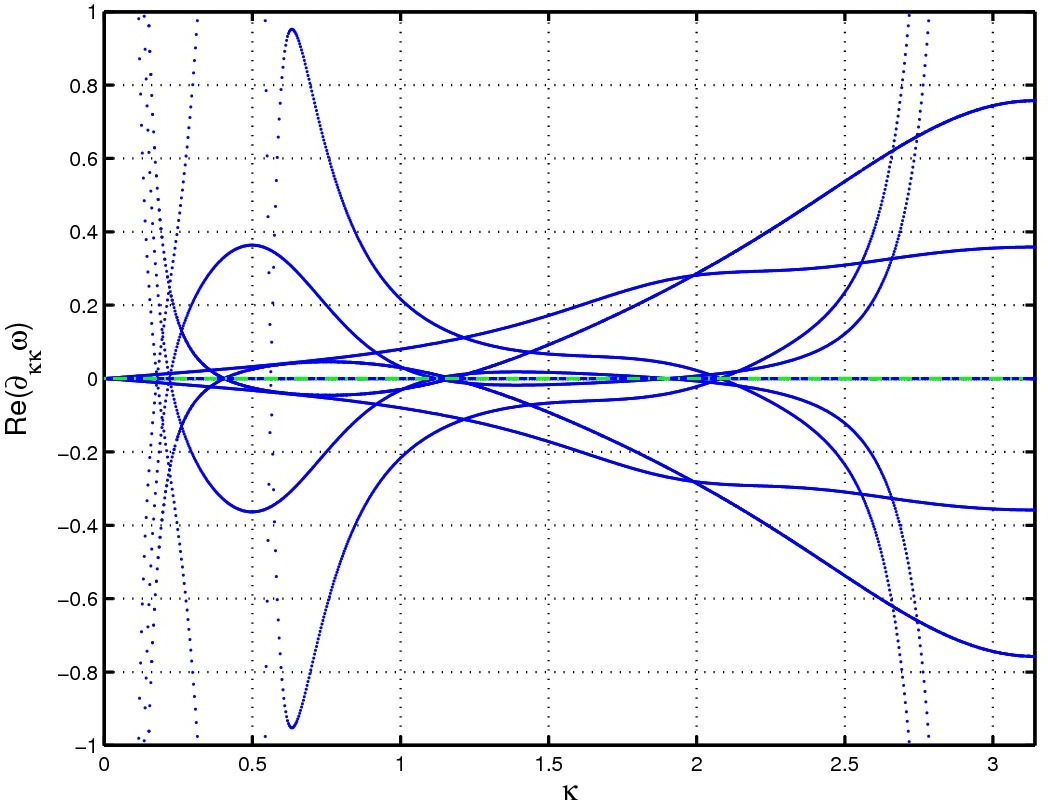}}\\
\caption{Local information for derivatives of $Re[\omega]$: (Left): $Re[\omega_{\kappa}]$; (Righ): $Re[\omega_{\kappa\kappa}]$. In the left subfigure, Green  lines denote $Re[\omega^{s}_{\kappa}]$; Red  line denotes $Re[\omega^{\pm}_{\kappa}]$.  In the right subfigure, Green  lines denote $Re[\omega^{\pm}_{\kappa\kappa}]$; Red  line denotes $Re[\omega^{s}_{\kappa\kappa}]$. The parameters are from the index 0 of Tabel \ref{table:2}.}\label{fig:16}
\end{center}
\end{figure}

In the above discussion, we focused on observing the sensitivity of $\omega$ with respect to $\kappa$ for the D2Q9 MRT-LBM. Now, we investigate the sensitivity of $\omega$ for the D3Q15 MRT-LBM.  From Figs. \ref{fig:11}$\sim$\ref{fig:16}, it is seen that there exist very significant changes for the first and second order derivatives of  the acoustic and shear modes with respect to $\kappa$. From Figs. \ref{fig:12}, \ref{fig:14} and \ref{fig:16}, it is observed that the computed $c_s$ has a significant dependence on the wave-number magnitudes, and  the computed $c_s$ deviates significantly from the exact $c_s$. From Figs. \ref{fig:11}, \ref{fig:13} and \ref{fig:15},  the computed $Im[\omega_{\kappa}]$ and $Im[\omega_{\kappa\kappa}]$ deviate  from the exact  $Im[\omega_{\kappa}]$ and $Im[\omega_{\kappa\kappa}]$ . Numerically,   $Im[\omega^{\pm}]$ is very sensitive to the wave number $\kappa$.

From the current 2D and 3D results, using the given free relaxation parameters in the literatures, the numerical dispersion and dissipation relations are sensitive to the wave number $\kappa$ for  $\widehat{\bf u\cdot k }=0$.  The behaviors of the numerical group velocity will become sensitive when the wave-number magnitude of $|{\bf k}|$ is close to 2. It is necessary to point out that if $\widehat{\bf u\cdot k }\neq 0$, we will obtain the most insensitive results. In order to save space, the investigations for $|{\bf u}|>0$  and  $\widehat{\bf u\cdot k }\neq 0$ are not shown here. With respect to other free parameters, we will address these problems in complex plane in the next part.

\subsubsection{The behaviors of eigenvalues of $M^{\rm mrt}$ in the complex plane}

From Eq. (\ref{stability:c}),  the linear stability of the MRT-LBM requires that the eigenvalues of $M^{\rm mrt}$ should lie in a unit circle whose origin is located at the point (0,0). 
The investigations of sensitivities will address the factors that have a significant influence on the distribution of the eigenvalues in the complex planes. From Eq. (\ref{th:d}), the distribution of the hydrodynamic modes are only dependent on  the parameter set 
$\widehat{\Upsilon}=\{\bf k$, $\bf u$, $\widehat{{\bf k}\cdot {\bf u}}$, $\nu$, $\eta$\}.  
That means the behavior of eigenvalues of the matrix $M^{\rm mrt}$ should be dependent on the parameter set $\widehat{\Upsilon}$.  However, the researches dealing with the relations of dispersion and dissipation of the linearized MRT-LBM indicate that there exists some significant dependence on the orientations of $\bf k$, $\bf u$ and free relaxation parameters in the matrix $M^{\rm mrt}$ \cite{lallemandluo,xusagaut}.  
The values of the free relaxation parameters are set according to the recommended values \cite{lallemandluo}.  In order to implement the analysis, we give the following definations for ${\bf k}$ and ${\bf u}$. For the 2D problems, ${\bf k}$ and ${\bf u}$ are defined by

\begin{equation}\label{2d:k}
k_x=|{\bf k}|{\rm cos}(\theta_k),\ k_y=|{\bf k}|{\rm sin}(\theta_k);\ u=|{\bf u}|{\rm cos}(\theta_u),\ v=|{\bf u}|{\rm sin}(\theta_u).
\end{equation}
The inner product of   ${\bf k}$ and ${\bf u}$ is given by
\begin{equation}
{\bf k\cdot u}=|{\bf k}||{\bf u}|{\rm cos}(\theta_k-\theta_u).
\end{equation}
For 3D problems,  ${\bf k}$ and ${\bf u}$ are defined as follows
\begin{equation}\label{3d:k}
\begin{array}{lll}
k_x=|{\bf k}|{\rm cos}(\theta_k) {\rm sin}
(\zeta_k),&\ k_y=|{\bf k}|{\rm sin}(\theta_k){\rm sin}(\zeta_k),&\ k_z=|{\bf k}|{\rm cos}(\zeta_k);\\
u=|{\bf u}|{\rm cos}(\theta_u) {\rm sin}(\zeta_u),&\ v=|{\bf u}|{\rm sin}(\theta_u){\rm sin}(\zeta_u),&\ w=|{\bf u}|{\rm cos}(\zeta_u).
\end{array}
\end{equation}
The inner product of   ${\bf k}$ and ${\bf u}$ is given by
\begin{equation}\label{uk3d}
{\bf k\cdot u}=|{\bf k}||{\bf u}|({\rm cos}(\theta_k-\theta_u) {\rm sin}(\zeta_k) {\rm sin}(\zeta_u)+ {\rm cos}(\zeta_k) {\rm cos}(\zeta_u)).
\end{equation}
In Eq. (\ref{uk3d}), if let $\theta_k=\theta_u$, we have
\begin{equation}
{\bf k\cdot u}=|{\bf k}||{\bf u}| {\rm cos}(\zeta_k-\zeta_u).
\end{equation}

For the 2D problems, the discrete values of the variable $|{\bf k}|$ are set as a sequence of number $\{i\cdot\delta k\}_{0\leq i\leq J }$ where $\delta k=\pi/J$.
First,  the parameters given in Index 4 in Table \ref{table:1} are used to implement our 2D investigations. Considering the wave-number vector ${\bf k}$ parallel to ${\bf e_x}$ and  $\widehat{{\bf k}\cdot {\bf e_x}}$ is equal to 0 or $\pi$, we consider  $\widehat{{\bf k}\cdot {\bf u}}$ as a random parameter in the interval $[0,\pi]$.  
Ten uniform random values of $\widehat{{\bf u}\cdot {\bf e_x}}$ for each value of ${\bf k}$ are generated in $[0,\pi]$ to investigate the influences of $\widehat{{\bf u}\cdot {\bf e_x}}$ on the distribution of the eigenvalues of the matrix $M^{\rm mrt}$. 
In order to address the influences of $\widehat{{\bf k}\cdot {\bf e_x}}$, we let ${{\bf u}\Vert{\bf e_x}}$ ( $\widehat{{\bf u}\cdot {\bf e_x}}$ is equal to 0 or $\pi$) , and ten uniform random values of $\widehat{{\bf k}\cdot {\bf e_x}}$  for each value of ${\bf k}$ are generated in the range $ [0,\pi]$. 
In Fig. \ref{fig:17}, the distribution of the eigenvalues of the matrix $M^{\rm mrt}$ is given. 
From the figure, it is clear that  $\widehat{{\bf k}\cdot {\bf e_x}}$ for   ${{\bf u}\Vert{\bf e_x}}$  has more significant influence than  $\widehat{{\bf u}\cdot {\bf e_x}}$ for  ${{\bf k}\Vert{\bf e_x}}$. 
When ${\bf u}$ is parallel to ${\bf e_x}$,  The distribution of the eigenvalues of the matrix $M^{\rm mrt}$ appears very random with respect to the random $\widehat{{\bf k}\cdot {\bf e_x}}$. In Fig. \ref{fig:18} , the distribution of the eigenvalues is figured for $\widehat{\bf k\cdot e_x}=\pi/4$ and $\widehat{\bf u\cdot e_x}=\pi/4$. $\widehat{{\bf u}\cdot {\bf k}}$ is regarded as a random variable.  In Fig. \ref{fig:19} , the distribution of the eigenvalues is displayed for $\widehat{\bf k\cdot e_x}=\pi/2$ and $\widehat{\bf u\cdot e_x}=\pi/2$.

According to Figs. \ref{fig:17} - \ref{fig:19}, we conclude that the distribution of the eigenvalues of the matrix $M^{\rm mrt}$ is very sensitive to the angle between ${\bf k}$ and ${\bf e_x}$ . Therefore,  ${\bf k}$ and ${\bf e_x}$ are expected to strongly govern the dissipation and dispersion relation of the linearized MRT-LBM. Furthermore, it is observed that the distribution  of the eigenvalues appears symmetrical  with respect to the  real axis. This symmetry property is obvious because of the symmetry property of the wave-number vectors. The angle  between the wave-number of the red color distribution and the wave-number of the blue color distribution is equal to $\pi$ corresponding to each ${\bf k}$.

 \begin{figure}[!htbp]
\begin{center}
\scalebox{0.1}[0.1]{\includegraphics[angle=0]{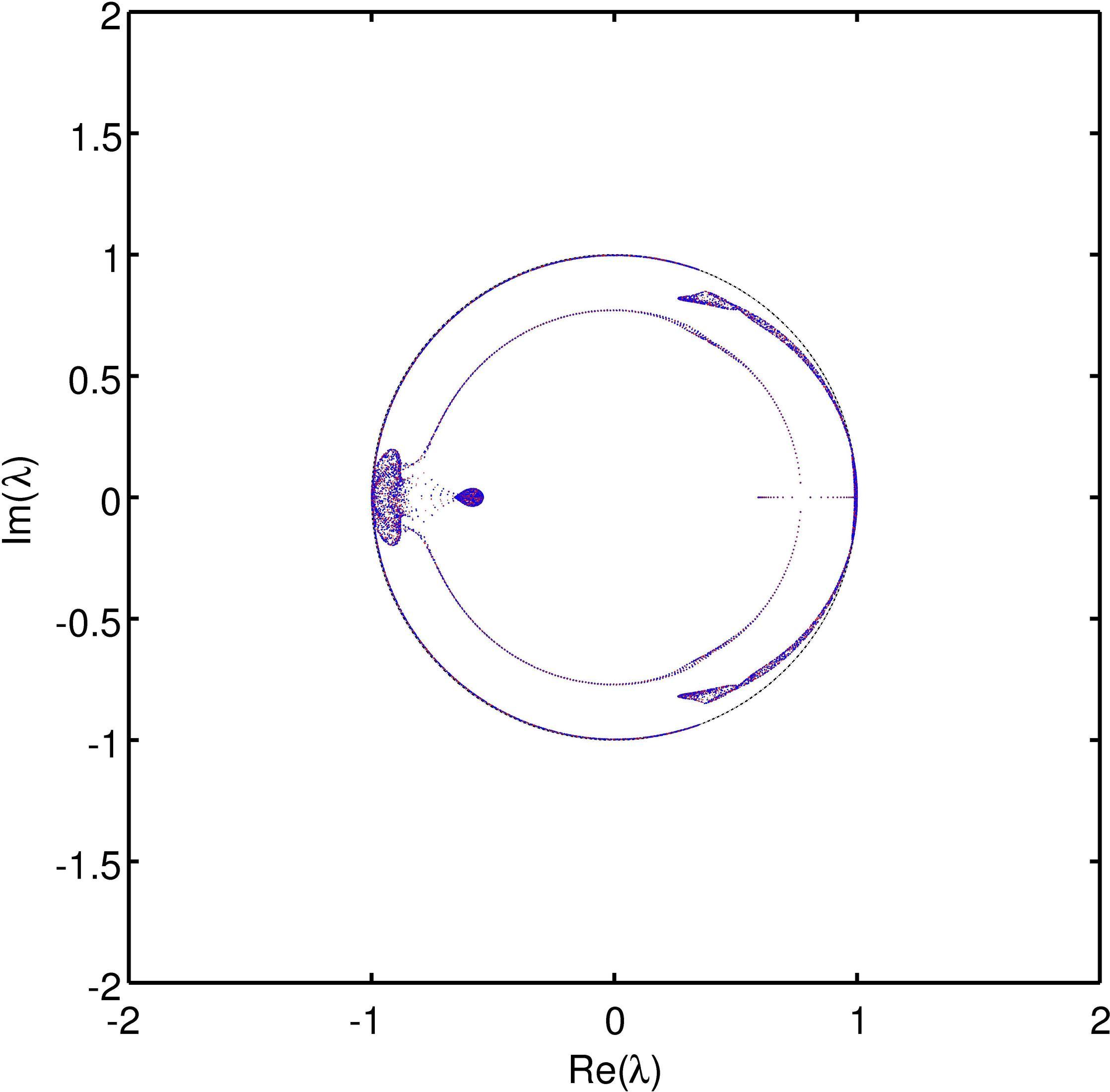}}
\scalebox{0.1}[0.1]{\includegraphics[angle=0]{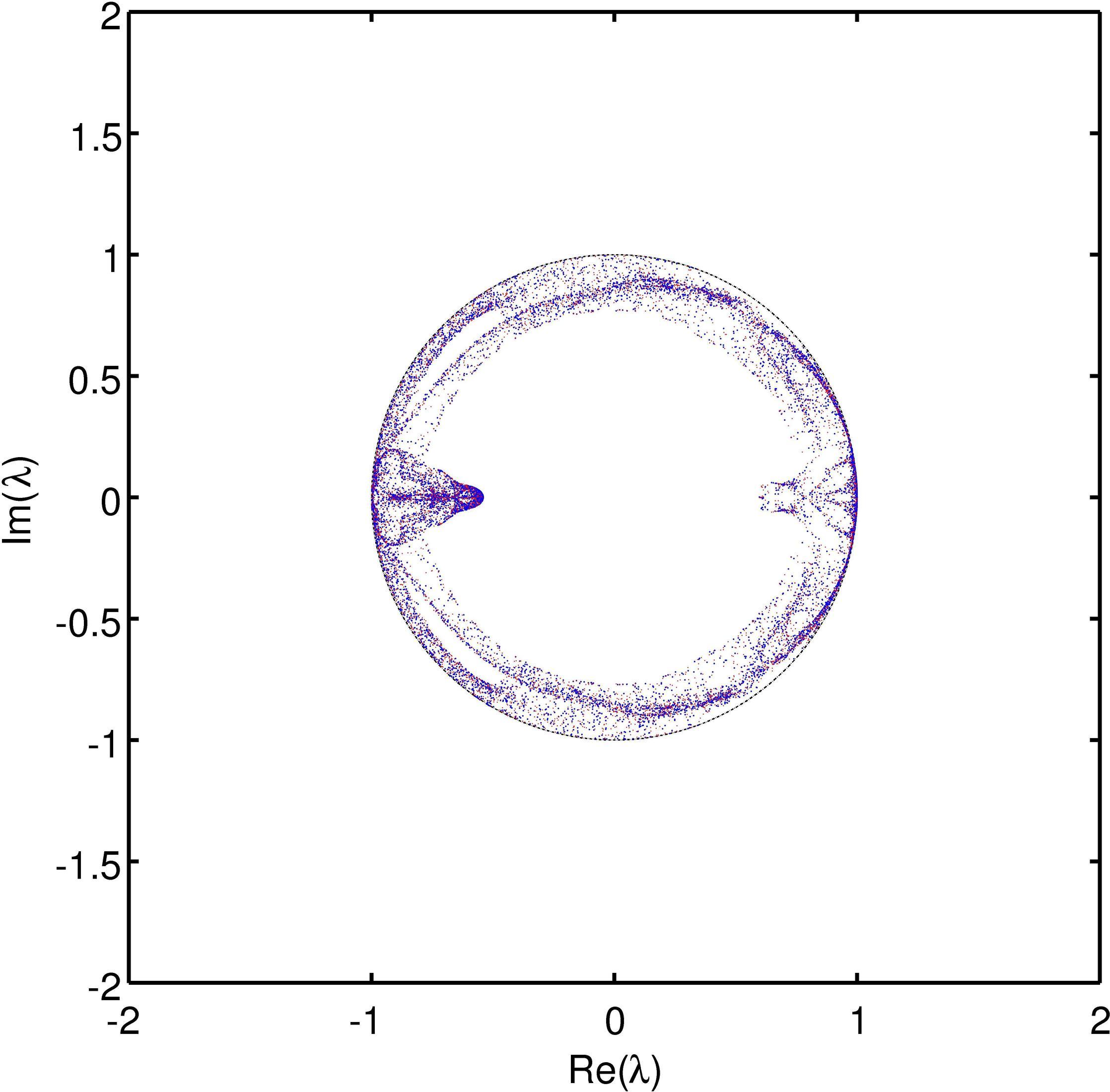}}
\caption{Distribution of the eigenvalues of the matrix $M^{\rm mrt}$: (Left) ${{\bf k}\Vert{\bf e_x}}$ , and $\widehat{{\bf u}\cdot {\bf e_x}}$ is regarded as a uniform  random variable in the range $ [0,\pi]$.  For each value of ${\bf k}$,  ten random numbers for $\widehat{{\bf u}\cdot {\bf e_x}}$ are generated in $[0,\pi]$. (Right) ${{\bf u}\Vert{\bf e_x}}$ , and $\widehat{{\bf k}\cdot {\bf e_x}}$ is regarded as a uniform  random variable in the range $ [0,\pi]$.  For each value of ${\bf k}$,  ten random numbers for  $\widehat{{\bf k}\cdot {\bf e_x}}$ are generated in $[0,\pi]$. The parameters are from the index 4 of Tabel \ref{table:1}.  (Blue Color) ${\bf k}=[k_x,k_y]$; (Red Color) $ {\bf k}=[-k_x,-k_y]$. The circles in the subfigures stand for the unitary circles with the origin located at (0,0), which are also the stable regions of the linearized MRT-LBM.}\label{fig:17}
\end{center}
\end{figure}

 \begin{figure}[!htbp]
\begin{center}
\scalebox{0.1}[0.1]{\includegraphics[angle=0]{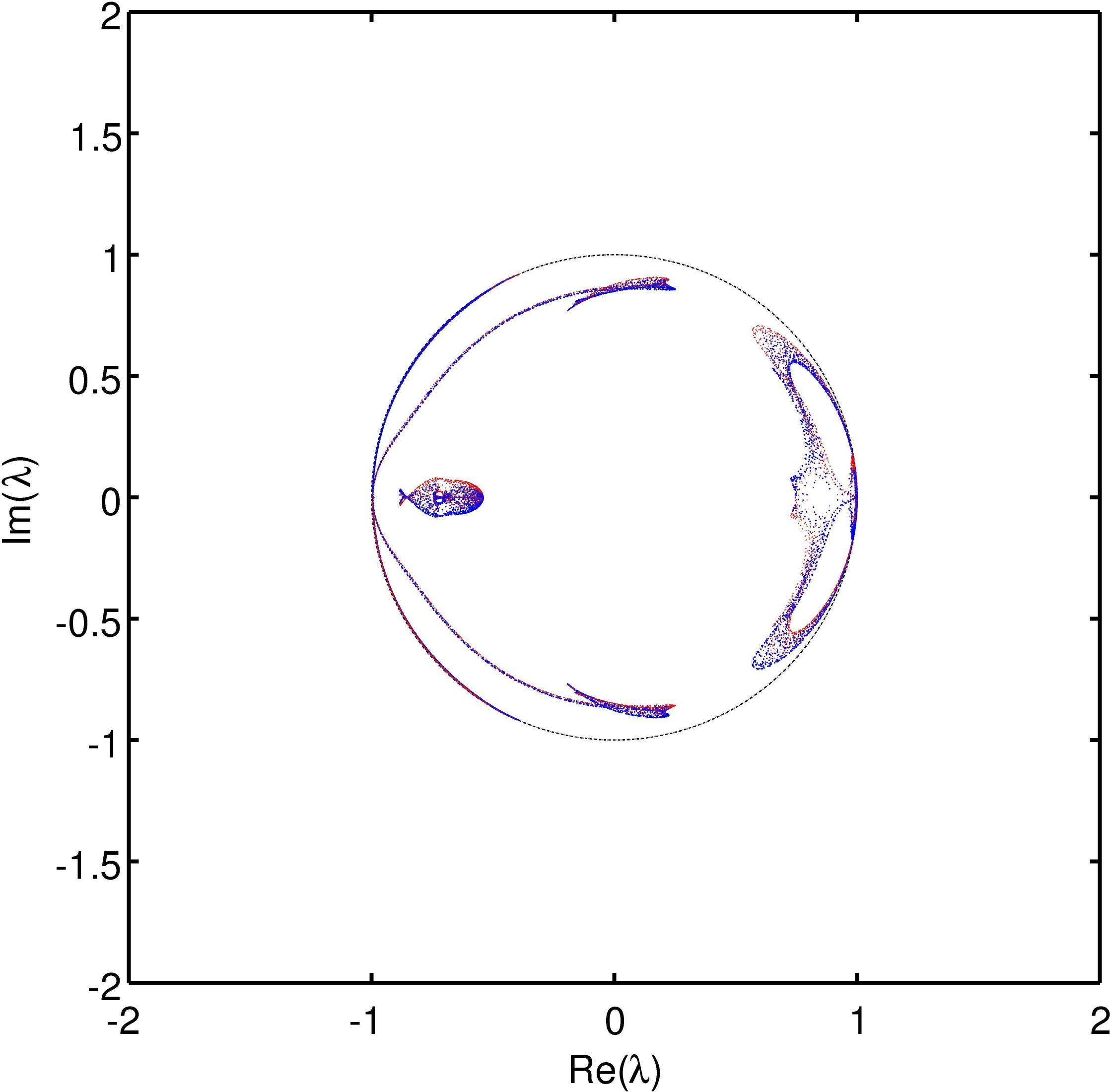}}
\scalebox{0.1}[0.1]{\includegraphics[angle=0]{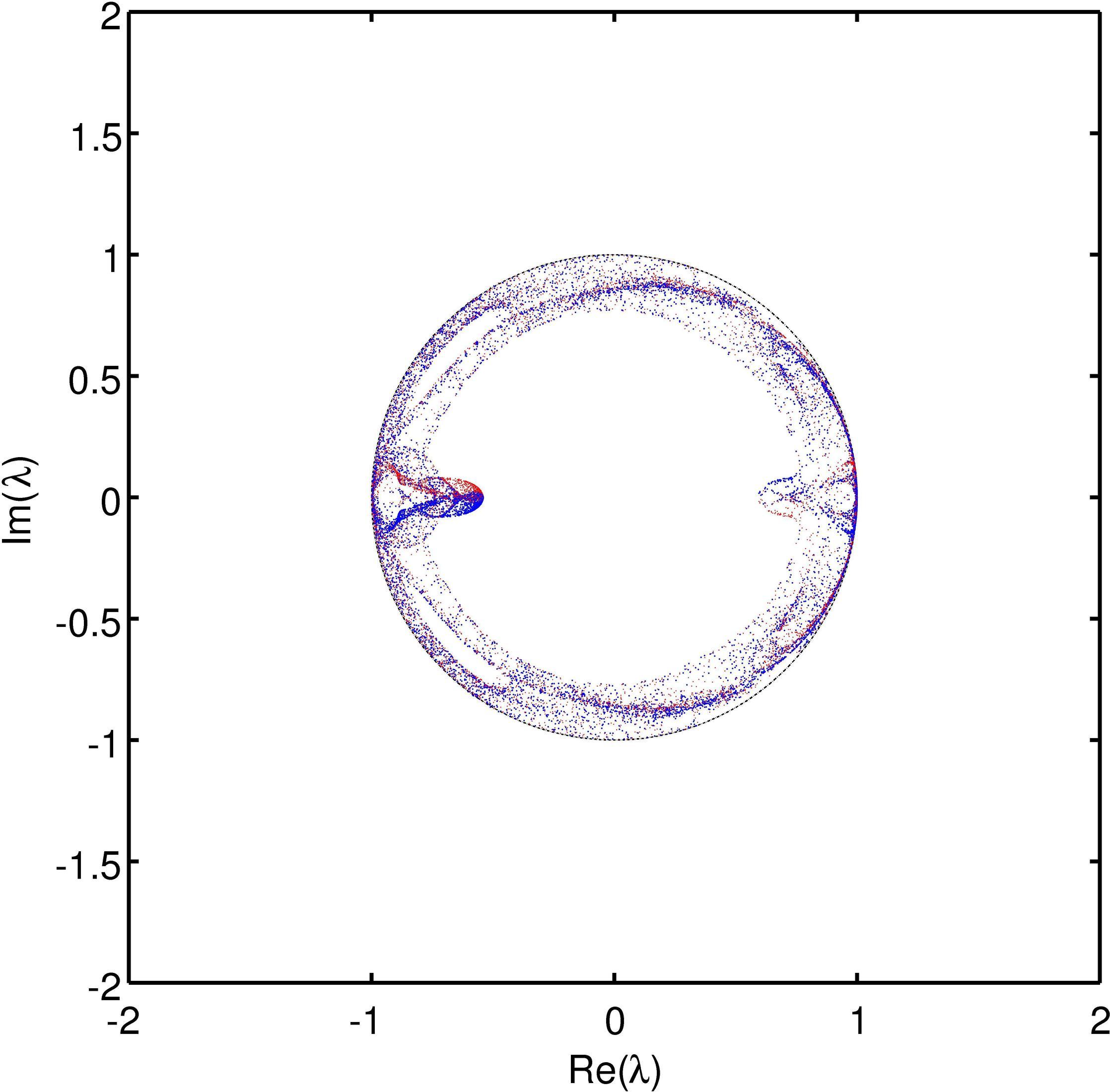}}
\caption{Distribution of the eigenvalues of the matrix $M^{\rm mrt}$: (Left) $\widehat{\bf k\cdot e_x}=\pi/4$; (Right) $\widehat{\bf u\cdot e_x}=\pi/4$ . $\widehat{{\bf u}\cdot {\bf k}}$ is regarded as a uniform  random variable in the range $ [0,\pi]$.  For each value of ${\bf k}$,  ten random numbers for $\widehat{{\bf u}\cdot {\bf k}}$ are generated in $[0,\pi]$. The parameters are from the index 4 of Tabel \ref{table:1}.  (Blue Color) ${\bf k}=[k_x,k_y]$; (Red Color) $ {\bf k}=[-k_x,-k_y]$. The circles in the subfigures stand for the unitary circles with the origin located at (0,0), which are also the stable regions of the linearized MRT-LBM.}\label{fig:18}
\end{center}
\end{figure}

 \begin{figure}[!htbp]
\begin{center}
\scalebox{0.1}[0.1]{\includegraphics[angle=0]{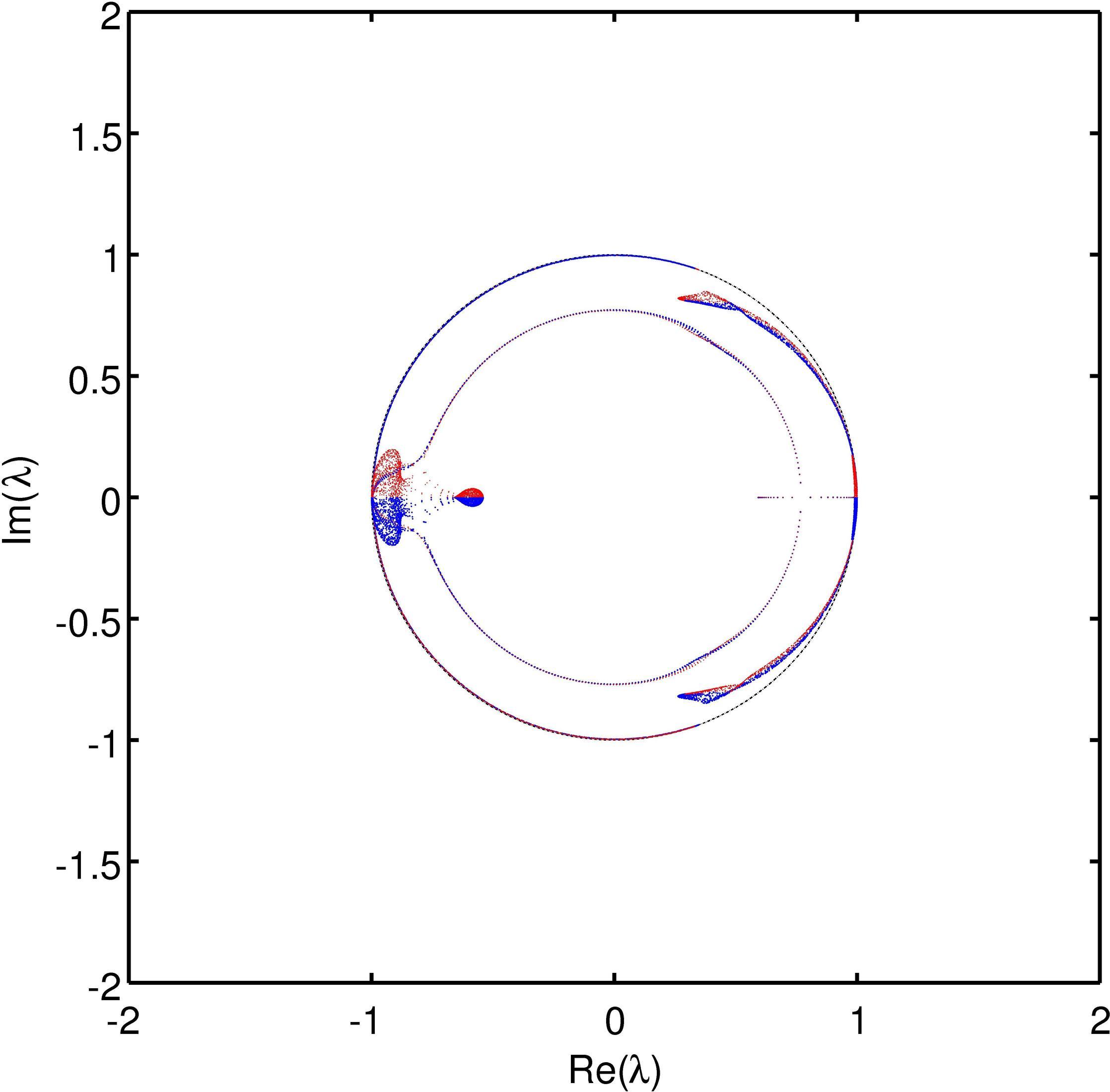}}
\scalebox{0.1}[0.1]{\includegraphics[angle=0]{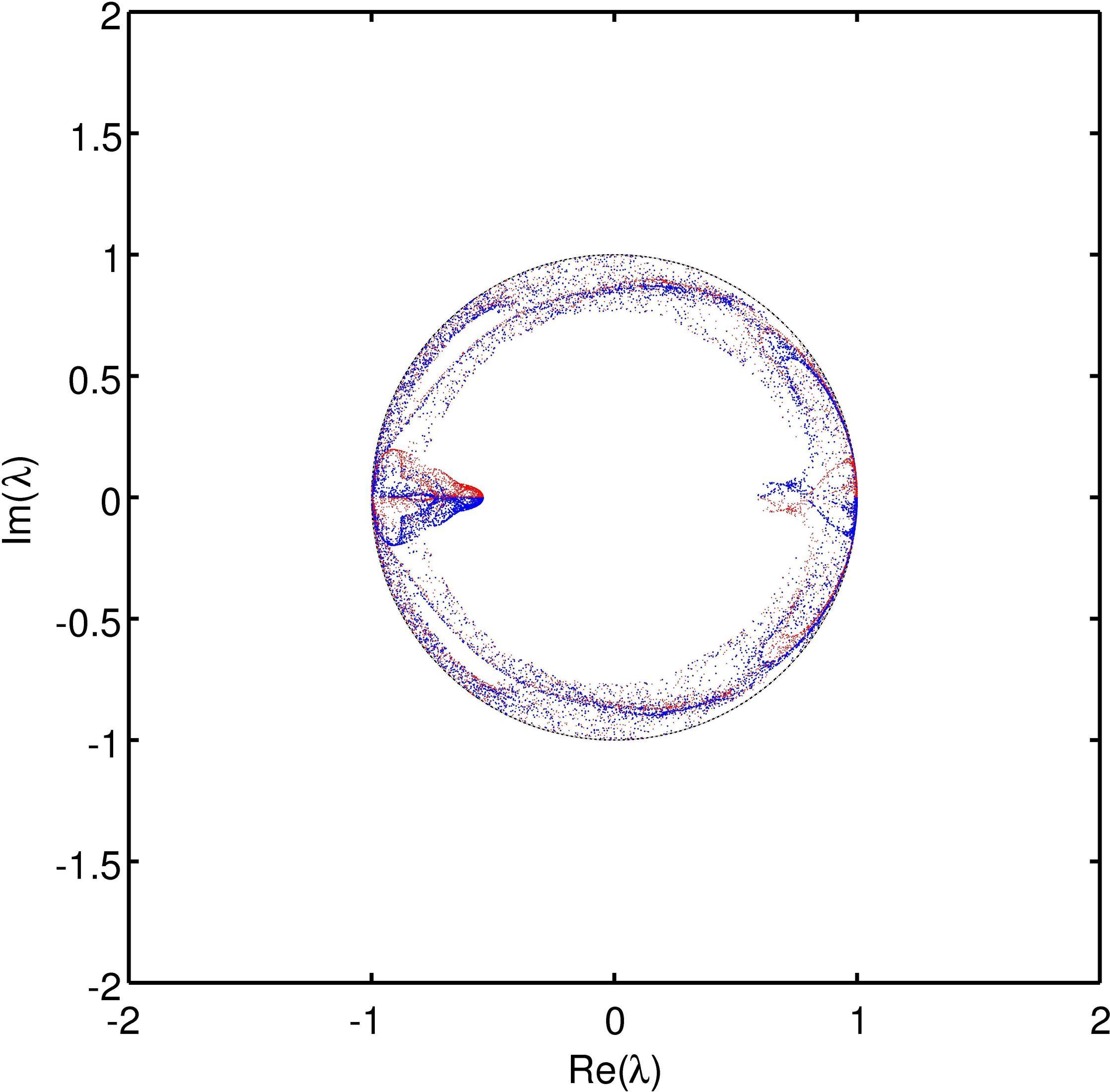}}
\caption{Distribution of the eigenvalues of the matrix $M^{\rm mrt}$: (Left) $\widehat{\bf k\cdot e_x}=\pi/2$; (Right) $\widehat{\bf u\cdot e_x}=\pi/2$ . $\widehat{{\bf u}\cdot {\bf k}}$ is regarded as a uniform  random variable in the range $ [0,\pi]$.  For each value of ${\bf k}$,  ten random numbers for $\widehat{{\bf u}\cdot {\bf k}}$ are generated in $[0,\pi]$. The parameters are from the index 4 of Table \ref{table:1}.  (Blue Color) ${\bf k}=[k_x,k_y]$; (Red Color) $ {\bf k}=[-k_x,-k_y]$. The circles in the sub figures stand for the unitary circles with the origin located at (0,0), which are also the stable regions of the linearized MRT-LBM.}\label{fig:19}
\end{center}
\end{figure}

 \begin{figure}[!htbp]
\begin{center}
\scalebox{0.45}[0.45]{\includegraphics[angle=0]{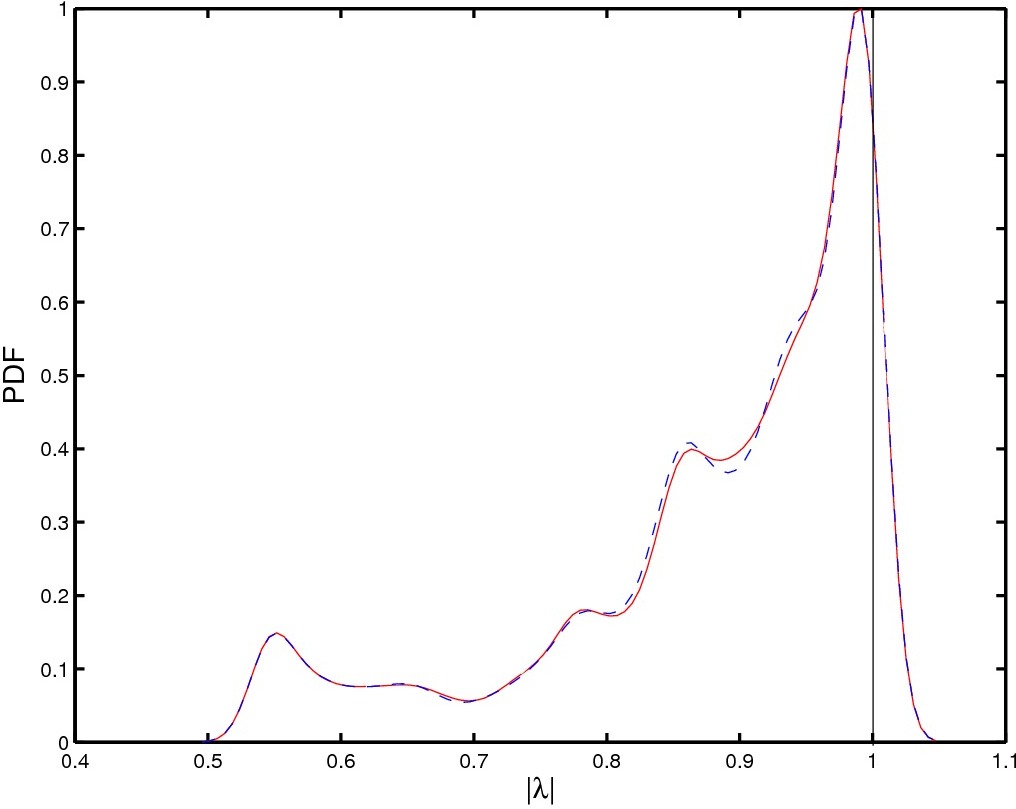}}
\caption{Probability density functions of eigenvalue modules (In Eq. (\ref{2d:k}), $\theta_k$ and $\theta_u$ are regarded as  random variables in the rang $[0,\pi]$ ) : (Red) $\widehat{\bf k\cdot e_x}=0$ and $\widehat{\bf u\cdot e_x}=0$; the difference of $\theta_k$ and $\theta_u$ is equal to 0 . (Blue) $\widehat{{\bf u}\cdot {\bf k}}=\pi/2$ ;  the difference of $\theta_k$ and $\theta_u$ is equal to $\pi/2$. The parameters are chosen from Index 4 in Table \ref{table:1}. }\label{fig:20}
\end{center}
\end{figure}

Here, we revisit the conclusion about the stability of the transverse modes and the longitude modes. It was indicated that the transverse modes is more stable than longitudinal modes and the sound waves  propagating in the direction of the mean flow are quite unstable for the BGK-LBM \cite{lallemandluo}. This  instability  instability can be reduced in the original MRT-LBM \cite{lallemandluo}.
In Fig. \ref{fig:20}, when $\theta_k$ and $\theta_u$ are regarded as the two  random variables in the range $[0,\pi]$,  the probability density function (PDF) profiles of of eigenvalue modules are given for  the transverse modes and the longitudinal modes.  From this figure, it is clear that for the original MRT-LBM, the PDF profiles of the  eigenvalue modules of the matrix $M^{\rm mrt}$ nearly has the same shape.  And it is also observed that there exist some modules larger than 1 (refer to the vertical line $x=1$ in the figure). That means the MRT-LBM is not always stable for any $|{\bf k}|\in [0,\pi]$ and any small shear viscosity. These instability phenomena agree with the results of the second-order sensitivity analysis in Sec. \ref{fisrtsecondder}. 

In Eq. (\ref{3d:k}), taking $\theta_k=\theta_u$, we regard $\theta_k$ and $\theta_u$ as random variables in the range $[0,2\pi]$. Furthermore, we consider two cases: (a) assuming $\zeta_k=0$, $\zeta_u$ is a random variable in the range $[0,\pi]$;  (b) assuming $\zeta_u=0$, $\zeta_k$ is a random variable in the range $[0,\pi]$. In Fig. \ref{fig:21}, the distribution of the eigenvalues of the matrix $M^{\rm mrt}$ is plotted for any $\widehat{\bf u\cdot k}$. In Fig.  \ref{fig:21}, the red points overlap with the blue points, meaning that  for  ${\bf k}=[k_x,k_y]$ and  ${\bf k}=[-k_x,-k_y]$, similar eigenvalue behaviors exist. 
The random phenomenon in the right sub figure of  Fig.  \ref{fig:21} indicates that the eigenvalue distribution is very sensitive to the orientation of the wave number ${\bf k}$. This phenomenon is identical with what was observed in the case of 2D linearized MRT-LBM.  
In Fig.\ref{fig:22}, we consider two additional cases: (a) $\theta_k=0$ and $\zeta_k=0$, $\theta_u$ and $\zeta_u$ are random in $[0,2\pi]$ and $[0,\pi]$, repectively; (b) $\theta_u=0$ and $\zeta_u=0$, $\theta_k$ and $\zeta_k$ are random in $[0,2\pi]$ and $[0,\pi]$, repectively. From the figure, it is clear that we can obtain the same conclusion as that we have in Fig.  \ref{fig:21} .

 \begin{figure}[!htbp]
\begin{center}
\scalebox{0.1}[0.1]{\includegraphics[angle=0]{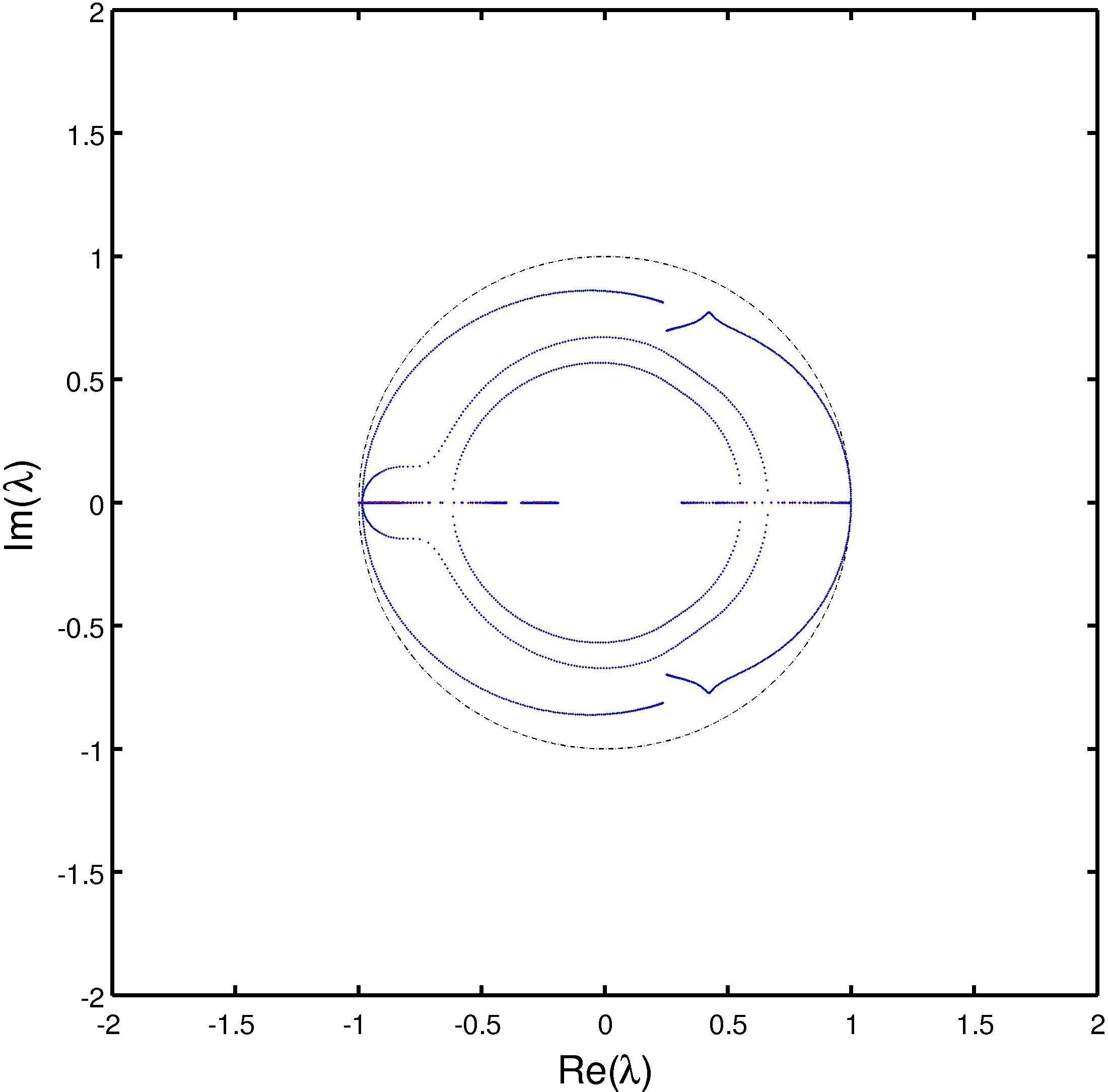}}
\scalebox{0.1}[0.1]{\includegraphics[angle=0]{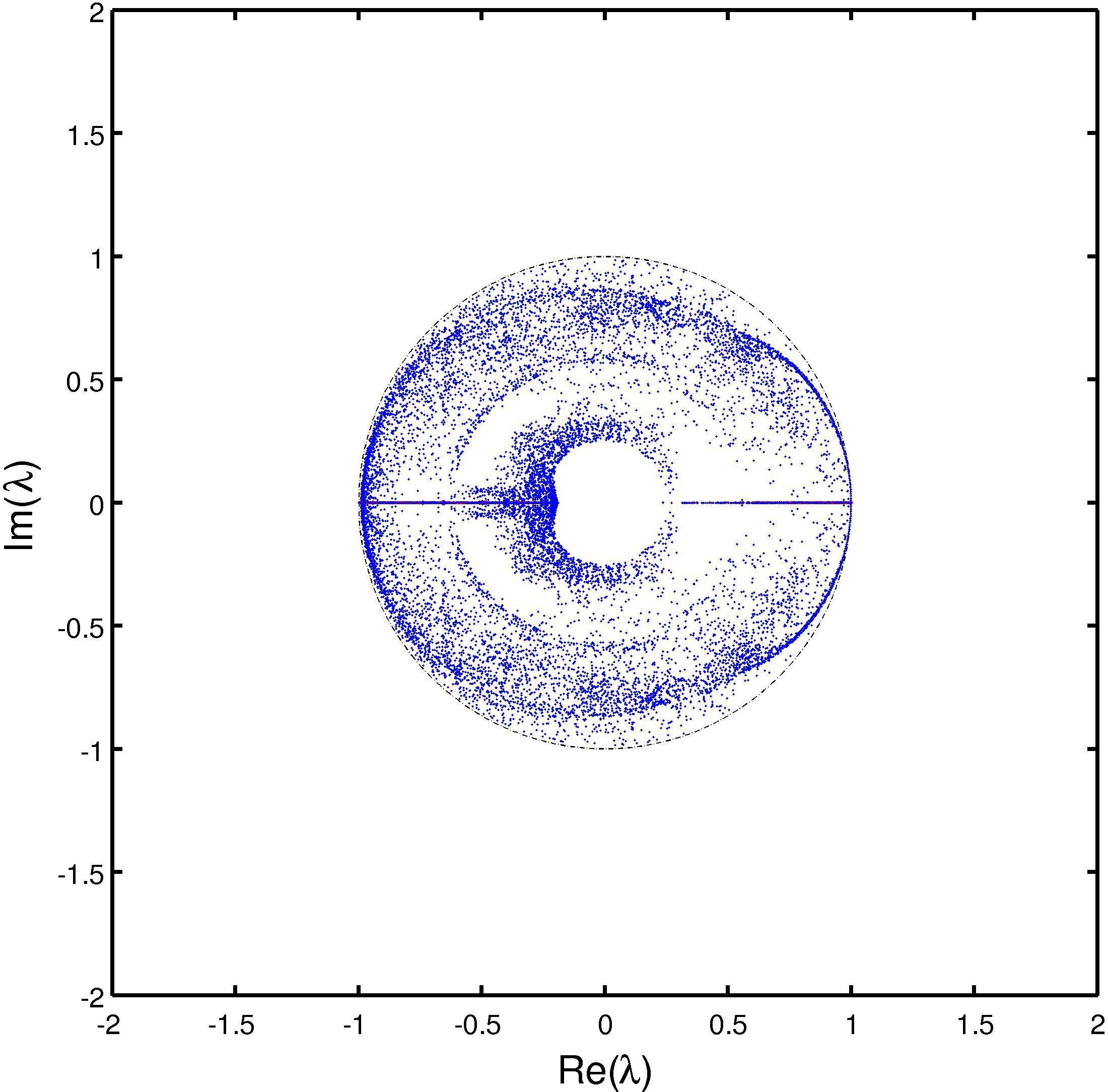}}
\caption{Trajectories of the eigenvalues of the matrix $M^{\rm mrt}$ for the 3D linearized MRT-LBM.  (Left) $\zeta_k=0$; (Right) $\zeta_u=0$. The parameters are from the index 3 of Tabel \ref{table:2}.  (Blue Color) ${\bf k}=[k_x,k_y]$; (Red Color) $ {\bf k}=[-k_x,-k_y]$. The circles in the subfigures stand for the unitary circles with the origin located at (0,0), which are also the stable regions of the linearized MRT-LBM.}\label{fig:21}
\end{center}
\end{figure}

 \begin{figure}[!htbp]
\begin{center}
\scalebox{0.1}[0.1]{\includegraphics[angle=0]{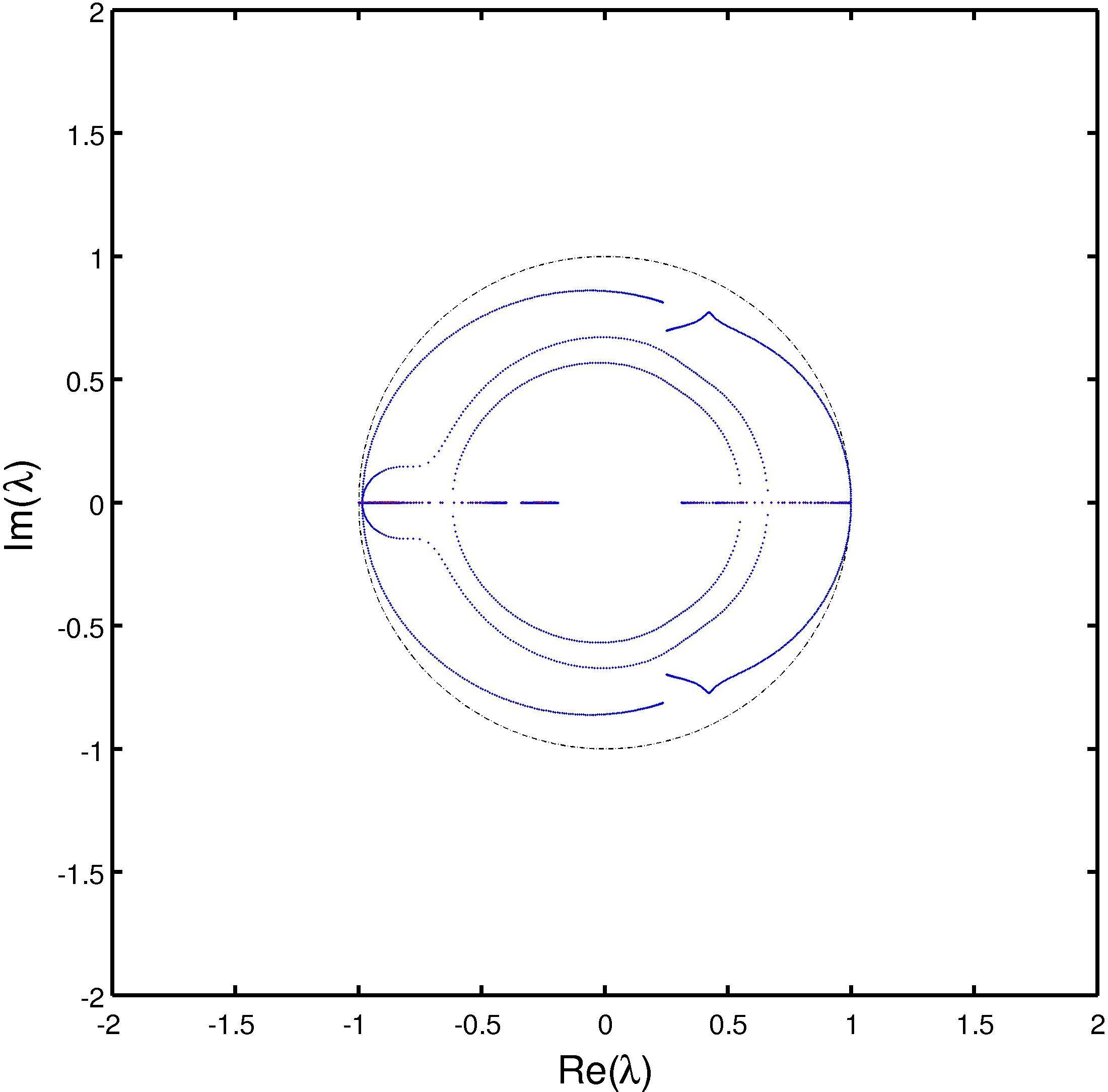}}
\scalebox{0.1}[0.1]{\includegraphics[angle=0]{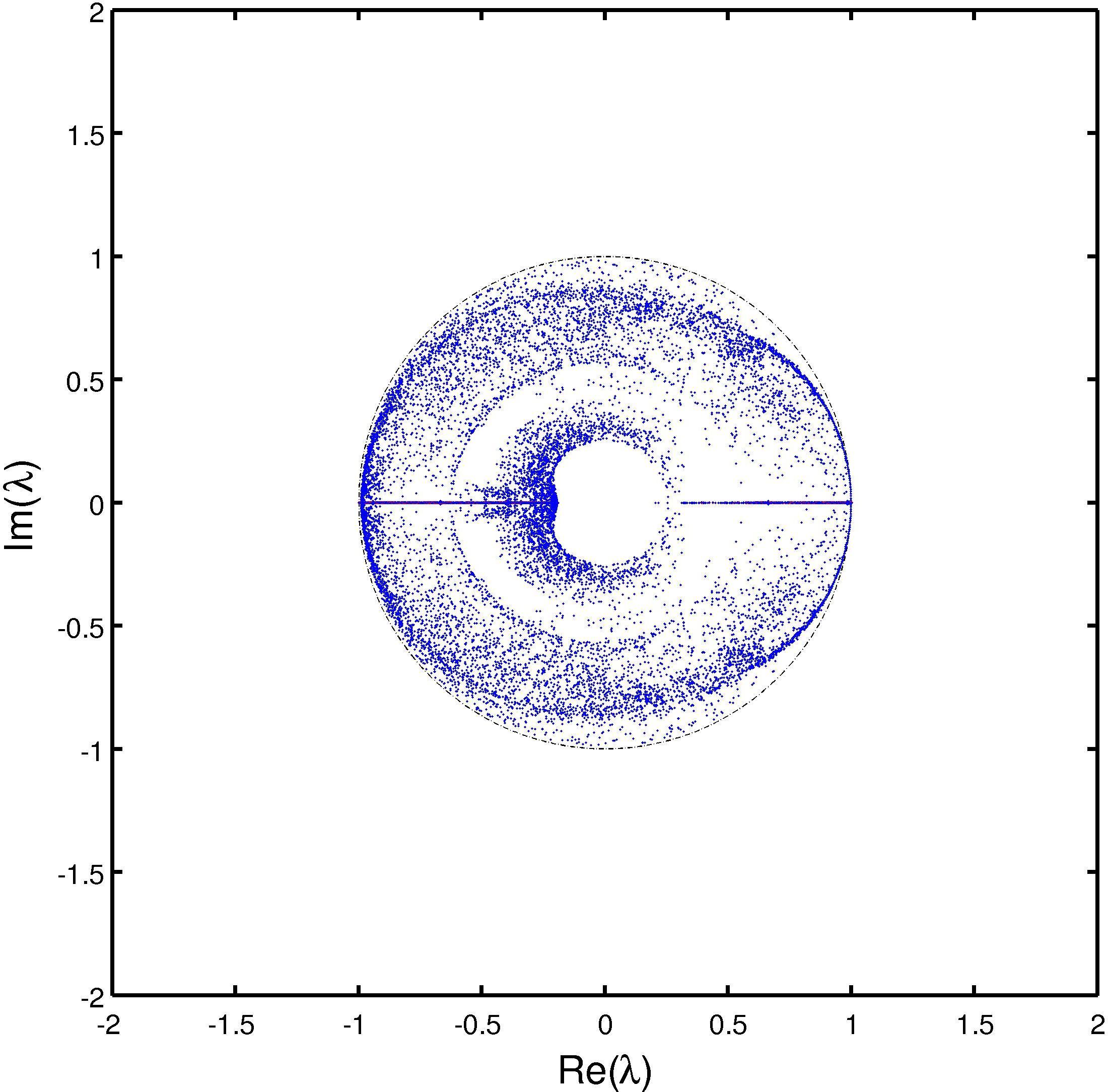}}
\caption{Trajectories of the eigenvalues of the matrix $M^{\rm mrt}$ for the 3D linearized MRT-LBM.  (Left) $\theta_k=0$ and $\zeta_k=0$, $\theta_u$ and $\zeta_u$ are random in $[0,2\pi]$ and $[0,\pi]$, repectively; (Right)  $\theta_u=0$ and $\zeta_u=0$, $\theta_k$ and $\zeta_k$ are random in $[0,2\pi]$ and $[0,\pi]$, repectively. The parameters are from the index 3 of Tabel \ref{table:2}.  (Blue Color) ${\bf k}=[k_x,k_y]$; (Red Color) $ {\bf k}=[-k_x,-k_y]$. The circles in the subfigures stand for the unitary circles with the origin located at (0,0), which are also the stable regions of the linearized MRT-LBM.}\label{fig:22}
\end{center}
\end{figure}

From the investigations of eigenvalue distribution of the matrix $M^{\rm mrt}$ for the  2D and 3D linearized MRT-LBM, it is concluded that for the given $\widehat{\bf u\cdot k}$, the dispersion and dissipation behaviors are mainly determined by the orientation of ${\bf k}$.  Therefore, the acoustic modes and shear modes of the linearized MRT-LBM corresponding to the analytical hydrodynamic modes (\ref{th:d}) are mainly sensitive to the wavenumber vector ${\bf k}$. 

\section{Simplified optimization strategies to determine the free relaxation parameters }

In this section, we will focus on simplifying the  optimization strategy proposed in \cite{xusagaut}. The simplification is based on the sensitivity analysis presented in Sec. \ref{sec:sensitivity} and the properties of the matrix $B^{(n)}$.

An optimization strategy to obtain the optimal free relaxation parameters was proposed in \cite{xusagaut}. 
This optimization strategy is well suited for the 2D/3D MRT-LBM.  
In order to handle low bulk viscosity flows, the researches in the present paper will be focused on dealing with the following minimization problem for 2D and 3D problems \cite{xusagaut}:

\begin{equation}\label{opt:2d}
{\rm Min}:\hskip 1cm \mathcal{G}_{\rm 2d}(\Xi_{\rm r})=\int_{0\leq\theta_k\leq 2\pi,\ \delta t|{\bf k}|\leq\pi,\ 0\leq\theta_u\leq 2\pi,\  |{\bf u}|\leq u_0 }\|\mathcal{M}_{\epsilon}^{(n)}\|_{F}^2{\rm d}\theta_k{\rm d}(\delta t k){\rm d}\theta_u{\rm d}\rm{u},
\end{equation}
and
\begin{equation}\label{opt:3d}
{\rm Min}:\hskip 1cm \mathcal{G}_{\rm 3d}(\Xi_{\rm r})=\int_{0\leq\theta_k\leq 2\pi,\ 0\leq\zeta_k\leq \pi,\  \delta
 t|{\bf k}|\leq\pi,\ 0\leq\theta_u\leq 2\pi,\ 0\leq\zeta_u\leq \pi,\ |{\bf u}|\leq u_0 }\|\mathcal{M}_{\epsilon}^{(n)}\|_{F}^2{\rm d}\theta_k{\rm d}{\zeta_k}{\rm d}(\delta t k){\rm d}\theta_u{\rm d}\zeta_u{\rm d}\rm{u},
\end{equation}
where $\rm{u}$ indicates $|\bf\rm{u}|$, $\Xi_{\rm r}$ denotes the free relaxation parameter set ( $\Xi_{\rm r}\subset \Xi$), $\|\cdot\|_F$ indicates the matrix Frobenius norm and the matrix $\mathcal{M}_{\epsilon}$ is dependent on $|{\bf u}|$, $|{\bf k}|$, $\delta t$ and free relaxation parameters, which is defined by
\begin{equation}\label{res:1}
\mathcal{M}_{\epsilon}^{(n)}= \mathcal{B}^{(n)}-\mathcal{B}+\delta t^2\widehat{\mathcal{B}}.
\end{equation}
In Eq. (\ref{res:1}), $\mathcal{B}^{(n)}$ and  $\mathcal{B}$  are defined by
\begin{equation}
\mathcal{B}^{(n)}=\delta t {B}^{(n)},\  \mathcal{B}=\delta t {B}.
\end{equation}
The matrix $\widehat{\mathcal{B}}$ in Eq. (\ref{res:1}) is given as follows for the 2D/3D L-MRT-LBM
\begin{equation}
 \left[ \begin {array}{ccc} 0&0&0\\\noalign{\medskip}0& -\frac{1}{3}{{k_x}}^{2}
\sigma_e&-\frac{1}{3}{k_x}{k_y}\sigma_e\\\noalign{\medskip}0&-\frac{1}{3}
{k_x}{k_y}\sigma_e&-\frac{1}{3}{{k_y}}^{2}\sigma_e\end {array}
 \right] ,
\end{equation}
and
\begin{equation}
  \left[ \begin {array}{cccc} 0&0&0&0\\\noalign{\medskip}0&-\frac{1}{9}{{ k_x}}
^{2}({ \sigma_\nu}+2{ \sigma_e})&-\frac{1}{9}{ k_y}{
 k_x} \left( { \sigma_\nu}+2{ \sigma_e} \right) &-\frac{1}{9}{ k_z}
{ k_x} \left( { \sigma_\nu}+2{ \sigma_e} \right) 
\\\noalign{\medskip}0&-\frac{1}{9}{ k_y}{ k_x} \left( { \sigma_\nu}+
2{ \sigma_e} \right) &-\frac{1}{9}{{ k_y}}
^{2}({ \sigma_\nu}+2{ \sigma_e})&-\frac{1}{9}{ k_z}{ k_y} \left( { \sigma_\nu}+2{
 \sigma_e} \right) \\\noalign{\medskip}0&-\frac{1}{9}{ k_z}{ k_x}
 \left( { \sigma_\nu}+2{ \sigma_e} \right) &-\frac{1}{9}{ k_z}{ k_y
} \left( { \sigma_\nu}+2{ \sigma_e} \right) &-\frac{1}{9}{{ k_z}}^{2}({ \sigma_\nu}+2{ \sigma_z})\end {array} \right] ,
\end{equation}
which are corresponding to the exact total bulk viscosity dissipation coefficient matrix of $\delta t$ (refer to \ref{App:coef} ).

According to the analysis displayed in Sec. \ref{ch:sensitivities},  the eigenvalues are mainly sensitive to the wave number ${\bf k}$. Therefore, taking  ${\bf u}$ to be parallel to ${\bf e_x}$ and considering the most sensitive factors,  we can simplify Eq.(\ref{opt:2d}) and Eq. (\ref{opt:3d}) as follows
\begin{equation}\label{opt:2do}
{\rm Min}:\hskip 1cm \mathcal{G}_{\rm 2d}(\Xi_{\rm r})=\int_{0\leq\theta_k\leq 2\pi,\ \delta t|{\bf k}|\leq\pi, \  |{\bf u\cdot e_x}|\leq u_0 }\|\mathcal{M}_{\epsilon}^{(n)}\|_{F}^2{\rm d}\theta_k{\rm d}(\delta t k){\rm d}\rm{u},
\end{equation}
and
\begin{equation}\label{opt:3do}
{\rm Min}:\hskip 1cm \mathcal{G}_{\rm 3d}(\Xi_{\rm r})=\int_{0\leq\theta_k\leq 2\pi,\ 0\leq\zeta_k\leq \pi,\ \delta t|{\bf k}|\leq\pi , \  |{\bf u\cdot e_x}|\leq u_0}\|\mathcal{M}_{\epsilon}^{(n)}\|_{F}^2{\rm d}\theta_k{\rm d}{\zeta_k}{\rm d}(\delta t k){\rm d}\rm{u},
\end{equation}
where $\rm{u}=|{\rm\bf u\cdot e_x}|$. 
More specifically, the above simplified formula Eq. (\ref{opt:3do}) is very effective for obtaining the free relaxation parameters of the 3D problems, which reduce the integral variables. 

In Table \ref{table:3}, illustrative examples are given for both 2D and 3D problems. The truncated error of Eq. (\ref{R-F}) is set to equal 4 for handling the  acoustic problems with  the low bulk viscosity and for guaranteeing stability \cite{xusagaut}.  
In order to investigate the dispersion and dissipation relations for the optimal MRT-LBM, the profiles of the transverse modes and the longitudinal modes are shown in Figs. \ref{fig:23}$\sim$\ref{fig:26}. 
From these figures, it is clear that the simplified optimization strategies  (\ref{opt:2do})$\sim($\ref{opt:3do}) can make the MRT-LBM schemes endowed with the low-dispersion low-dissipation  properties. 
From the viewpoint of  stability, the optimal MRT-LBM appears more stable than the original MRT-LBM \cite{lallemandluo,xusagaut}. Especially, for the optimal MRT-LBM, the lower bulk dissipation parameter can be chosen for  aeroacoustic problems. 
In the original MRT-LBM, because of sustaining the stability of the MRT-LBM schemes, the low bulk viscosity is forbidden or impossible  \cite{lallemandluo,xusagaut}. From Figs. \ref{fig:23} - \ref{fig:25}, it is observed that for the optimal MRT-LBM, the acoustic modes with  ${\bf u}$ perpendicular to ${\bf k}$  are more stable than the acoustic modes with  ${\bf u}$ parallel to ${\bf k}$, and the shear modes with  ${\bf u}$ perpendicular to ${ \bf k}$  are more accurate  than the shear modes with  ${\bf u}$ parallel to ${\bf k}$.

\begin{table}[!htbp]
\caption{Optimized free relaxation parameters corresponding to the specified $s_\nu$ and $s_e$ for the MRT-LBM under the condition ${\bf u \Vert e_x}$ ($\widehat{\bf u\cdot e_x}=0$) . The upper bound $u_0$ of  $|{\bf u}|$ is equal to 0.2.   The symbol ``$\diagdown$ " indicates that the corresponding fields are inapplicable. For D2Q9, ${\bf e_x}=(1,0)$. For D3Q19, ${\bf e_x}=(1,0,0)$. The truncated error in Eq. (\ref{R-F}) is equal to $\delta t^4$. }\label{table:3}
\begin{tabular*}{\textwidth}{@{\extracolsep{\fill}}llllllll}\toprule[1pt]
Model &${\rm Index}$&$s_{\nu}$& $s_e$ & $s_\epsilon$ & $s_q$ & $s_\pi$& $s_t$\\ \midrule[1pt]
D2Q9& 0 &1.99960008   &1.99960008 & 1.997623852 & 1.999448768 & $\diagdown$& $\diagdown$ \\
D3Q19& 1 &1.99960008   &1.99960008 & 1.994155318 & 1.998982400 &1.992025433& 1.999363601  \\
D2Q9& 2 & 1.886792453   &1.886792453 & 1.470217043 & 1.842711354 & $\diagdown$& $\diagdown$ \\
D3Q19&3 & 1.999960001 & 1.9999800002& 1.9993910550&1.999916523&1.9996934332&1.999886136\\ 
D2Q9 & 4&1.995389843 & 1.9953898430 &  1.9729857787 & 1.9875516898&$\diagdown$& $\diagdown$ \\
& 5&1.999996000& 1.9999960008& 1.9999762501& 1.999994487&$\diagdown$& $\diagdown$\\
& 6&1.9999988&1.9999988&1.9999928163& 1.9999992651&$\diagdown$& $\diagdown$\\
\bottomrule[1pt]
\end{tabular*}
\end{table}

 \begin{figure}[!htbp]
\begin{center}
\scalebox{0.55}[0.55]{\includegraphics[angle=0]{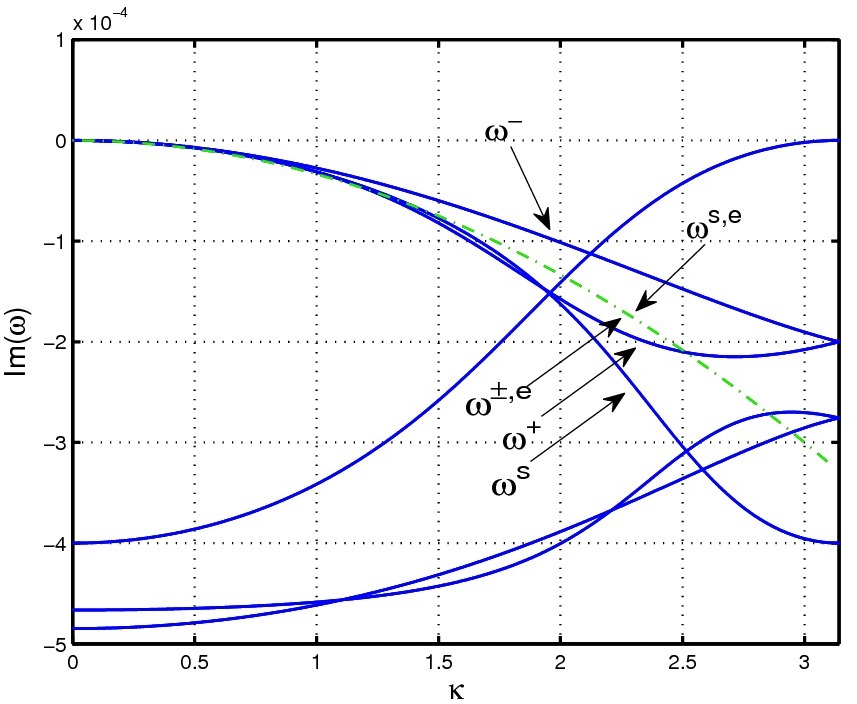}}
\scalebox{0.55}[0.55]{\includegraphics[angle=0]{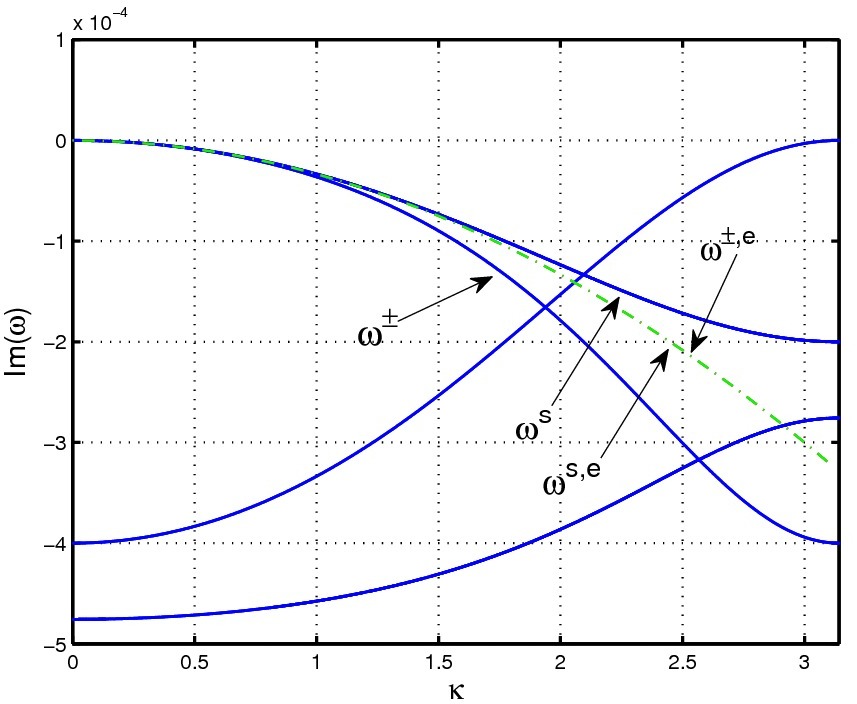}}\\
{\centering (a) ${\bf u\Vert k}$\hspace{7cm} (b) ${\bf u\bot k}$}
\caption{The dissipation relations for the optimized 2D MRT-LBM: (a) $\widehat{\bf u\cdot e_x}=0$; (b)  $\widehat{\bf u\cdot e_x}=0$,  $\widehat{\bf k\cdot e_x}=\pi/2$. The relaxation parameters are given by Index 0 in Table \ref{table:3}. The magnitude of ${\bf u}$ is equal to 0.2. $\omega^{\pm,e}$ and $\omega^{s,e}$ denote the exact acoustic modes and shear modes respectively. $\omega^{\pm}$ and $\omega^{s}$ denote the numerical acoustic modes and shear modes respectively.}\label{fig:23}
\end{center}
\end{figure}
 \begin{figure}[!htbp]
\begin{center}
\scalebox{0.55}[0.55]{\includegraphics[angle=0]{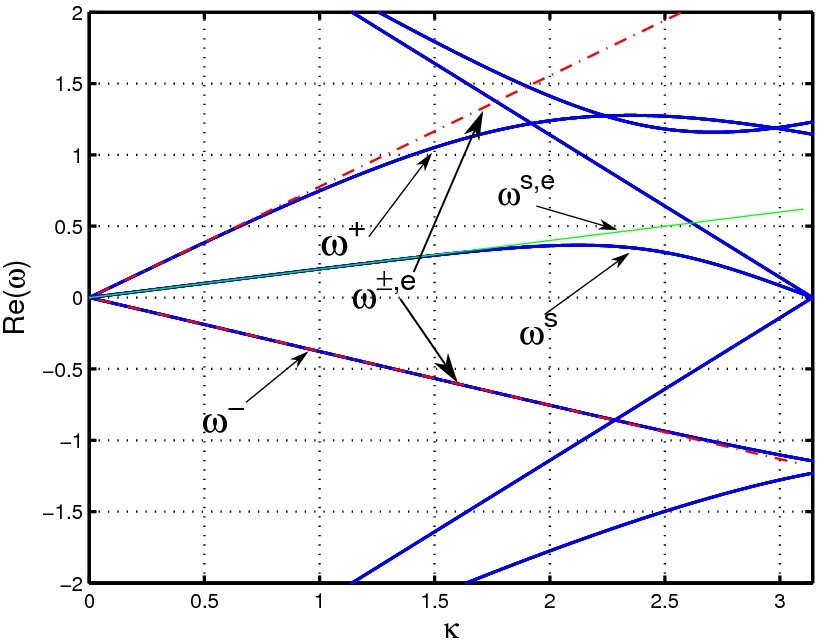}}
\scalebox{0.55}[0.55]{\includegraphics[angle=0]{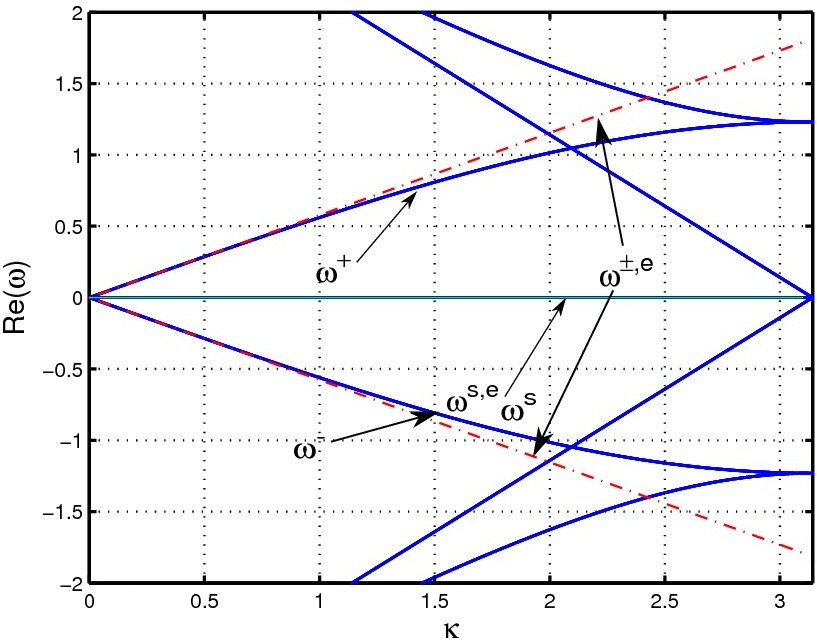}}\\
{\centering (a) ${\bf u\Vert k}$\hspace{7cm} (b) ${\bf u\bot k}$}
\caption{The dispersion relations for the optimized 2D MRT-LBM: (a) $\widehat{\bf u\cdot e_x}=0$; (b)  $\widehat{\bf u\cdot e_x}=0$,  $\widehat{\bf k\cdot e_x}=\pi/2$. The relaxation parameters are given by Index 0 in Table \ref{table:3}. The magnitude of ${\bf u}$ is equal to 0.2. (Green line) Exact shear modes ${\rm Re}[\omega^{s}]$;  (Red line) Exact acoustic modes ${\rm Re}[\omega^{\pm}]$. $\omega^{\pm,e}$ and $\omega^{s,e}$ denote the exact acoustic modes and shear modes respectively. $\omega^{\pm}$ and $\omega^{s}$ denote the numerical acoustic modes and shear modes respectively.}\label{fig:24}
\end{center}
\end{figure}

 \begin{figure}[!htbp]
\begin{center}
\scalebox{0.55}[0.55]{\includegraphics[angle=0]{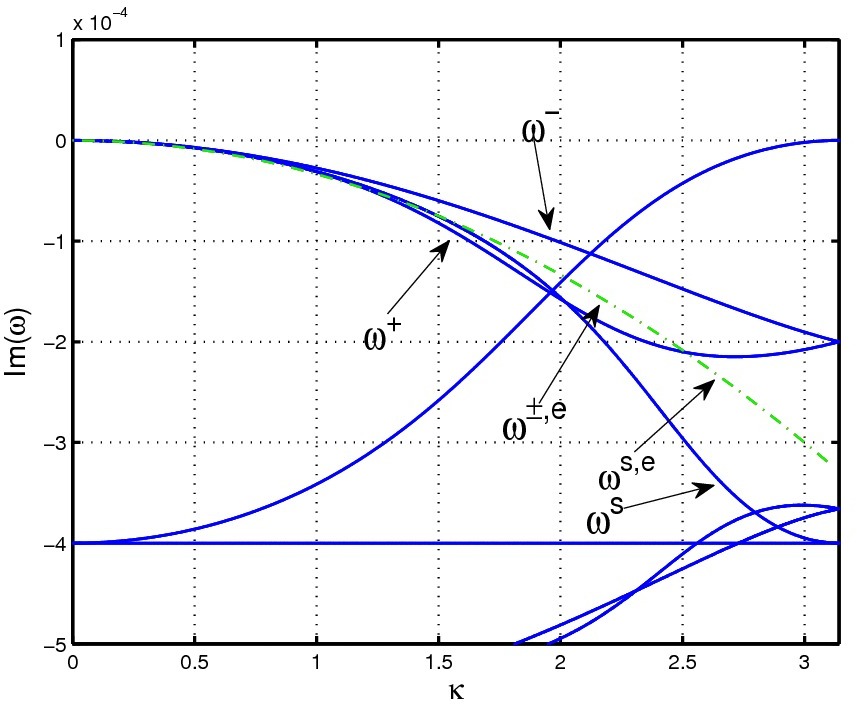}}
\scalebox{0.55}[0.55]{\includegraphics[angle=0]{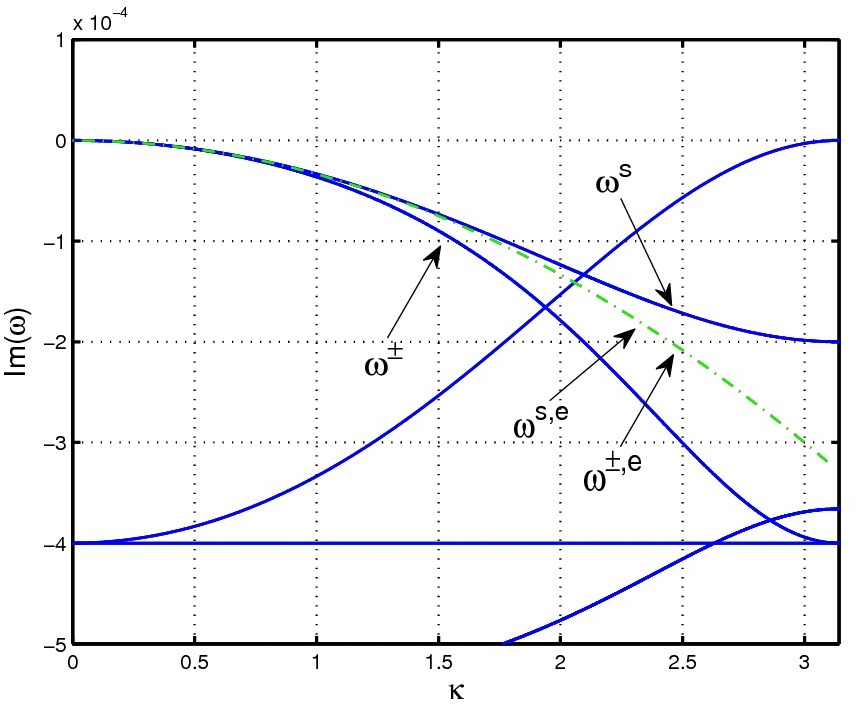}}\\
{\centering (a) ${\bf u\Vert k}$\hspace{7cm} (b) ${\bf u\bot k}$}
\caption{The dissipation relations for the optimized 3D MRT-LBM: (a) $\widehat{\bf u\cdot e_x}=0$; (b)  $\widehat{\bf u\cdot e_x}=0$, $\theta_k=\theta_u=0$ and $\zeta_k=\pi/2$ in Eq. (\ref{3d:k}). The relaxation parameters are given by Index 1 in Table \ref{table:3}. The magnitude of ${\bf u}$ is equal to 0.2. $\omega^{\pm,e}$ and $\omega^{s,e}$ denote the exact acoustic modes and shear modes respectively. $\omega^{\pm}$ and $\omega^{s}$ denote the numerical acoustic modes and shear modes respectively.}\label{fig:25}
\end{center}
\end{figure}
 \begin{figure}[!htbp]
\begin{center}
\scalebox{0.55}[0.55]{\includegraphics[angle=0]{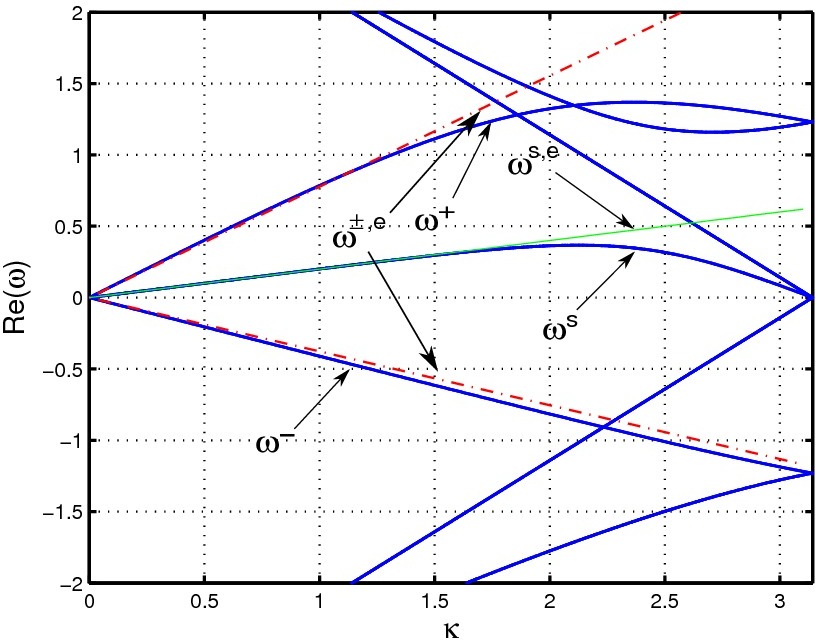}}
\scalebox{0.55}[0.55]{\includegraphics[angle=0]{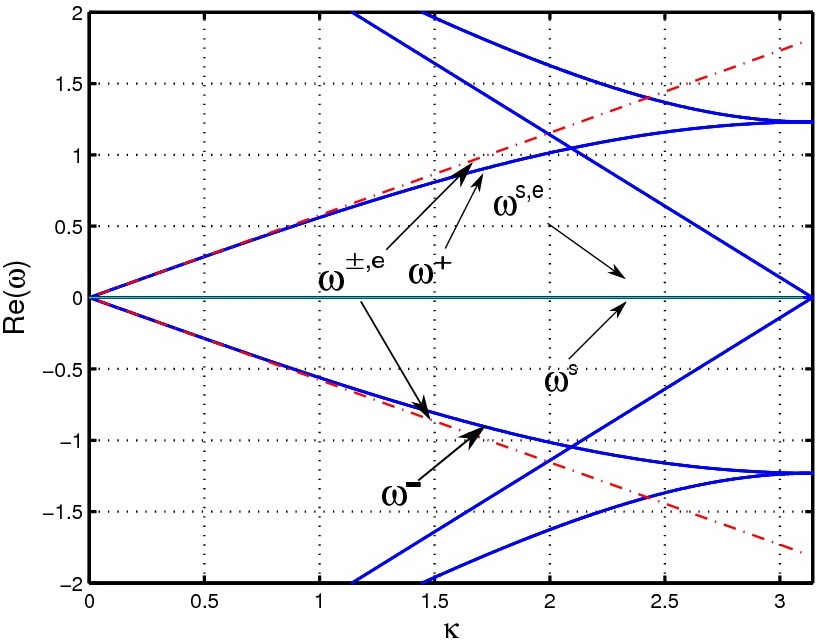}}\\
{\centering (a) ${\bf u\Vert k}$\hspace{7cm} (b) ${\bf u\bot k}$}
\caption{The dispersion relations for the optimized 3D MRT-LBM: (a) $\widehat{\bf u\cdot e_x}=0$; (b)  $\widehat{\bf u\cdot e_x}=0$, $\theta_k=\theta_u=0$ and $\zeta_k=\pi/2$ in Eq. (\ref{3d:k}). The relaxation parameters are given by Index 1 in Table \ref{table:3}. The magnitude of ${\bf u}$ is equal to 0.2. (Green line) Exact shear modes ${\rm Re}[\omega^{s}]$;  (Red line) Exact acoustic modes ${\rm Re}[\omega^{\pm}]$. $\omega^{\pm,e}$ and $\omega^{s,e}$ denote the exact acoustic modes and shear modes respectively. $\omega^{\pm}$ and $\omega^{s}$ denote the numerical acoustic modes and shear modes respectively.}\label{fig:26}
\end{center}
\end{figure}

\section{Numerical examples}

In this section, we perform some numerical simulations to validate the optimal  MRT-LBM on 2D/3D benchmark problems .   
All simulations are performed using an ad hoc version of PalaBos \cite{palabos}. 
 For the current optimization procedures,  the truncated error term is only up to the fourth order so that we can improve the dissipation relation errors of the MRT-LBM and the isotropic errors \cite{xusagaut}. In this part, we only focus on the case with the fourth-order truncated error terms. If you want to further improve  both the dispersion and dissipation errors, the higher-order truncated error terms are needed \cite{xusagaut}. 

\subsection{A 2D acoustic point source}

In this part, we consider propagation of the 2D acoustic wave generated by a point source. These investigations only focus on comparing the optimal MRT-LBM  with the BGK-LBM and the quasi-incompressible BGK-LBM. The time-dependent acoustic point source is set as 
\begin{equation}
\left\{\begin{array}{ll}
\rho(\cdot,t)&=1+\rho\prime{\rm sin}(2\pi/T\cdot t)\\
u(\cdot,t)&=0\\
v(\cdot,t)&=0\\
\end{array}\right.
\end{equation} 
where $\rho\prime=0.01$. 
The compuational domain  is $\Omega=[0,L]^2$ ($L=1$). 
The number of the lattice points is equal to $200\times 200$.
 The period $T$ is equal to $4\cdot \delta x$ ($\delta x=1/200$). 
 The relaxation parameters in the standard LBM are set by Index 5 in Table \ref{table:3}.  
 In Fig. \ref{fig:comp}, the contour lines for three different methods are shown based on the same contour level values. 
 It is very clear that the contour line result obtained by the optimal MRT-LBM is the best one. 
 The BGK-LBM  has nearly  the same results as the quasi-incompressible BGK-LBM. This results are  similar to that in former researches \cite{xusagaut}. Here, we point out that, with respect to the optimal MRT-LBM, the contour line result of the acoustic point source in our former research is better than the result of the optimal MRT-LBM in the current paper. It is needed to be clarified that  in the current optimization strategy, the mean flow influence is considered, but in the former research, the optimization strategy for the kind of  problem was implemented with the zero mean flows. The optimization strategies in this paper appear in a more general form. From the results of this problem, the well performances of the optimization strategies are further demonstrated. 

 \begin{figure}[!htbp]
\begin{center}
\scalebox{0.35}[0.35]{\includegraphics[angle=0]{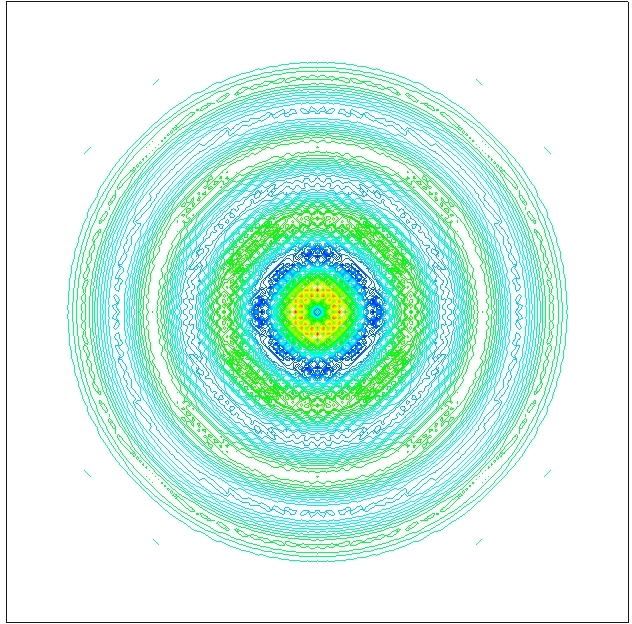}}
\scalebox{0.35}[0.35]{\includegraphics[angle=0]{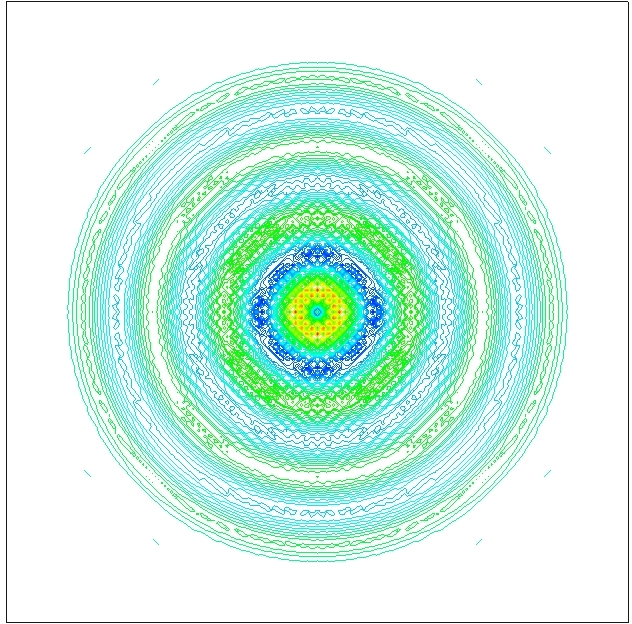}}\\
\centering{\hspace{2cm}(a) BGK-LBM \hspace{4cm}(b) Quasi-incompressible BGK-LBM}\\
\scalebox{0.35}[0.35]{\includegraphics[angle=0]{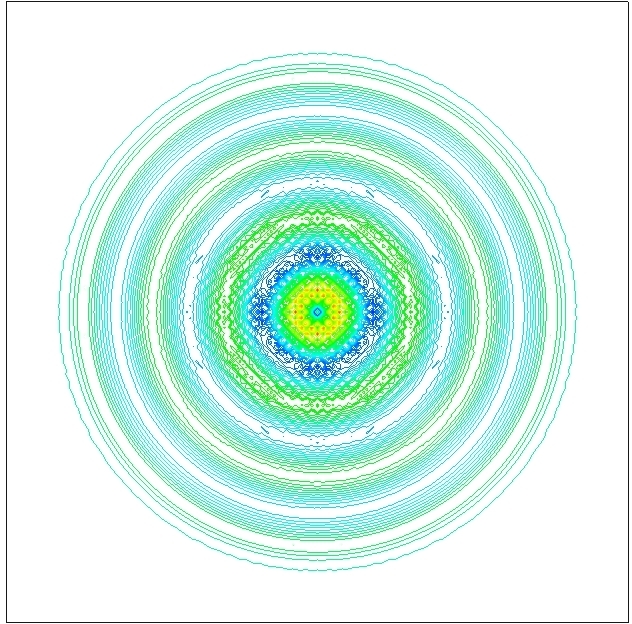}}\\
\centering{(c) Optimal MRT-LBM}\\
\caption{Density contour lines for the BGK-LBM, the quasi-incompressible BGK-LBM and the optimal MRT-LBM at $t=0.7$. The viscosity $\nu=1.66667\times 10^{-5}$.}\label{fig:comp}
\end{center}
\end{figure}

\subsection{The 2D acoustic pulse source}
We now consider the 2D acoustic pulse source for assessing the simplified 2D optimization strategy.
Assuming that  the viscosity effect is negligible on acoustic waves, the acoustic pulse problem possesses an analytical solution \cite{xusagaut,tam}.  The initial profile is given as follows
\begin{equation}
\left\{\begin{array}{ll}
\rho_0&=1+\rho\prime\\
u_0 &=U_0\\
v_0 & =0\\
\end{array}\right.
\end{equation}
where $\rho\prime$, $\epsilon$, $\alpha$, $r$ and $U_0$ are defined by 
\begin{equation}\label{3:para0}
\rho\prime=\epsilon{\rm exp}(-\alpha\cdot r^2),\ \epsilon=10^{-3},\ \alpha={\rm ln}(2)/b^2,\ r=\sqrt{(x-x_0)^2+(y-y_0)^2},\ U_0=0.01.
\end{equation}
The parameter $b$ in Eq. (\ref{3:para0}) is equal to 0.03. The exact solution of $\rho\prime$ (if $(x_0,y_0)=(0,0)$) is given by \cite{tam}
\begin{equation}
\rho\prime(x,y,z)=\frac{\epsilon}{2\alpha}\int_{0}^{\infty}{\rm exp}\left(-\frac{\xi^2}{4\alpha}\right){\rm cos}(c_st\xi){\rm J}_0(\xi\eta)\xi{\rm d}\xi
\end{equation}
where $\eta=\sqrt{(x-U_0t)^2+y^2+z^2}$ and ${\rm J}_0(\cdot)$ is the Bessel zero-order function of the first kind. The computational domain $\Omega=[0,L]^2$ ($L=1$).  The relaxation parameters are determined by Index 5 in Table \ref{table:3}.   In order to validate the precision of the optimal MRT-LBM, the mesh convergence and the time convergence are calculated. The adopted lattice resolutions are $100^2$, $200^2$, $300^2$ and $400^2$.  The evolution of the acoustic pulse is restricted in the computational domain in order to avoid the effects of boundaries. From here on, the $L^2$ relative error is defined by 
\begin{equation}
E_{L^2}(t)=\sqrt{\frac{\sum_{i=1}^{N_{\rm nodes}}(\rho\prime_{\rm th}({\rm x}_i,t)-\rho\prime_{\rm num}({\rm x}_i,t))^2}{\sum_{i=1}^{N_{\rm nodes}}(\rho\prime_{\rm th}({\rm x}_i,t))^2}},
\end{equation}
where $\rho\prime_{\rm th}({\rm x}_i,t)$ and $\rho\prime_{\rm num}({\rm x}_i,t)$ denote the theoretical and numerical solutions at the lattice nodes ${\rm x}_i$, respectively. First, in Fig. \ref{fig:27}, the density profiles for different LBS are given at $t=0.4$. 
It is clear that the results obtained by the original MRT-LBM are less satisfactory because of the large bulk viscosity. 
In order to show the quantitative comparisons,  the $L^2$ relative errors are given in Figs. \ref{fig:28}$\sim$\ref{fig:29} with respect to the mesh resolution and the evolution time, respectively.  
From Fig. \ref{fig:28}, it is clear that the results obtained by the optimal MRT-LBM have the lowest $L^2$ relative errors and the best convergence property. Further, some super-convergence properties are exhibited for the optimal MRT-LBM.  Let us observe Fig. \ref{fig:29} and we can conclude that when the mesh resolutions are fixed, the original MRT-LBM has the best precision. With respect to the evolution time, three methods have nearly the same time convergence order.

 \begin{figure}[!htbp]
\begin{center}
\scalebox{0.44}[0.44]{\includegraphics[angle=0]{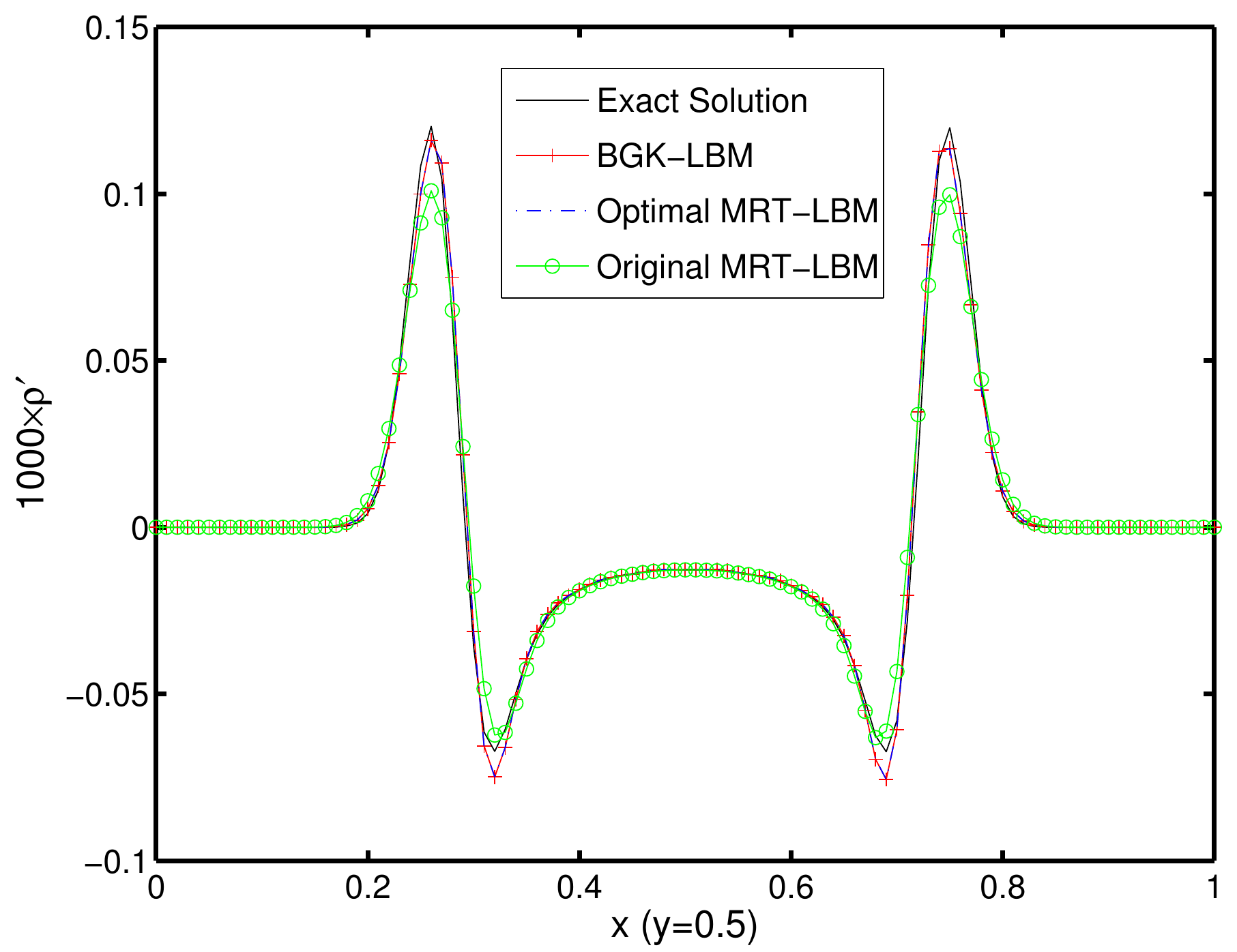}}
\scalebox{0.44}[0.44]{\includegraphics[angle=0]{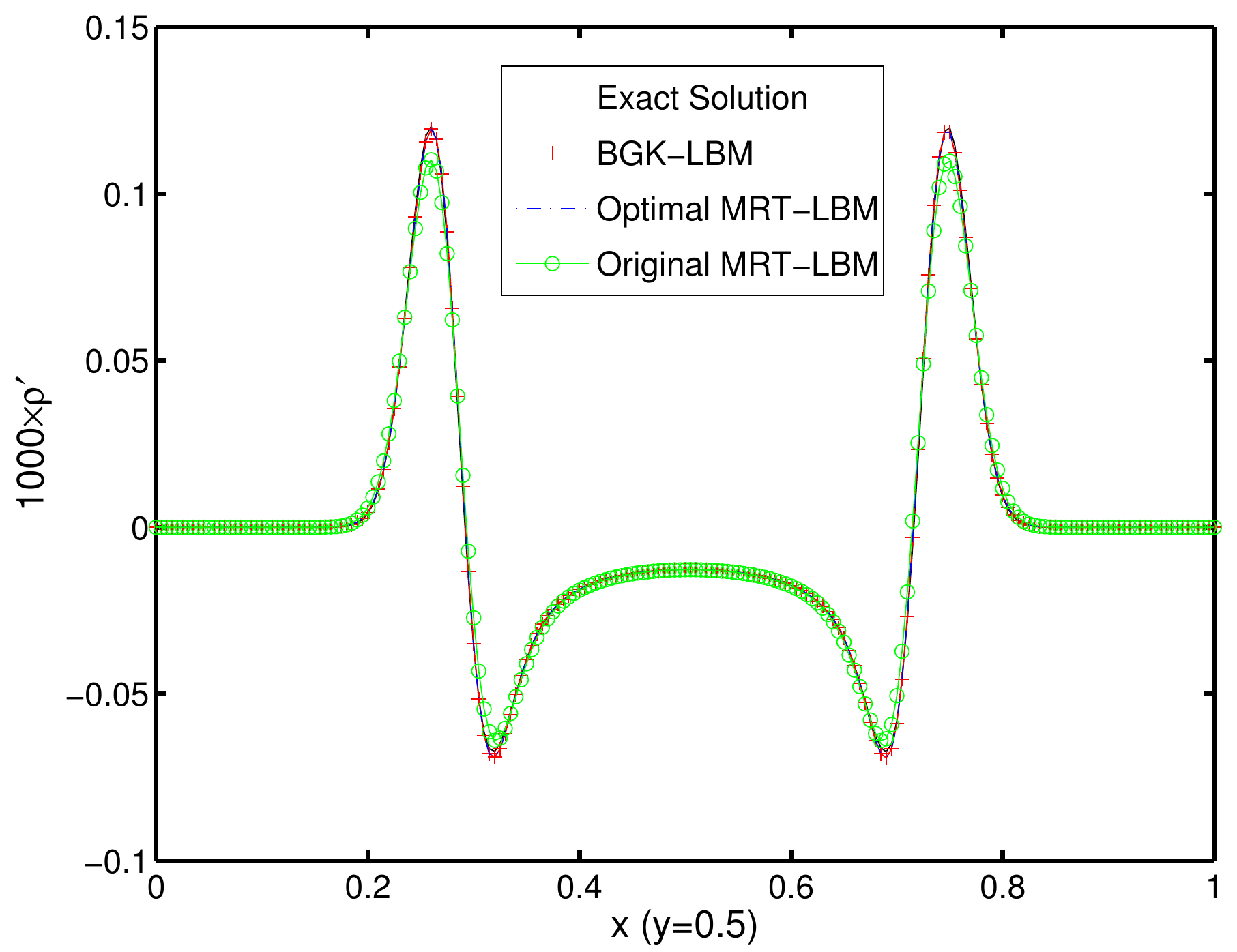}}\\
{\centering (a) Lattice $100^2$\hspace{6cm}(b)Lattice $200^2$}\\
\scalebox{0.44}[0.44]{\includegraphics[angle=0]{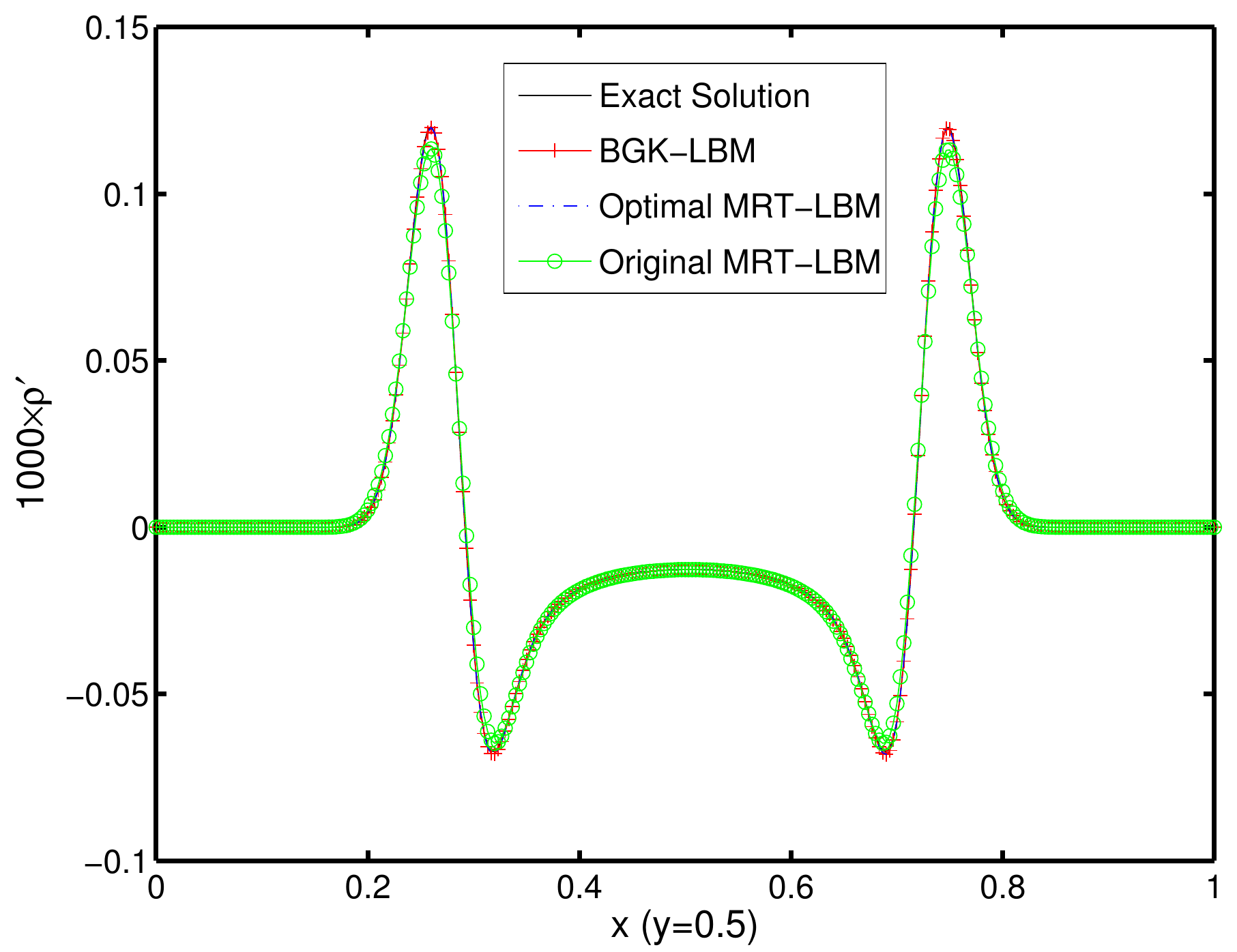}}
\scalebox{0.44}[0.44]{\includegraphics[angle=0]{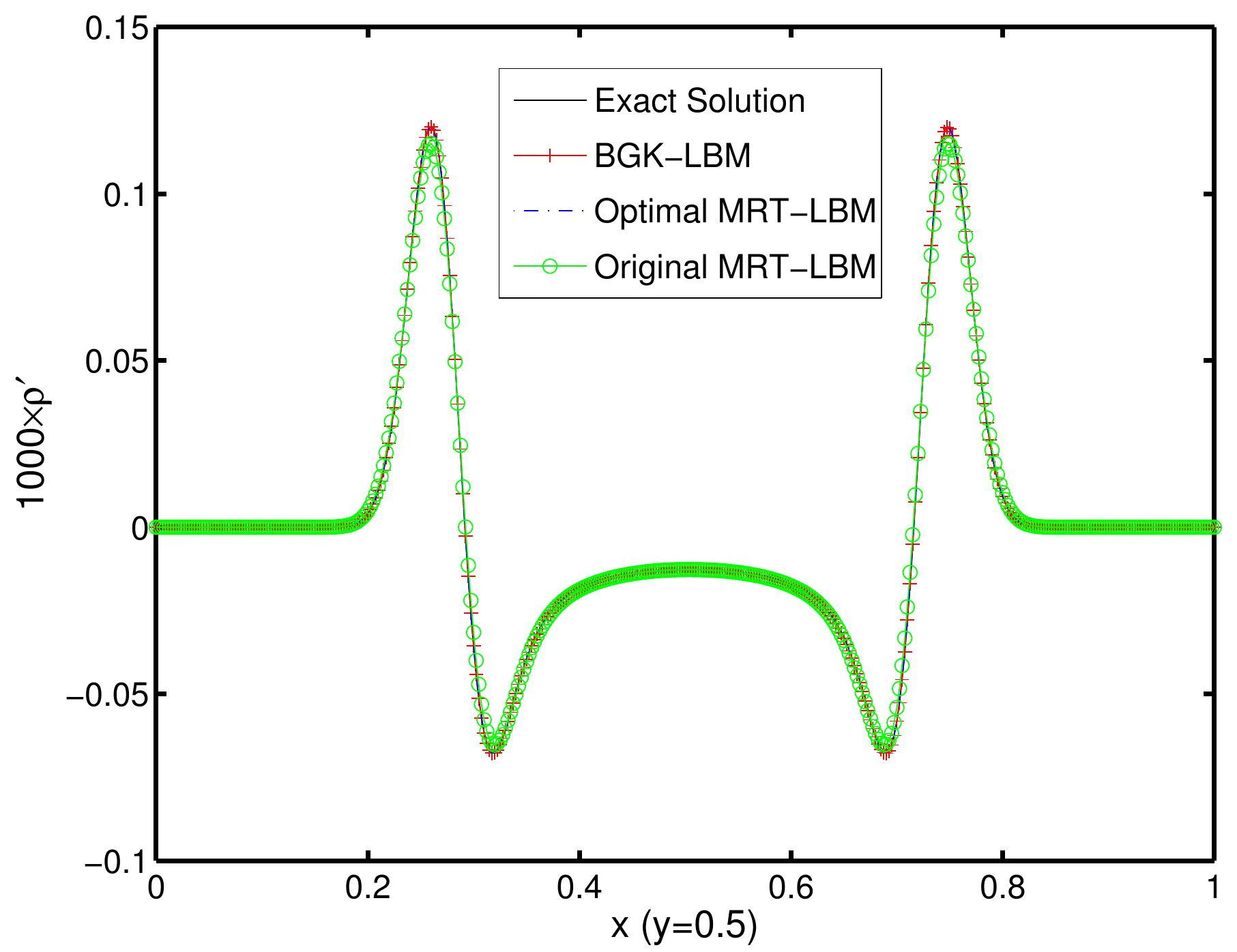}}\\
{\centering (c) Lattice $300^2$\hspace{6cm}(d)Lattice $400^2$}\\
\caption{Fluctuation density profiles of $1000\rho\prime$ for three different 2D LBS along the line y=0.5 at t=0.4: (a) Lattice number $100^2$; (b) Lattice number $200^2$; (c) Lattice number $300^2$; (d) Lattice  number $400^2$. }\label{fig:27}
\end{center}
\end{figure}

 \begin{figure}[!htbp]
\begin{center}
\scalebox{0.44}[0.44]{\includegraphics[angle=0]{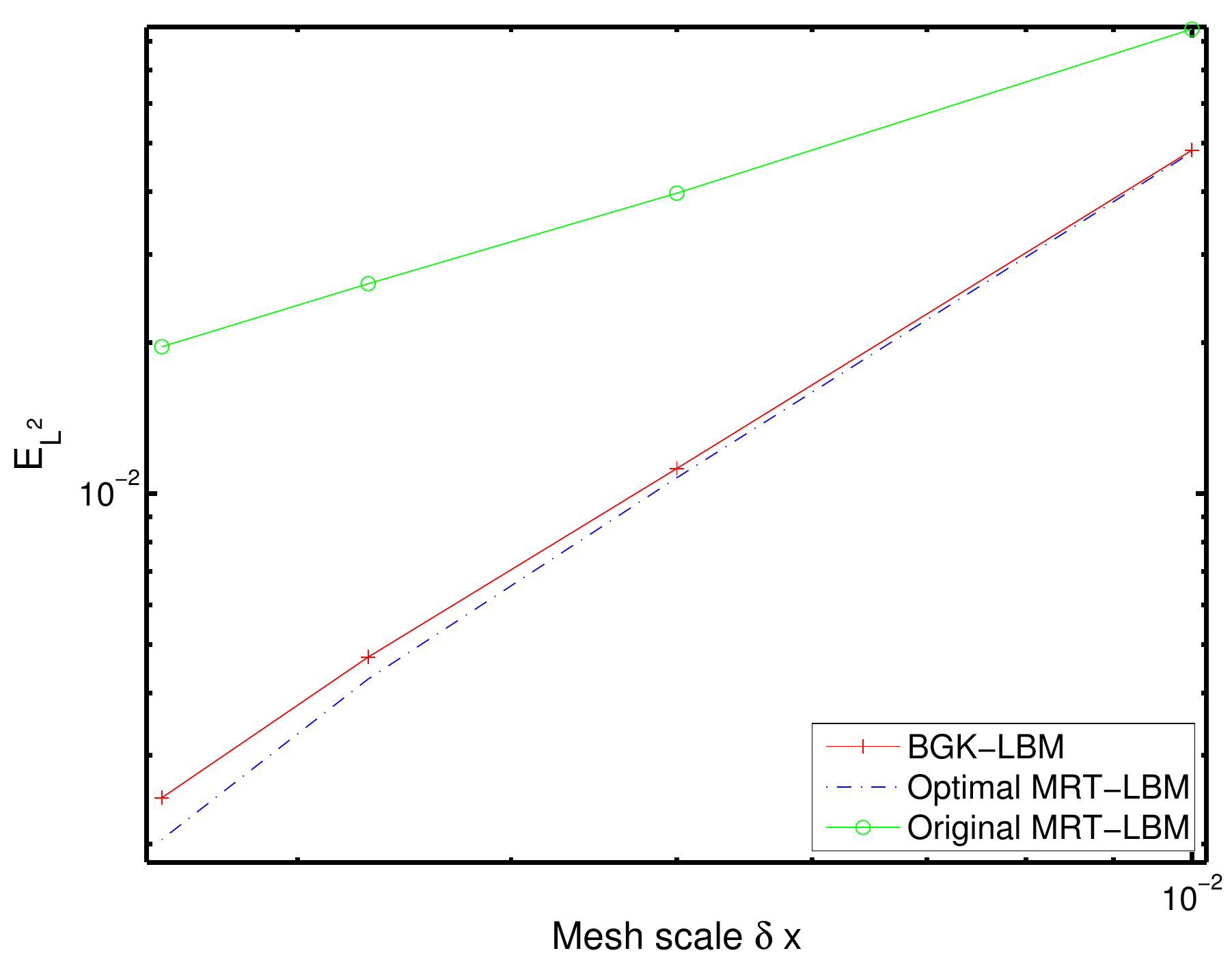}}
\scalebox{0.44}[0.44]{\includegraphics[angle=0]{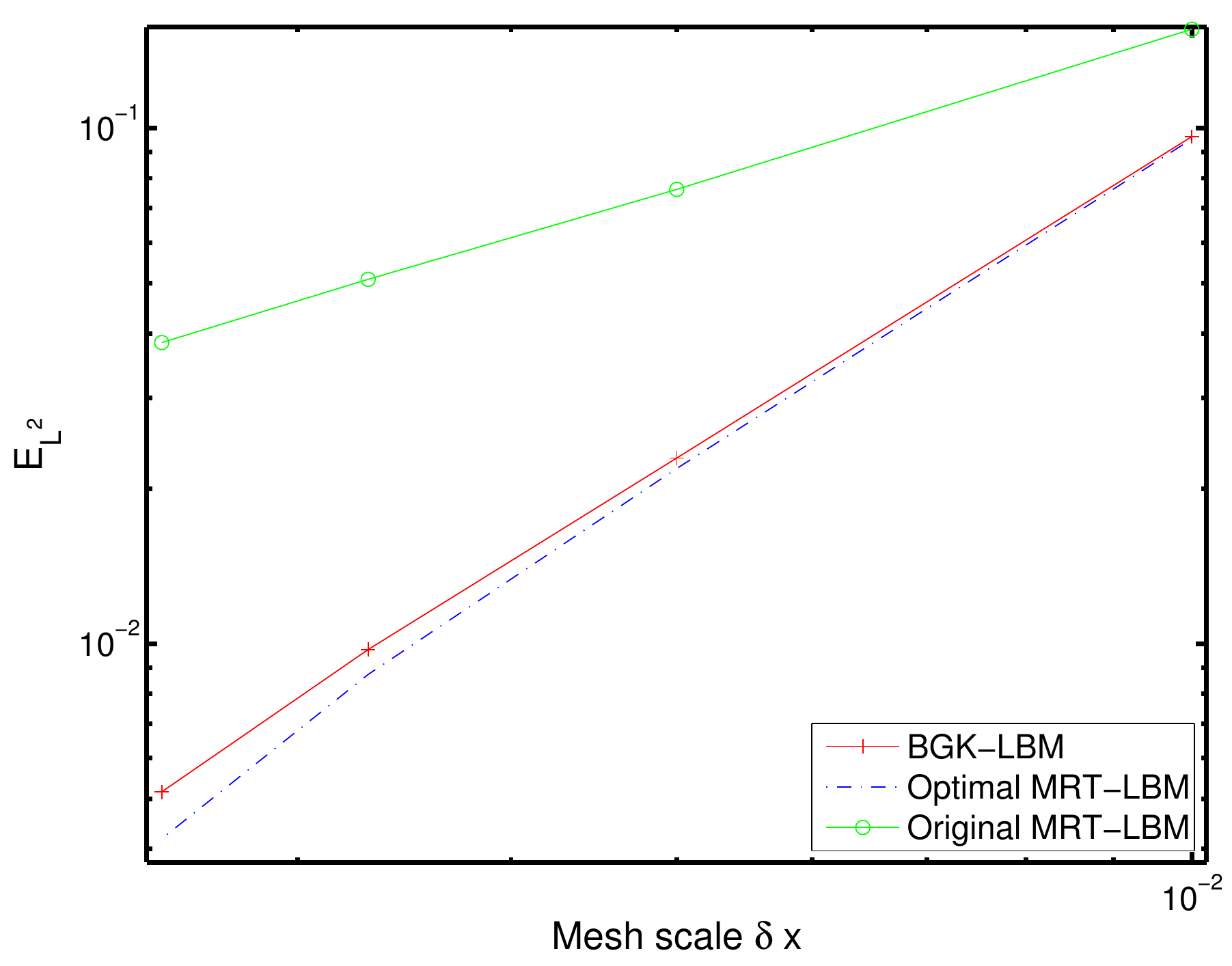}}\\
{\centering (a) at t=0.2 \hspace{6cm}(b)at t=0.4 }\\
\scalebox{0.44}[0.44]{\includegraphics[angle=0]{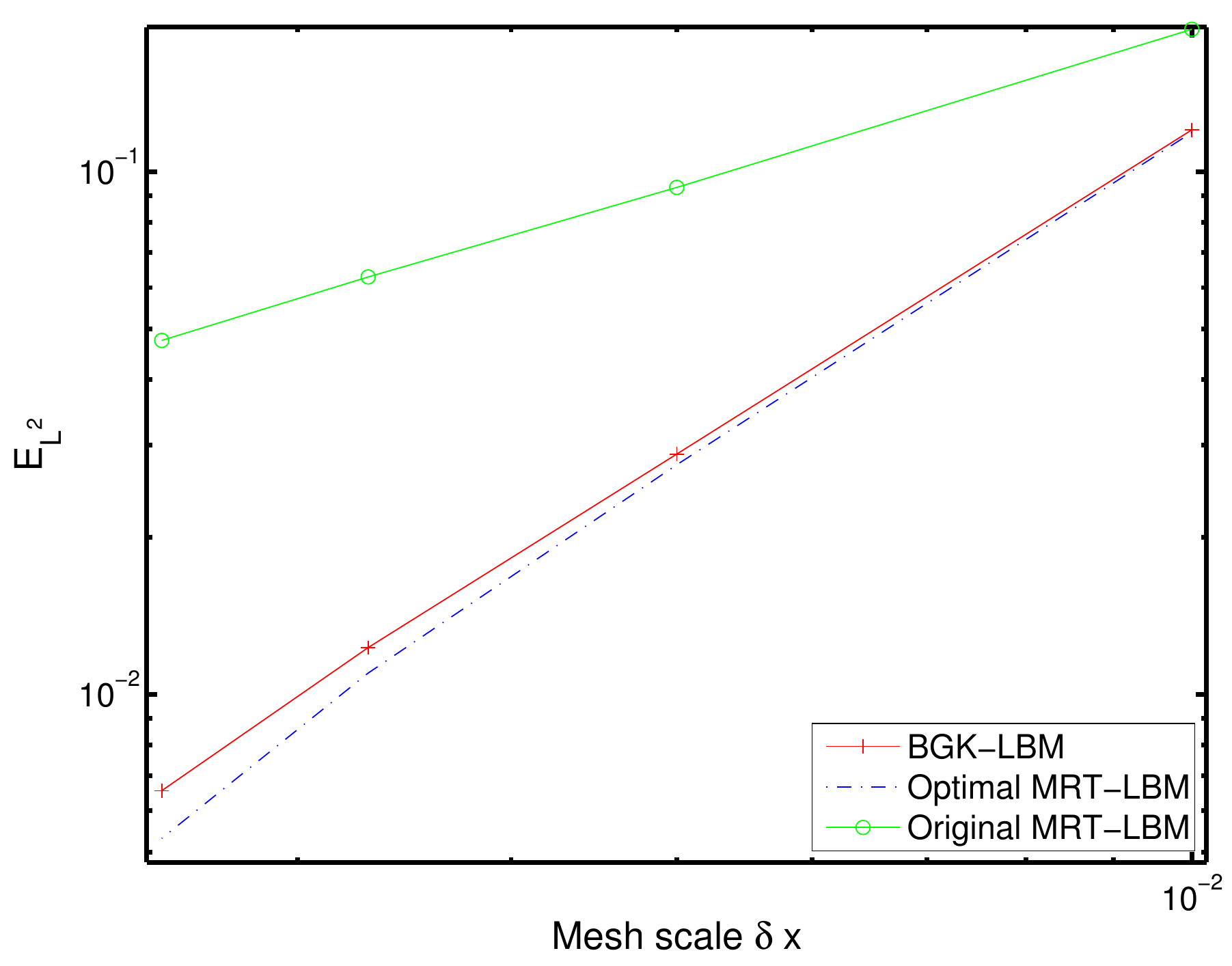}}
\scalebox{0.44}[0.44]{\includegraphics[angle=0]{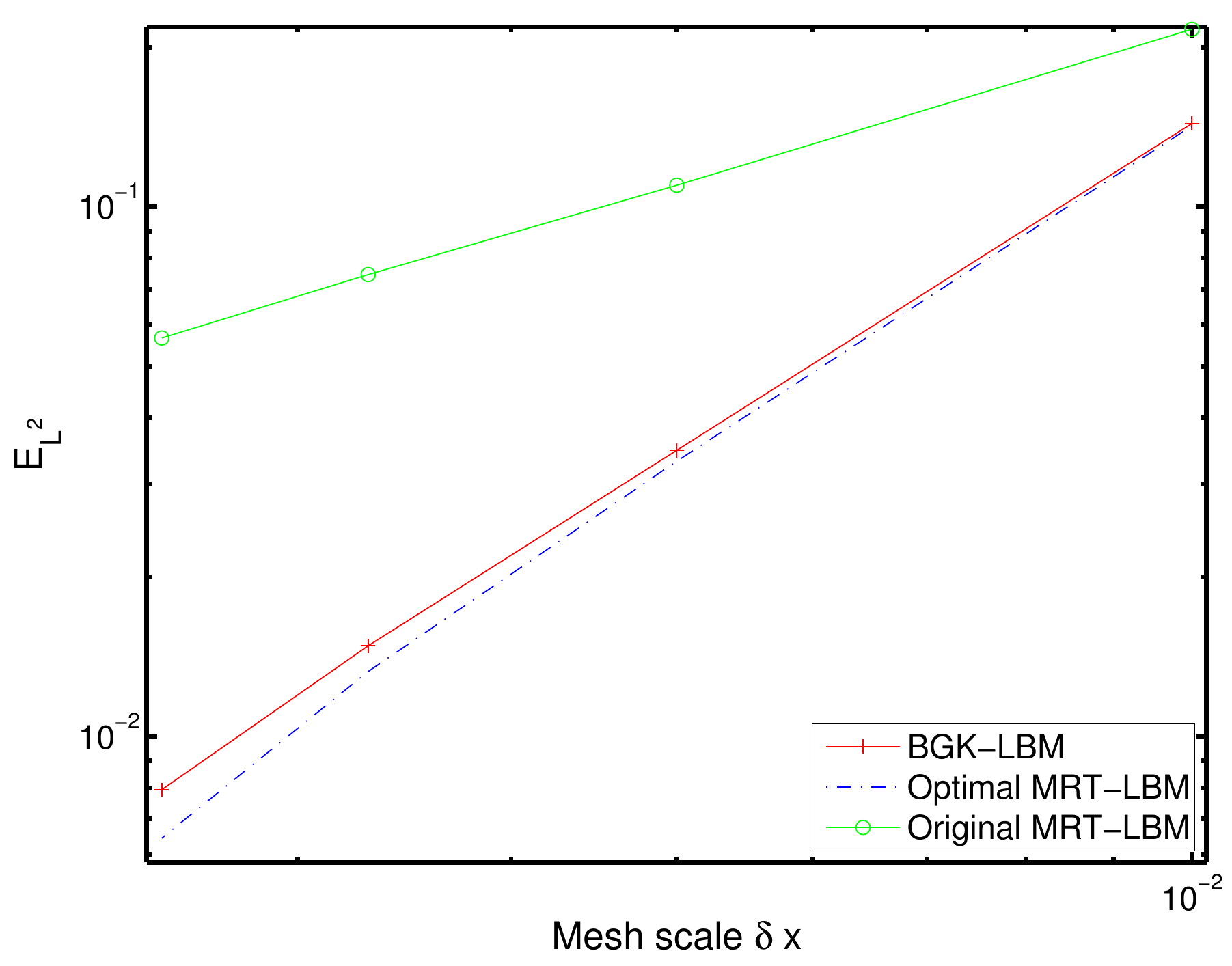}}\\
{\centering (c) at t=0.5 \hspace{6cm}(d)at t=0.6 }\\
\caption{The $L^2$ relative errors for three different 2D LBS with respect to the mesh scale $\delta x$: BGK-LBM, Optimal MRT-LBM, Original MRT-LBM.}\label{fig:28}
\end{center}
\end{figure}

 \begin{figure}[!htbp]
\begin{center}
\scalebox{0.44}[0.44]{\includegraphics[angle=0]{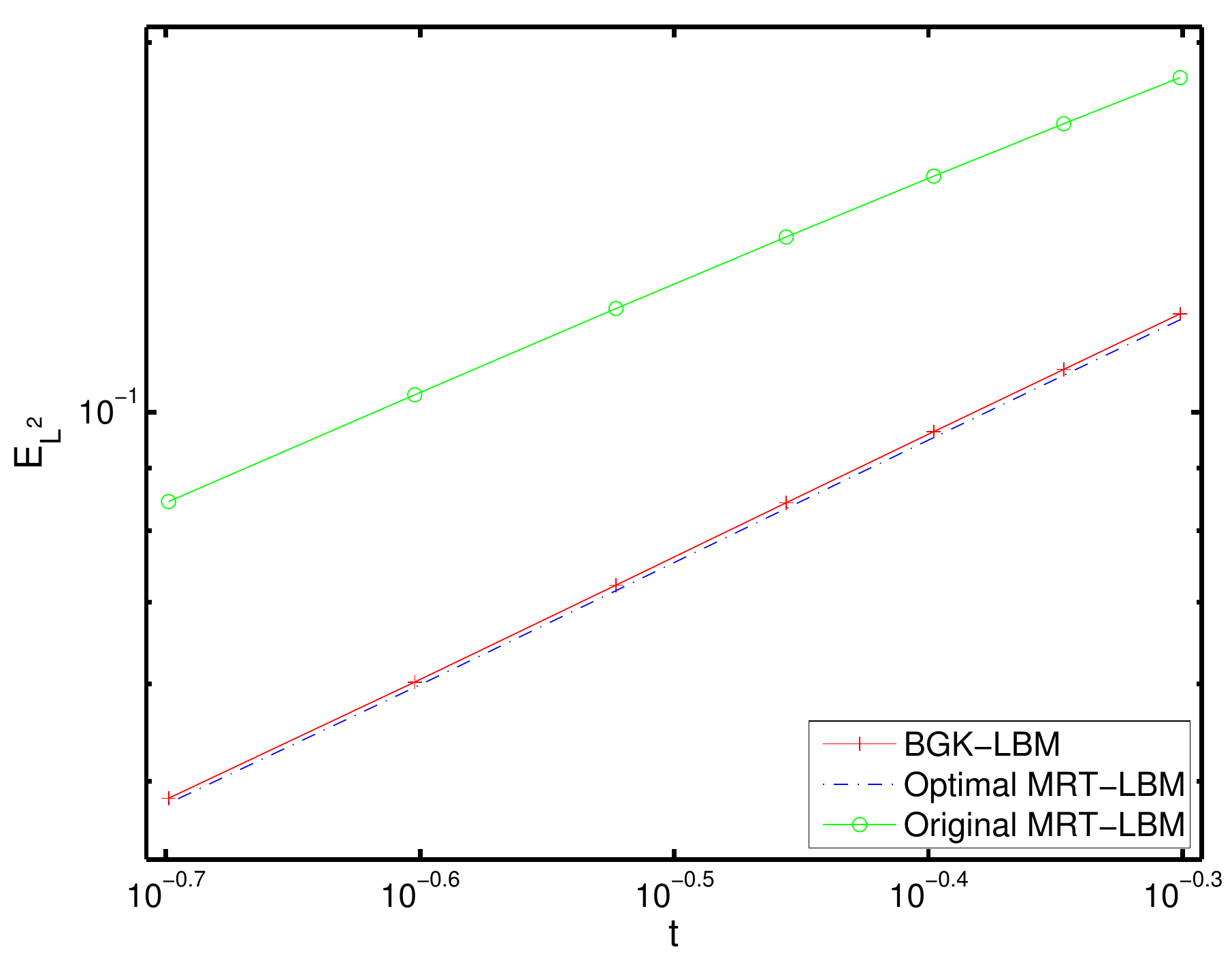}}
\scalebox{0.44}[0.44]{\includegraphics[angle=0]{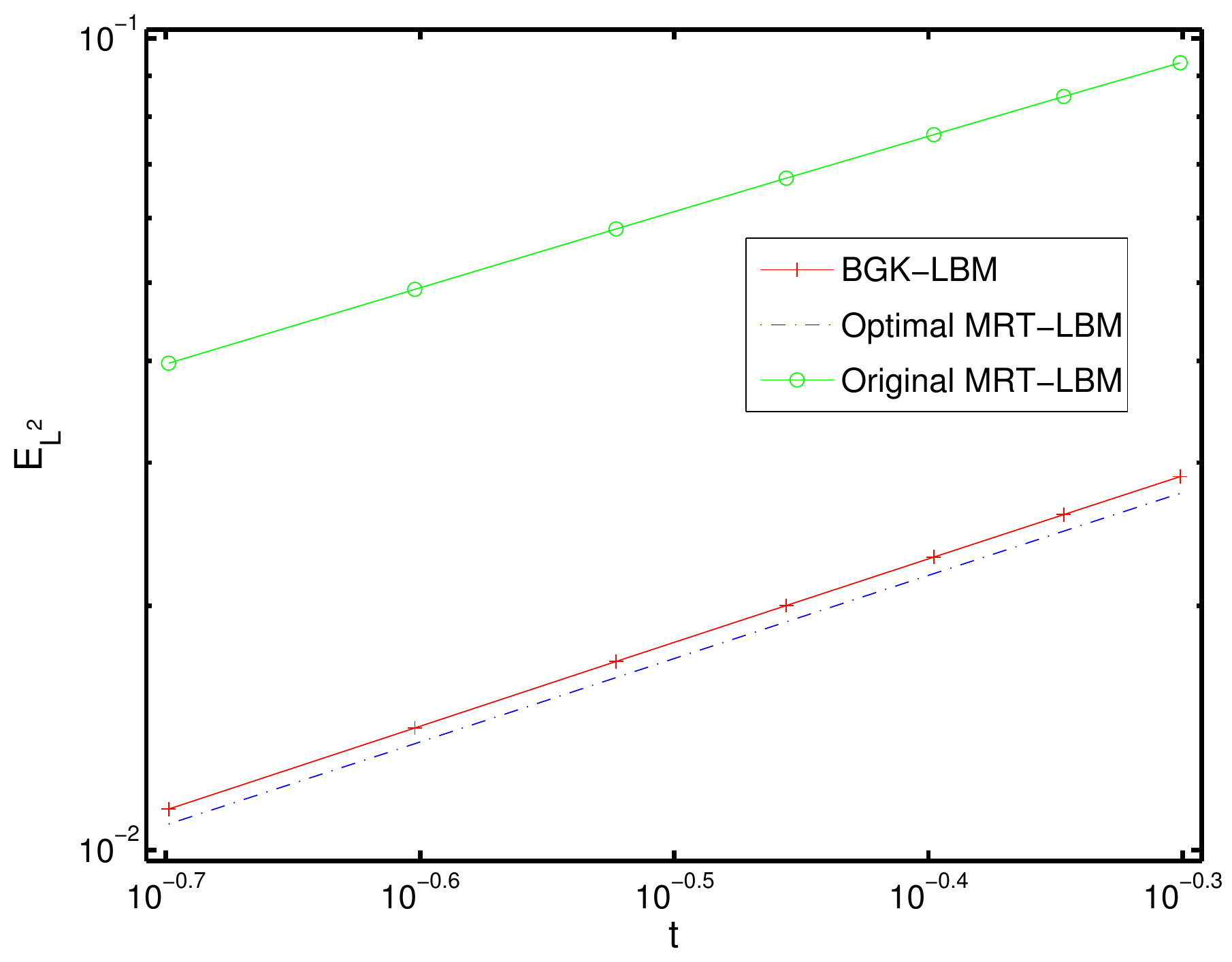}}\\
{\centering (a) Lattice $100^2$\hspace{6cm}(b)Lattice $200^2$}\\
\scalebox{0.44}[0.44]{\includegraphics[angle=0]{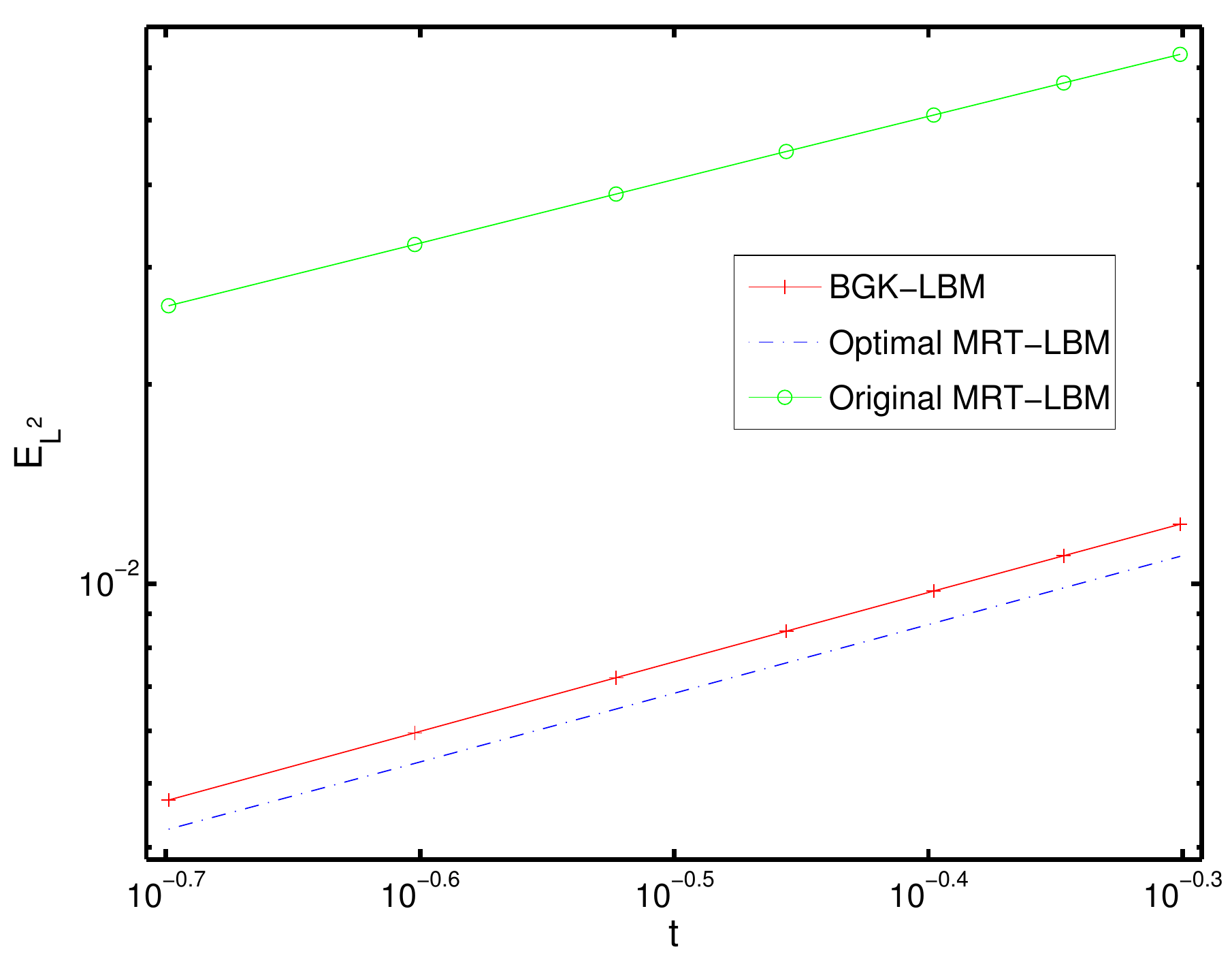}}
\scalebox{0.44}[0.44]{\includegraphics[angle=0]{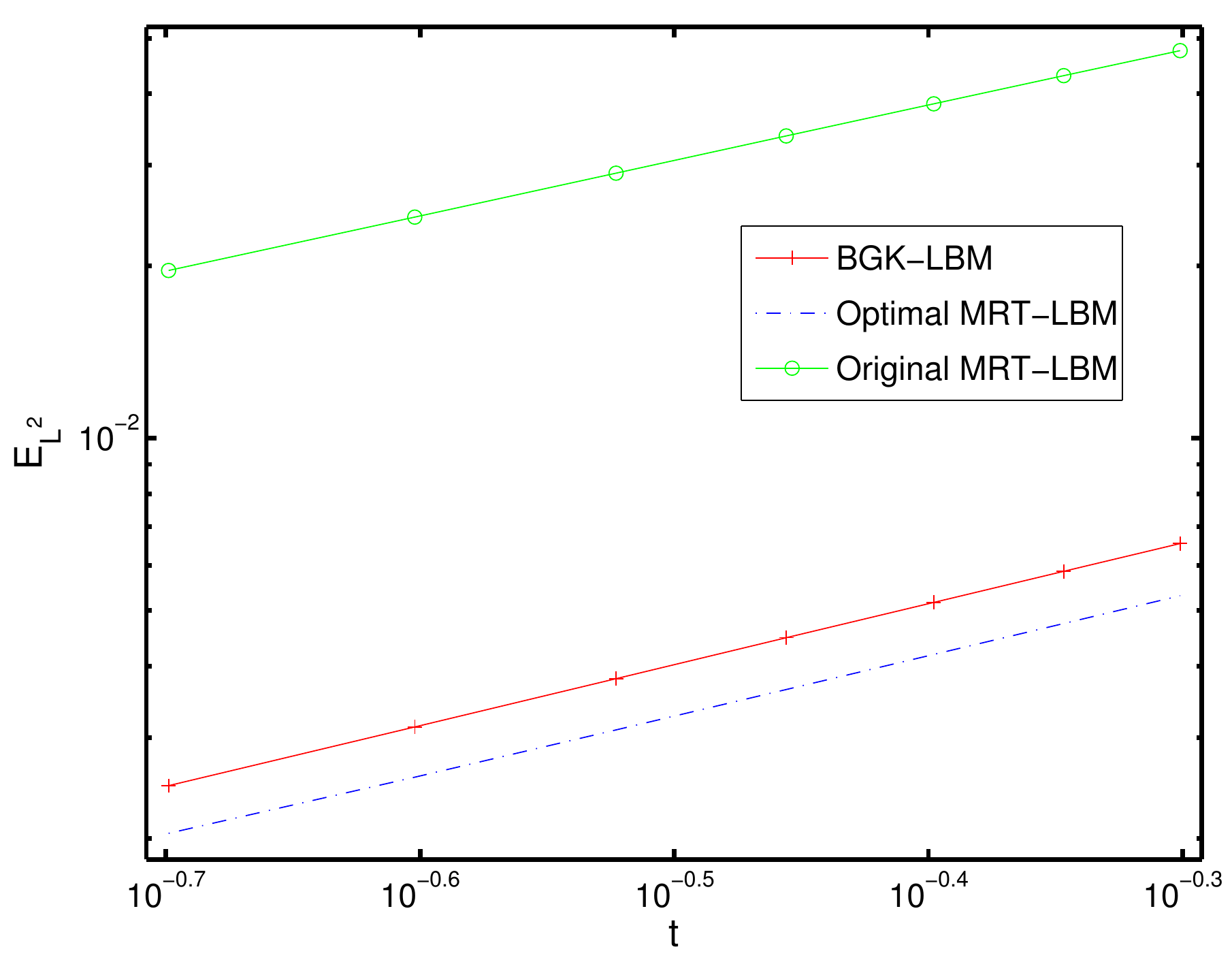}}\\
{\centering (c) Lattice $300^2$\hspace{6cm}(d)Lattice $200^2$}\\
\caption{The $L^2$ relative errors for three different 2D LBS with respect to the evolution time $t$: BGK-LBM, Optimal MRT-LBM, Original MRT-LBM. Lattice number: (a) $100^2$; (b)  $200^2$; (c) $300^2$; (d)   $400^2$.}\label{fig:29}
\end{center}
\end{figure}

\subsection{The 3D acoustic spherical pulse source}
In this part, we continue to  investigate the 3D acoustic spherical pulse source in order to obtain a further validation for the simplified 3D optimization strategies. Assuming the viscosity influence is ignored, the acoustic pulse problem possesses the analytical solution \cite{tamckw}.  The initial profile is given as follows
\begin{equation}
\left\{\begin{array}{ll}
\rho_0&=1+\rho\prime\\
u_0 &=U_0\\
v_0 & =0\\
w_0&=0
\end{array}\right.
\end{equation}
where $\rho\prime$, $\epsilon$, $\alpha$, $r$ and $U_0$ are defined by 
\begin{equation}\label{3:para}
\rho\prime=\epsilon{\rm exp}(-\alpha\cdot r^2),\ \epsilon=10^{-3},\ \alpha={\rm ln}(2)/b^2,\ r=\sqrt{(x-x_0)^2+(y-y_0)^2+(z-z_0)^2},\ U_0=0.01.
\end{equation}
The parameter $b$ in Eq. (\ref{3:para}) is equal to 0.03. The exact solution of $\rho\prime$ (if $(x_0,y_0,z_0)=(0,0,0)$) is given by \cite{tamckw}
\begin{equation}
\rho\prime(x,y,z)=\frac{\epsilon}{2\alpha\sqrt{\pi\alpha}}\int_{0}^{\infty}{\rm exp}\left(-\frac{\xi^2}{4\alpha}\right){\rm cos}(c_st\xi)\frac{{\rm sin}(\xi\eta)}{\xi\eta}\xi^2{\rm d}\xi
\end{equation}
where $\eta=\sqrt{(x-U_0t)^2+y^2+z^2}$. The computational domain $\Omega=[0,L]^3$ ($L=1$). 
 The simulation results of two cases are shown. The relaxation parameters are determined by Index 3 in Table \ref{table:3}. Because of the very low viscosity, the corresponding viscosity dissipation can be ignored.   The adopted lattice resolutions are $100^2$, $200^2$ and $300^2$. In Fig. \ref{fig:30}, the density profiles are shown for different LBS. It is observed that the profiles obtained by the original MRT-LBM have the worst precision. Quantitatively, the $L^2$ relative errors with respect to the mesh resolutions and the evolution time are  presented in Figs. \ref{fig:31}$\sim$\ref{fig:32}.  The conclusions are similar to that of  the 2D problems.  The optimal MRT-LBM has the best precision with respect to the mesh resolutions and the evolution time. At the same time, some super-convergence property of the optimal MRT-LBM is observed.

In all, from the numerical investigations, the optimal MRT-LBM not only improves the computational precision for acoustic problems, but also present some super-convergence properties compared with the BGK-LBM.

 \begin{figure}[!htbp]
\begin{center}
\scalebox{0.44}[0.44]{\includegraphics[angle=0]{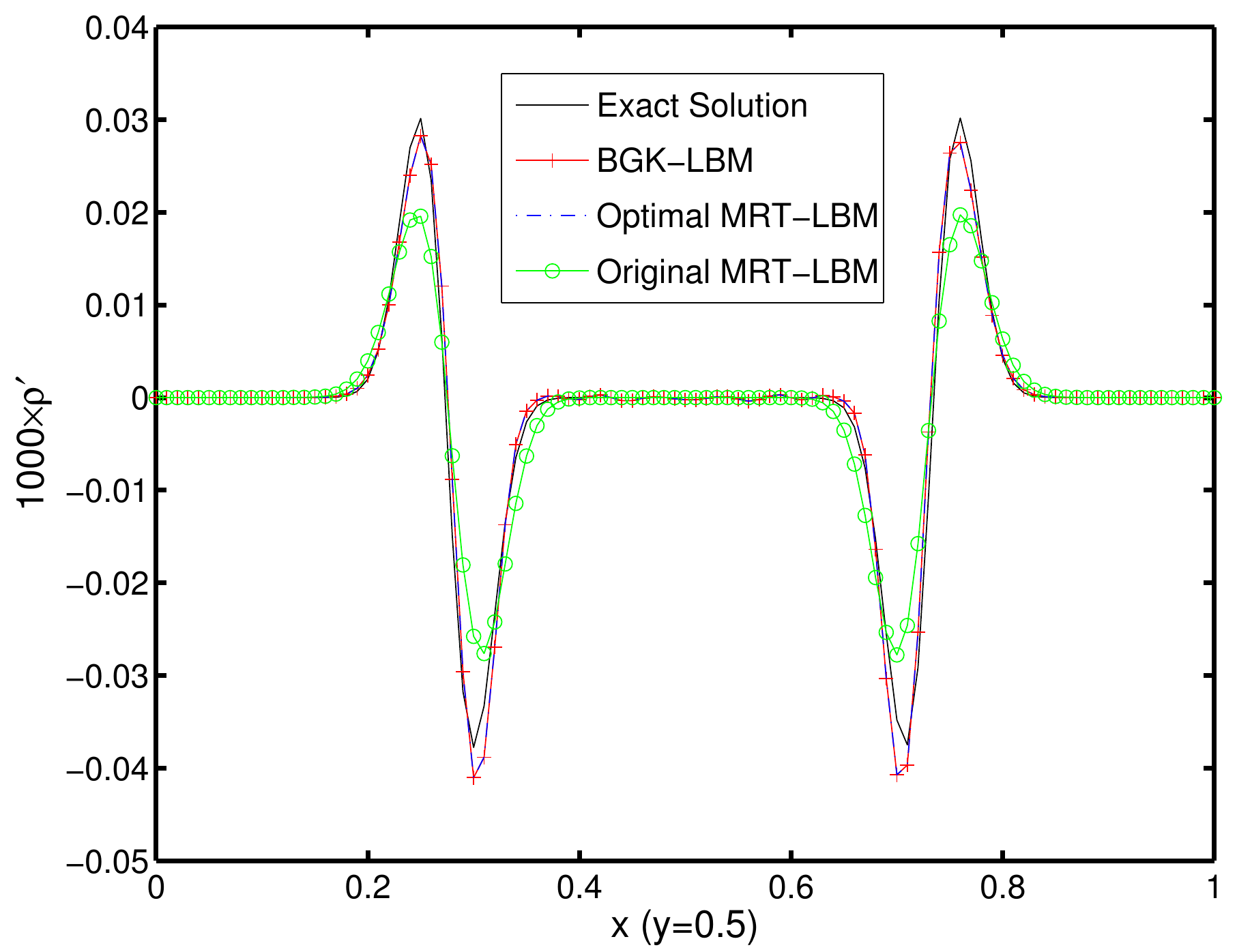}}
\scalebox{0.44}[0.44]{\includegraphics[angle=0]{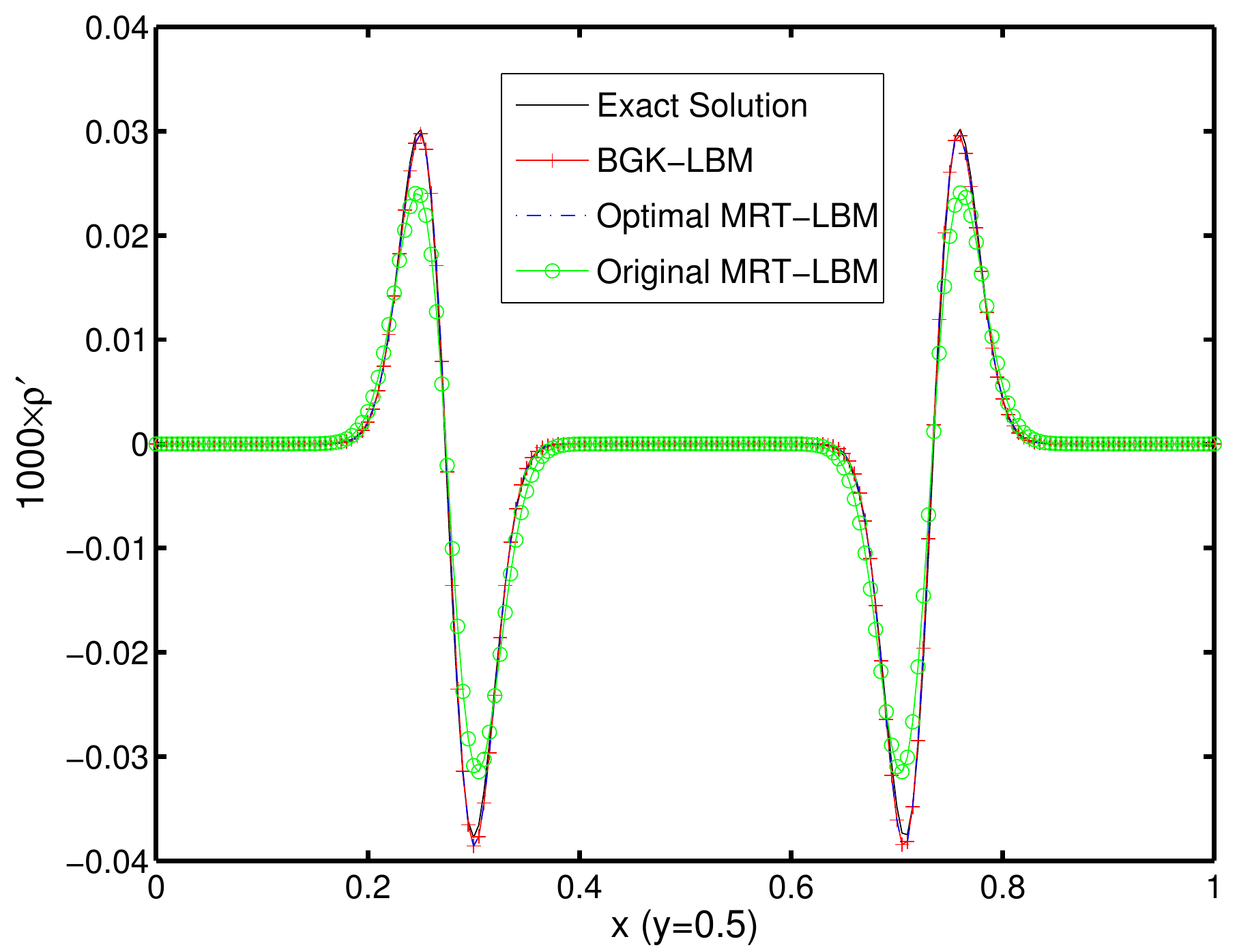}}\\
{\centering (a) Lattice $100^3$\hspace{6cm}(b)Lattice $200^3$}\\
\scalebox{0.44}[0.44]{\includegraphics[angle=0]{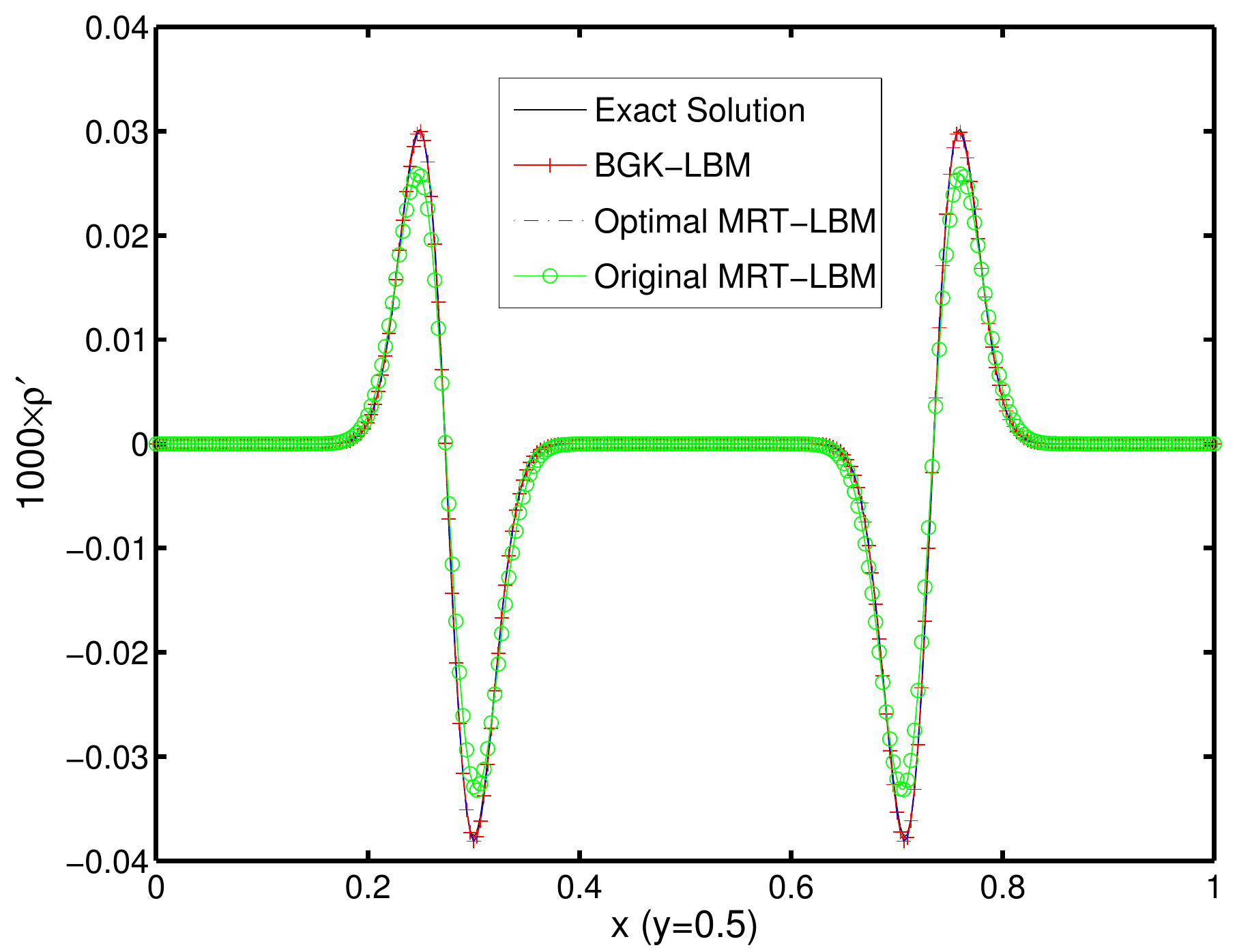}}\\
{\centering (c) Lattice $300^3$}\\
\caption{Fluctuation density profiles of $1000\rho\prime$ for three different 3D LBS along the line y=z=0.5 at t=0.4:  (a) Lattice number $100^3$; (b) Lattice number $200^3$; (c) Lattice number $300^3$.}\label{fig:30}
\end{center}
\end{figure}

 \begin{figure}[!htbp]
\begin{center}
\scalebox{0.44}[0.44]{\includegraphics[angle=0]{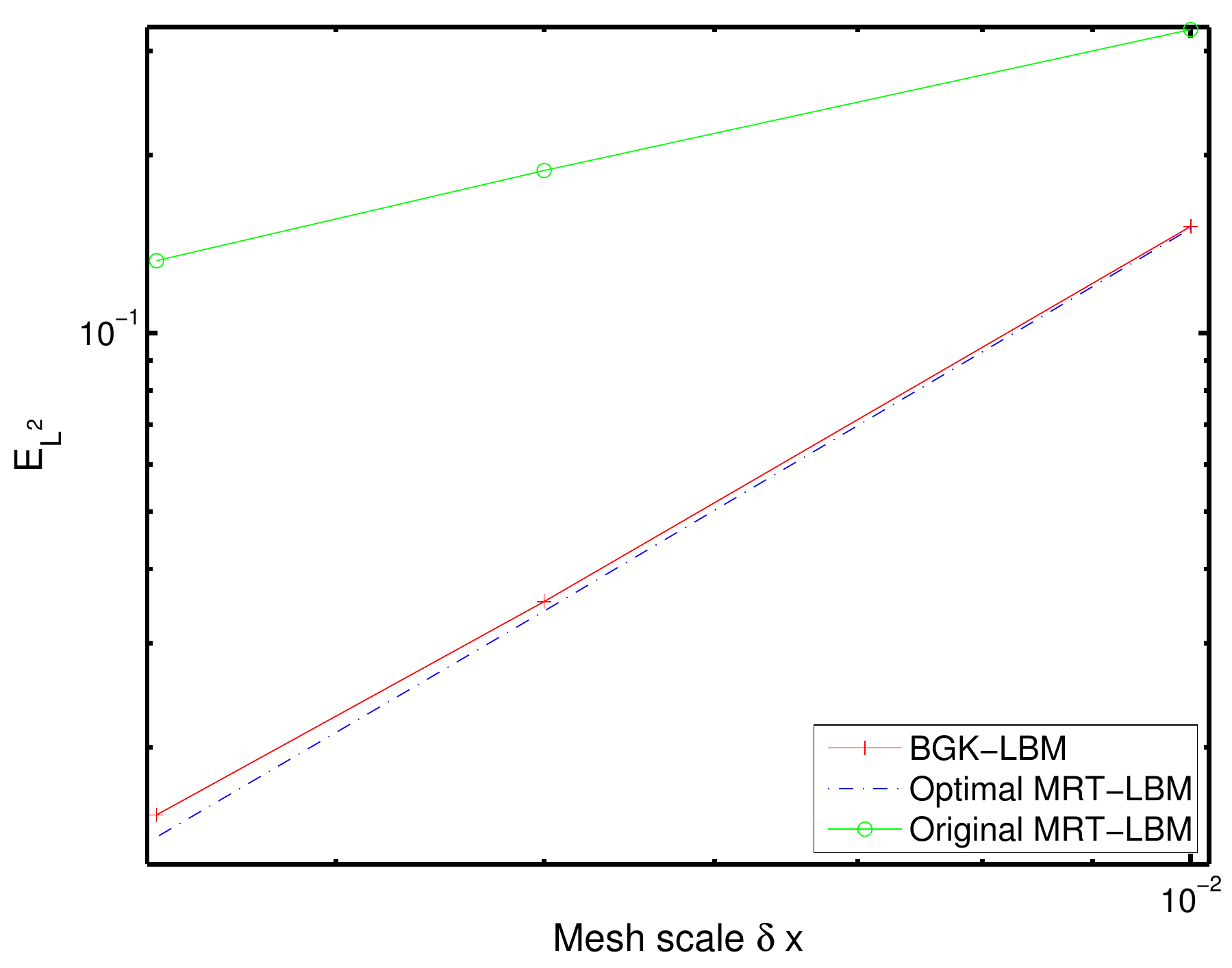}}
\scalebox{0.44}[0.44]{\includegraphics[angle=0]{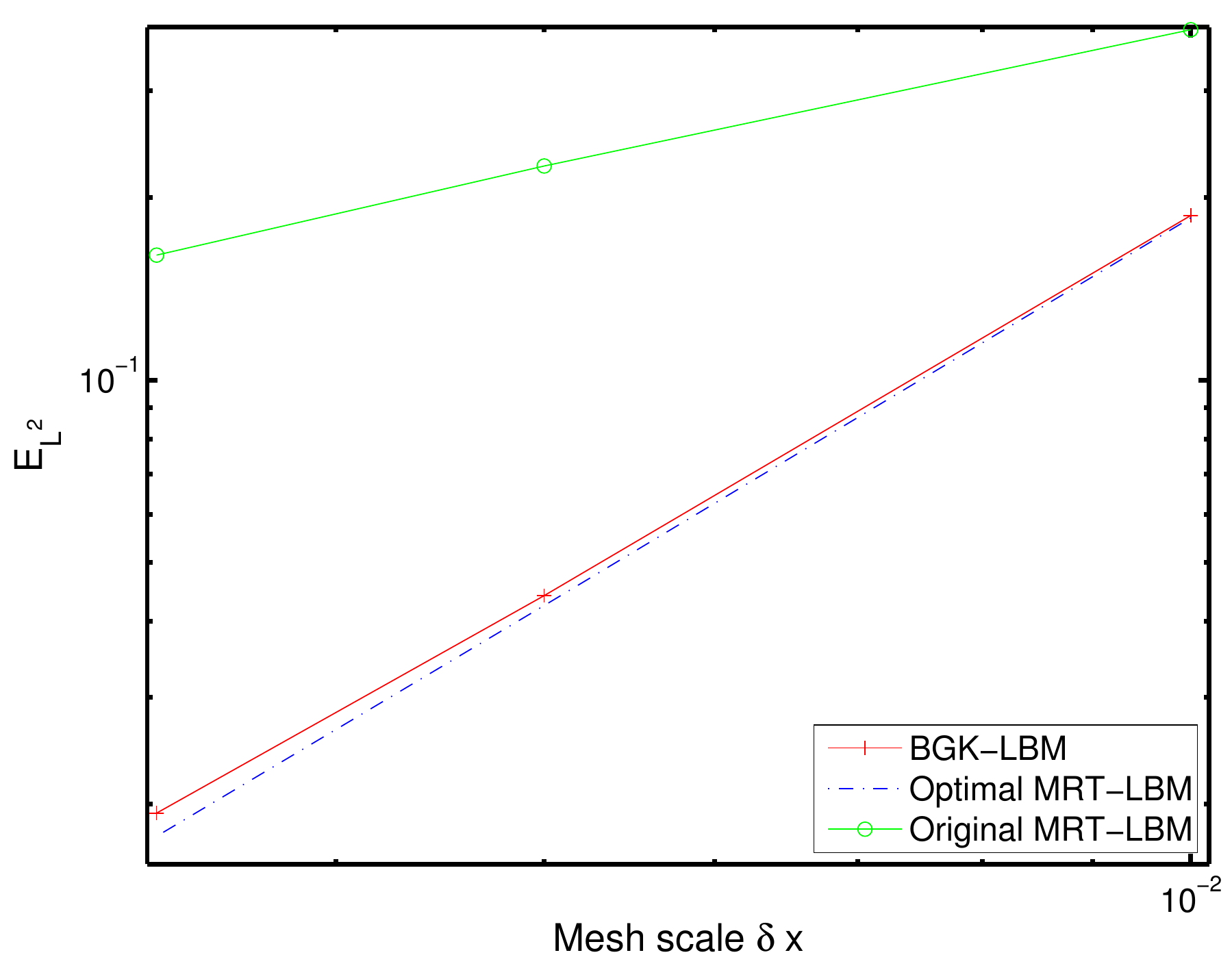}}\\
{\centering (a) at t=0.2 \hspace{6cm}(b)at t=0.4 }\\
\scalebox{0.44}[0.44]{\includegraphics[angle=0]{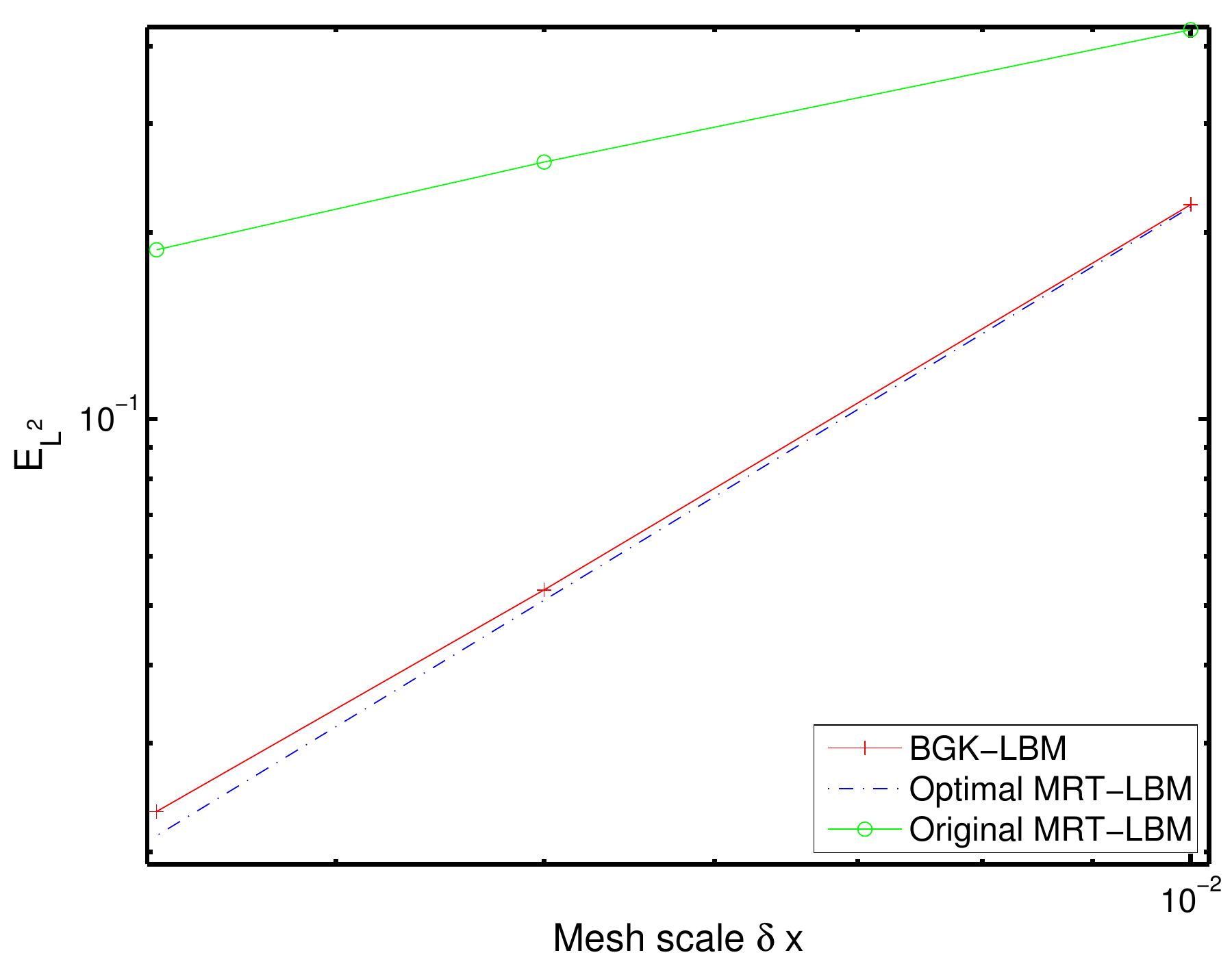}}\\
{\centering (c) at t=0.5 }\\
\caption{The $L^2$ relative errors for three different 3D LBS with respect to the mesh scale $\delta x$: BGK-LBM, Optimal MRT-LBM, Original MRT-LBM.}\label{fig:31}
\end{center}
\end{figure}

 \begin{figure}[!htbp]
\begin{center}
\scalebox{0.44}[0.44]{\includegraphics[angle=0]{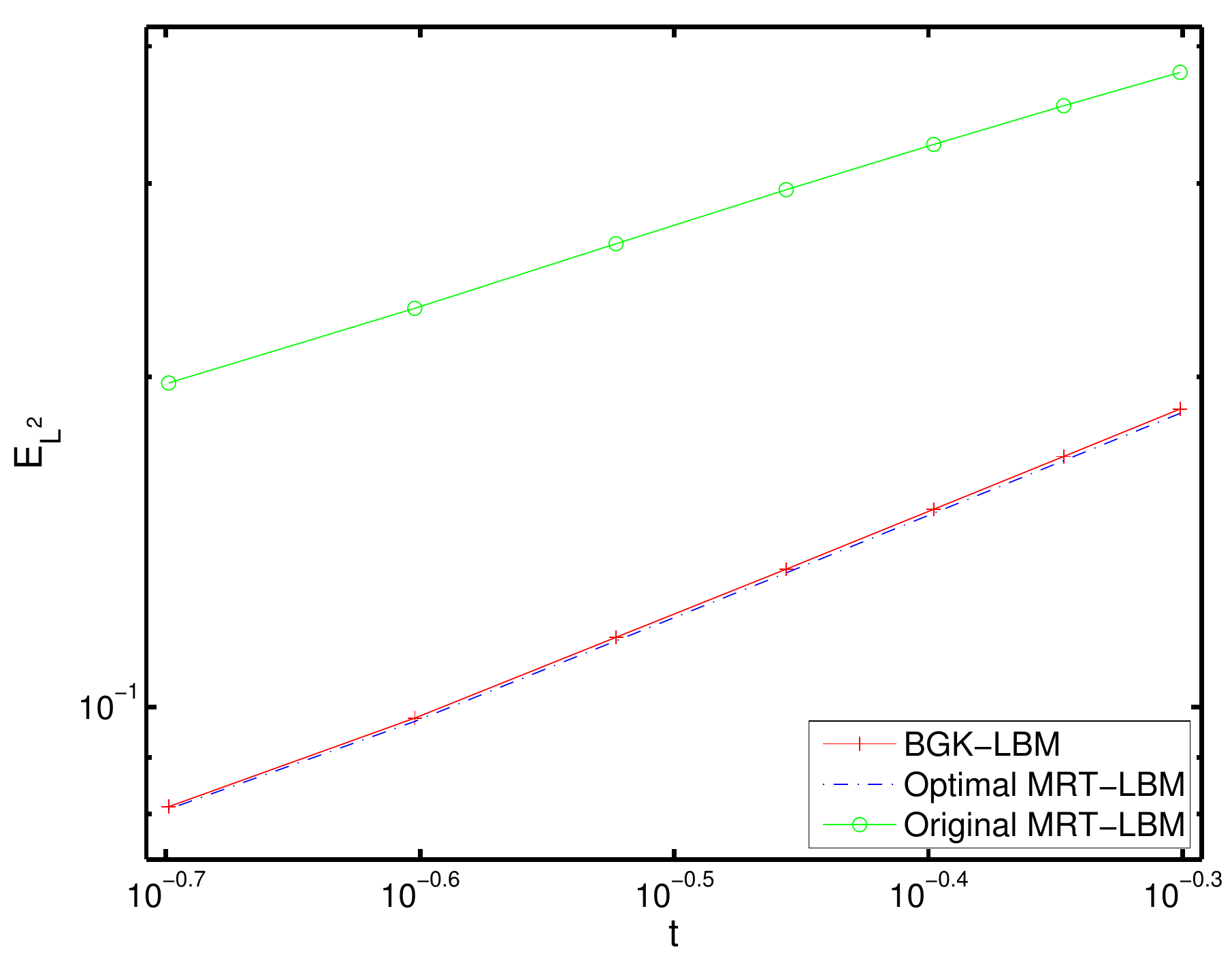}}
\scalebox{0.44}[0.44]{\includegraphics[angle=0]{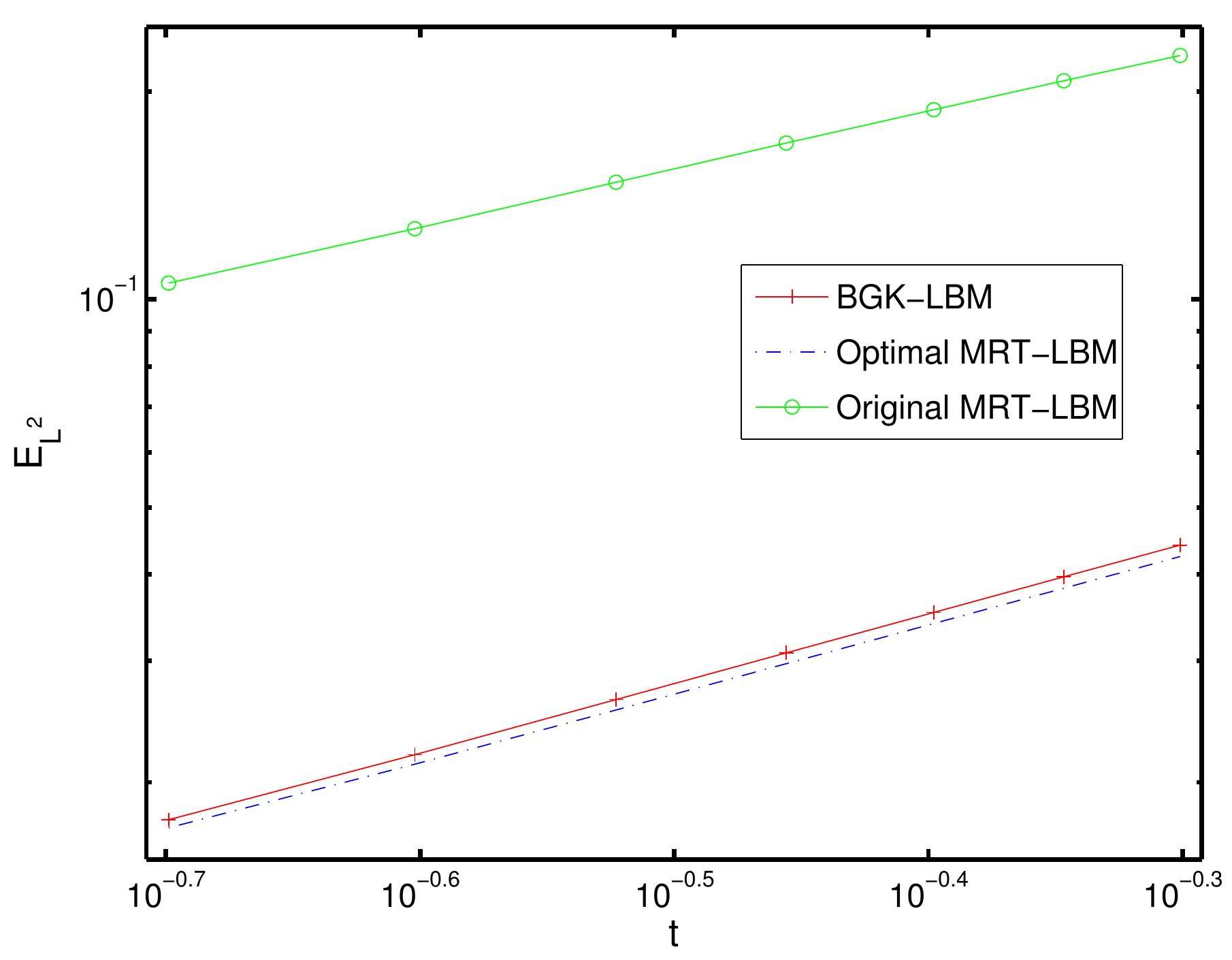}}\\
{\centering (a) Lattice $100^3$\hspace{6cm}(b)Lattice $200^3$}\\
\scalebox{0.44}[0.44]{\includegraphics[angle=0]{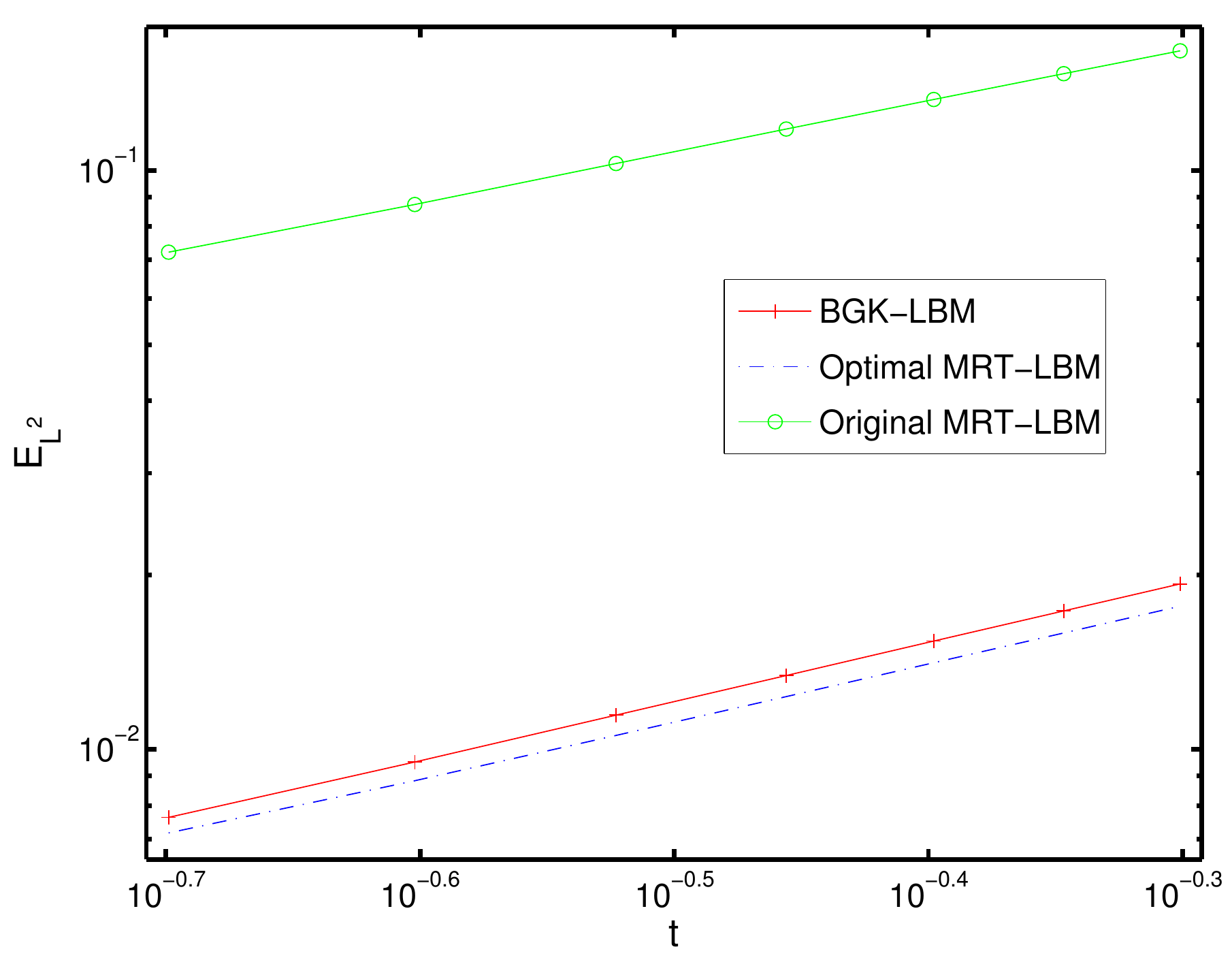}}\\
{\centering (c) Lattice $300^3$}\\
\caption{The $L^2$ relative errors for three different 3D LBS with respect to the evolution time $t$: BGK-LBM, Optimal MRT-LBM, Original MRT-LBM. Lattice number: (a) $100^3$; (b)  $200^3$; (c) $300^3$.}\label{fig:32}
\end{center}
\end{figure}

\subsection{The sound generation by a turbulent airflow over a moderately deep rectangular cavity}

The problem is to study aerodynamic sound generation by a turbulent air flow over a moderately deep cavity.
Despite the real flow is expected to be 3D and turbulent, it is known that dominant acoustic tonal modes are generated by nearly 2D structures which can be captured thanks to 2D simulations. Therefore, 2D simulations may be useful in predicting main frequencies of the radiated acoustic field.
This benchmark problem  is taken from the ``Fourth Computational Aeroacoustics (CAA) Workshop on Benchmark Problems". 
The ratio of the cavity width and the depth is 0.3124. In Fig. \ref{geo:cavity}, the schematic view of the computational domain is given along with the corresponding physical scales.
The inflow at the far left boundary was set as uniform with a velocity magnitude of 50 m/s along the horizontal direction. The outflow boundary condition and free-slip boundary condition are applied at the outlet and the top boundary, respectively. In order to avoid the influence of the sound wave reflection from the left boundary, the outlet boundary and the top boundary,   buffer (sponge) regions are used. 
The buffer region thickness at the inlet and outlet is equal to one-twentieth of the domain length along x-direction. The buffer region thickness for the top boundary is one-twentieth of the rectangle region. The lattice resolution along the thickness of the cavity tongue B is equal to 50. The optimal relaxation parameters are set by Index 4 in Table \ref{table:3}. It is necessary to point out that for the current configuration, the computation by the BGK-LBM diverged after about $1.4\times10^5$ time steps. For the current investigation, the computation by the optimal MRT-LBM was implemented for $6.48\times 10^5$ time steps and the optimal MRT-LBM was still stable.  Figs.\ref{spl:left}$\sim$\ref{spl:right} show the computed Fourier spectra of sound pressures with the measured sound pressure spectra and the CFD (TIDAL, Time Iterative Density/pressure based Algorithm) sound pressure spectra. The tonal frequencies and amplitudes of the computed spectra coincide with those of the experimental spectra very well. It is clear the obtained spectra by the optimal MRT-LBM is better than that by TIDAL.  
 \begin{figure}[!htbp]
\begin{center}
\scalebox{0.7}[0.7]{\includegraphics[angle=0]{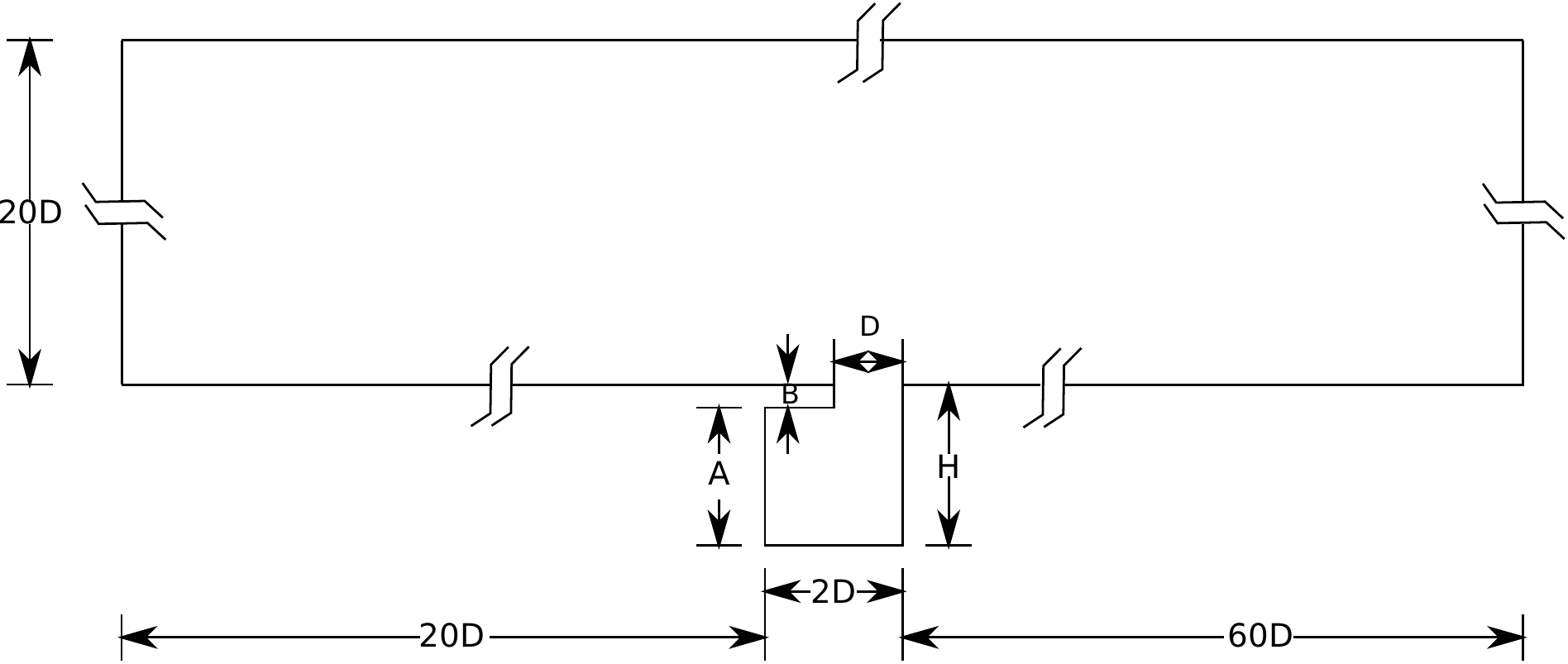}}
\end{center}
\caption{Schematic computational geometry: $A=25.42$mm, $D=7.94$mm, $B=3.18$mm.}\label{geo:cavity}
\end{figure}
 \begin{figure}[!htbp]
\begin{center}
\scalebox{0.65}[0.65]{\includegraphics[angle=0]{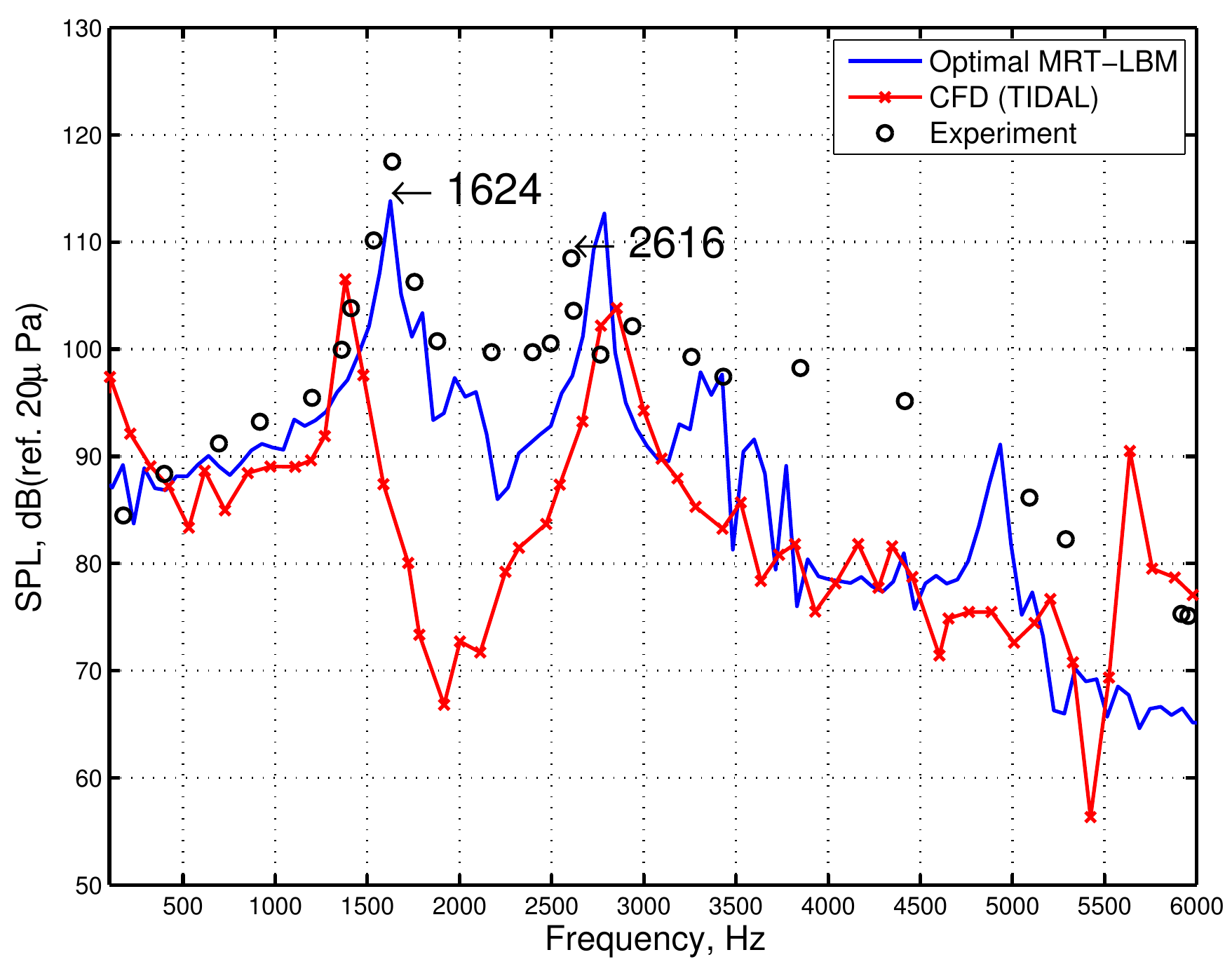}}
\end{center}
\caption{Comparison of sound pressure spectra at the centers of the cavity left wall }\label{spl:left}
\end{figure}
 \begin{figure}[!htbp]
\begin{center}
\scalebox{0.65}[0.65]{\includegraphics[angle=0]{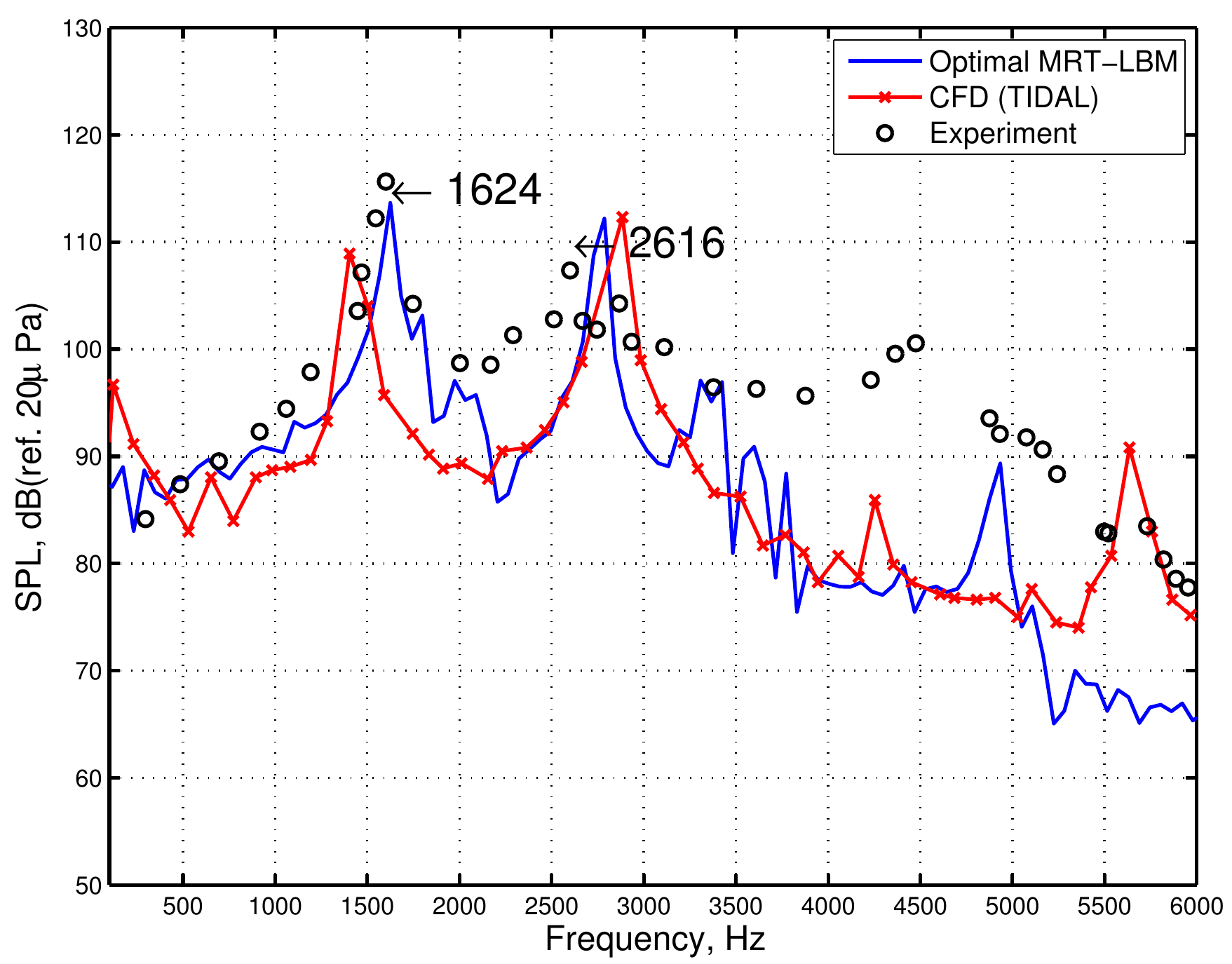}}
\end{center}
\caption{Comparison of sound pressure spectra at the centers of the cavity floor  }\label{spl:floor}
\end{figure}

 \begin{figure}[!htbp]
\begin{center}
\scalebox{0.65}[0.65]{\includegraphics[angle=0]{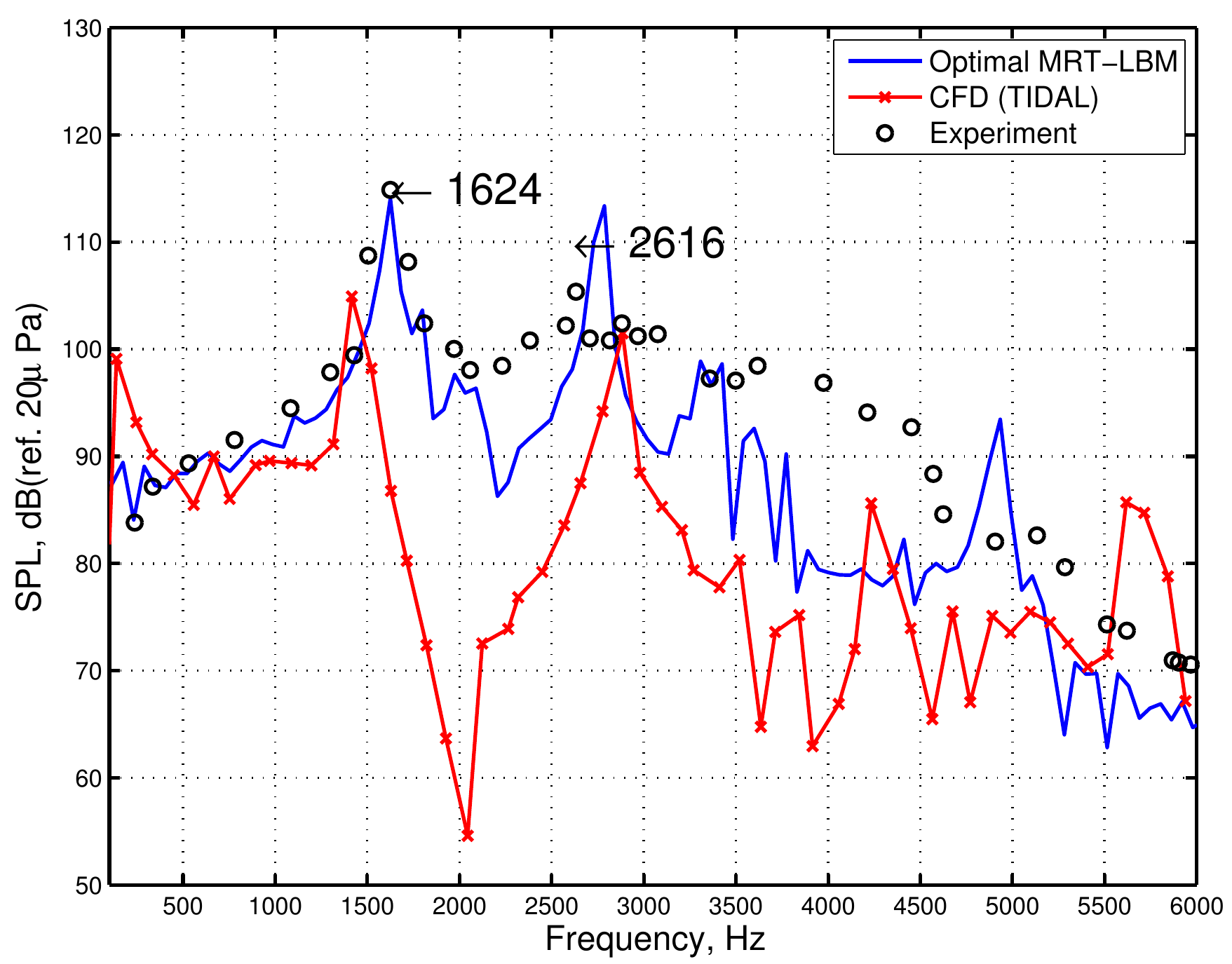}}
\end{center}
\caption{Comparison of sound pressure spectra at the centers of the cavity right wall  }\label{spl:right}
\end{figure}

\subsection{Application to the acoustic scattering from a cylinder solid body}

The acoustic scattering from a cylinder solid body was presented in the second CAA workshop on benchmark problems \cite{tamaiaa}. The objective of this problem focused on addressing the propeller generated sound field scattered off the fuselage of an aircraft.  By this benchmark, we address the performance comparisons among the optimal MRT-LBM, the DRP schemes and the high-order compact schemes for solving the linear acoustic problems (the DRP schemes and the high-order compact schemes are used to solve the linearized Euler equations).  The schematic geometry is given in Fig. \ref{scattering:schematic}. The left big black disc indicates the cylinder and the right small black disc denotes the acoustic source. The two-dimensional cylinder radius $R$ is equal to $D/2$ ($D=1$ for current researches). The centre of the cylinder is located at the origin. At the initial time ($t=0$), the initial profile of the acoustic source is defined by 
\begin{equation}\label{acoustic:profile}
p(x,y,0)=p_{\infty}+\delta {\rm exp}\left[-{\rm 2}\left(\frac{(x-x_sD)^2+y^2}{w^2}\right)\right],
\end{equation} 
where $w=0.2$, $\delta=c_s^2\times 10^{-4}$, $x_s=4$ and $p_{\infty}=c_s^2\rho_\infty$. The low $\delta$ is to ensure a linear response and guarantee the linear analysis theory. The problem  was propsed as the second problem in Category 1 in \cite{tamaiaa}, but the given analytic solution is incorrect \cite{Daniel}.  The correct analytical solution was given as follows \cite{Daniel,tamaiaa2}
\begin{equation}\label{correctanalytical}
p_{e}(x,y,t)=p_{\infty}+{\rm Re}\left\{\int_{0}^{\infty}\left(A_i(x,y,\omega)+A_r(x,y,\omega)\right)\omega{\rm exp}(-{\bf i}\omega t){\rm d}\omega\right\}.
\end{equation}
The incident wave amplitude is 
\begin{equation}
A_i(x,y,\omega)=\frac{1}{2b}{\rm exp}\{-\omega^2/(4b)\}
J_0(\omega r_s)
\end{equation}
where $r_s=\sqrt{(x-x_sD)^2+y^2}$, $b={\rm ln}2/w^2$ and $J_0$ is the Bessel function of order zero. And the corresponding reflected wave portion is given by
\begin{equation}
A_r(x,y,\omega)=A_r(r,\omega)=\sum_{k=0}^{\infty}C_{k}(\omega)H_k^{(1)}(r\omega){\rm cos}(k\omega),
\end{equation}
where $H_k^{(1)}$ is the $k$th order Hankel function of the first kind and 
\begin{equation}
C_k(\omega)=\frac{\omega}{2b}{\rm exp}\{-\omega^2/(4b)\}\frac{\epsilon_k}{\pi\omega\left[H_k^{(1)}\right]^\prime(r_0\omega)}\int_0^\infty J_1(\omega r_{s_0})\frac{r_0-x_s D{\rm cos}(\theta)}{r_{s_0}}{\rm cos}(k\theta){\rm d}\theta,
\end{equation}
where $\epsilon_0=1$, $\epsilon_k=2$ for $k\neq 0$, $r_0=1/2$ and $r_{s_0}=\sqrt{0.25+x_s^2D^2-x_sD{\rm cos}(\theta)}$.

In order to compare numerical results with the exact solution, the computational time is non-dimensionalized as $tc_s/D$ . The pressure samples are recorded at three different points $A(r=5D,\theta=\pi/2)$, $A(r=5D,\theta= (3/4)\pi)$ and $C(r=5D,\theta=\pi)$. The pressure signals are sampled between $5\leq tc_s/D\leq 10$. The acoustic pressure $p_{\rm osc}$ is non-dimensionalized as $p^\prime =p_{\rm osc}/\left[(\rho_\infty c_s^2)\delta\right]$ . For the current computations, the relaxation parameters are determined by Index 6 in Table \ref{table:3} and the lattice resolution along the diameter of the cylinder is equal to 50. The corresponding values of the shear and bulk viscosity  are very small,  so that the dissipation influence on the acoustic waves can be ignored. In Fig. \ref{scattering:ompthighorder}, the comparisons between the optimal MRT-LBM, the DRP schemes and the high-order compact schemes are presented. For the schemes of the DRP schemes and the high-order compact schemes, the temporal integration is performed using the 4-6 low dispersion and dissipation Runge-Kutta (LDDRK) optimized scheme \cite{xinzhang,Hu}. From the results, it is clear that the best results are obtained by the optimal MRT-LBM. This conclusion  guarantees that the optimal MRT-LBM is better than the fourth-order DRP scheme and sixth/fourth-order compact schemes of the linearized Euler equations. In Fig. \ref{scattering:optimalmrt}, the comparisons between the optimal MRT-LBM and the exact solution are shown. From the results, it is observed that  the computational error at the points B and C  is larger than that at the point A. The observed deviations from the exact solution are mainly from the non-linear influence of the lattice Boltzmann schemes, because the  exact solution is obtained by the linearized Euler equations. The another possible reason is attributed to the step approximation of the the cylinder boundary. However, the results of the optimal MRT-LBM are the best ones. In Fig. \ref{scattering:propagation}, the acoustic wave patterns are shown at $tc_s/D=5.77$ and $tc_s/D=8.66$. Three wave fronts are observed. The farthest wave front from the cylinder is generated by the initial condition. The next front is the wave reflected off the right surface of the cylinder. The wave front closest to the cylinder is produced by the collision and merging of the two parts of the initial wave front on the left side of the cylinder. 

 \begin{figure}[!htbp]
\begin{center}
\scalebox{0.5}[0.5]{\includegraphics[angle=0]{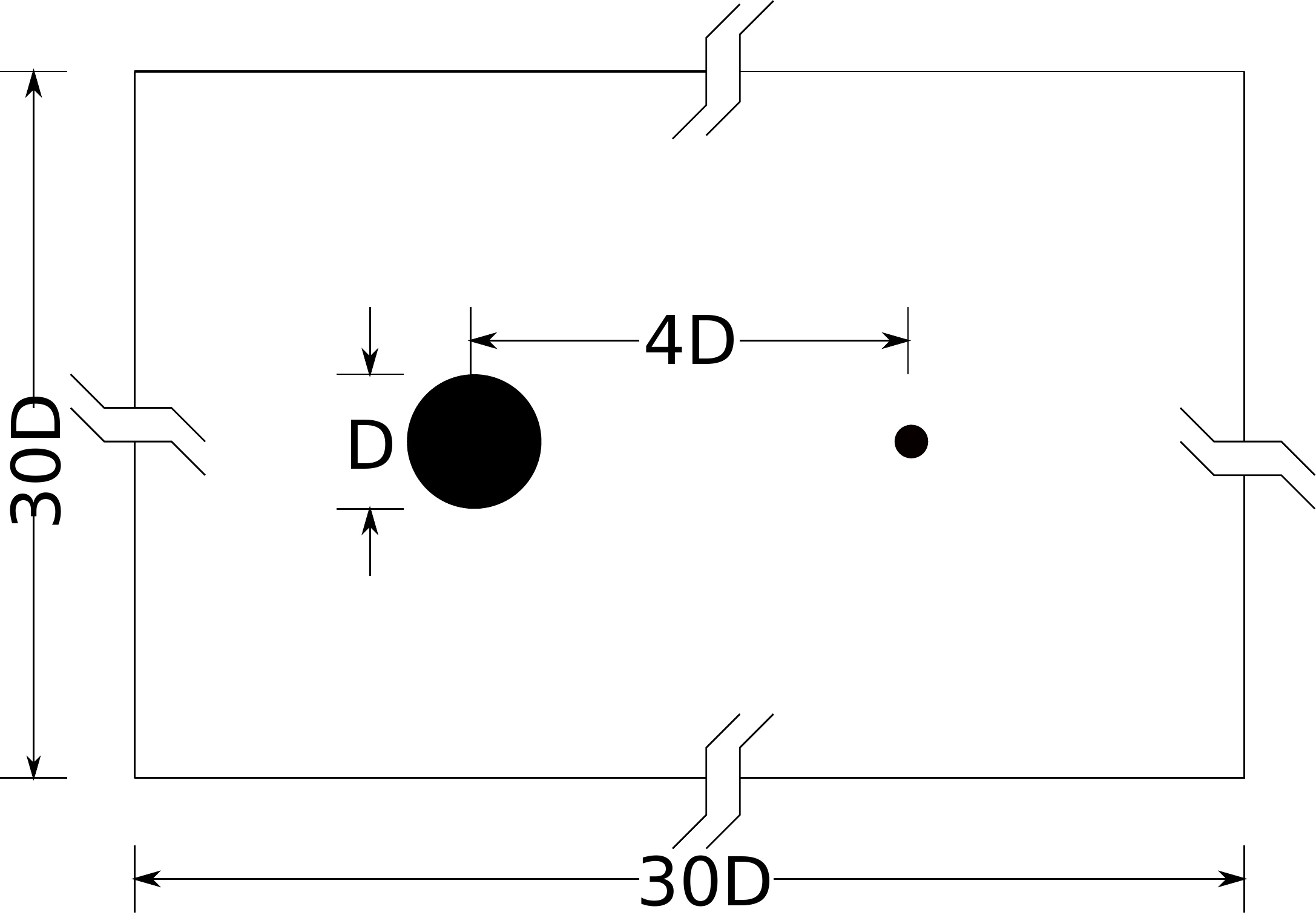}}
\end{center}
\caption{Schematic figure of the acoustic scattering problem.}\label{scattering:schematic}
\end{figure}

 \begin{figure}[!htbp]
\begin{center}
\scalebox{0.8}[0.8]{\includegraphics[angle=0]{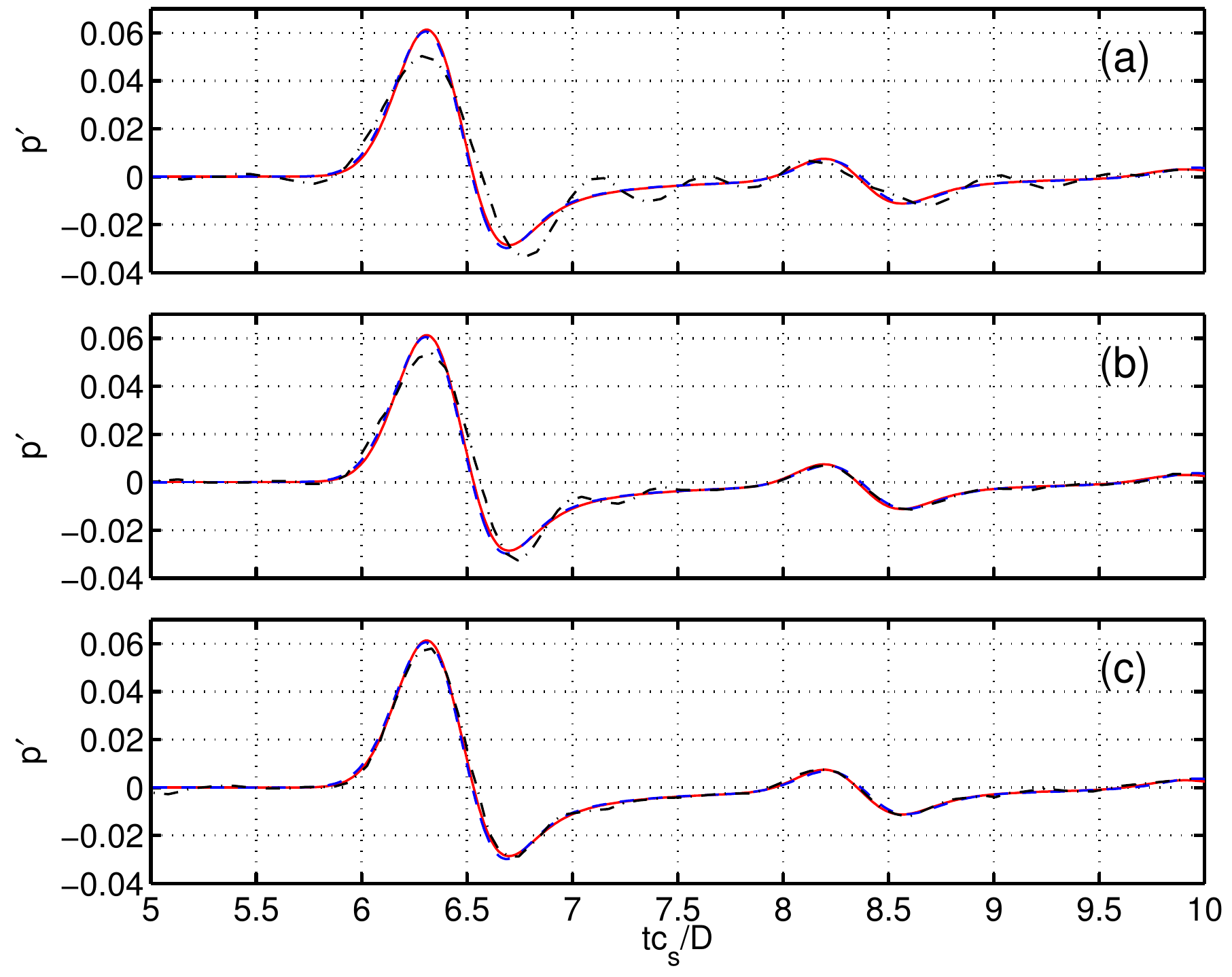}}
\end{center}
\caption{Comparison between computed (the optimal MRT-LBM, the DRP schemes and the high-order compact schemes) and exact solutions of two-dimensional scattering problem at the point A . (a) Compared with the fourth-order DRP scheme (seven-point stencil). (b) Compared with the sixth-order compact scheme (three-point stencil). (c) Compared with the fourth-order compact scheme (three-point stencil). Line ({\color{red}---}): Exact solution. Dashed line ({\color{blue}- - -}): optimal MRT-LBM solution. Dash-dot line (- $\cdot$ -): solutions of the high-order/DRP schemes.
}\label{scattering:ompthighorder}
\end{figure}

 \begin{figure}[!htbp]
\begin{center}
\scalebox{0.8}[0.8]{\includegraphics[angle=0]{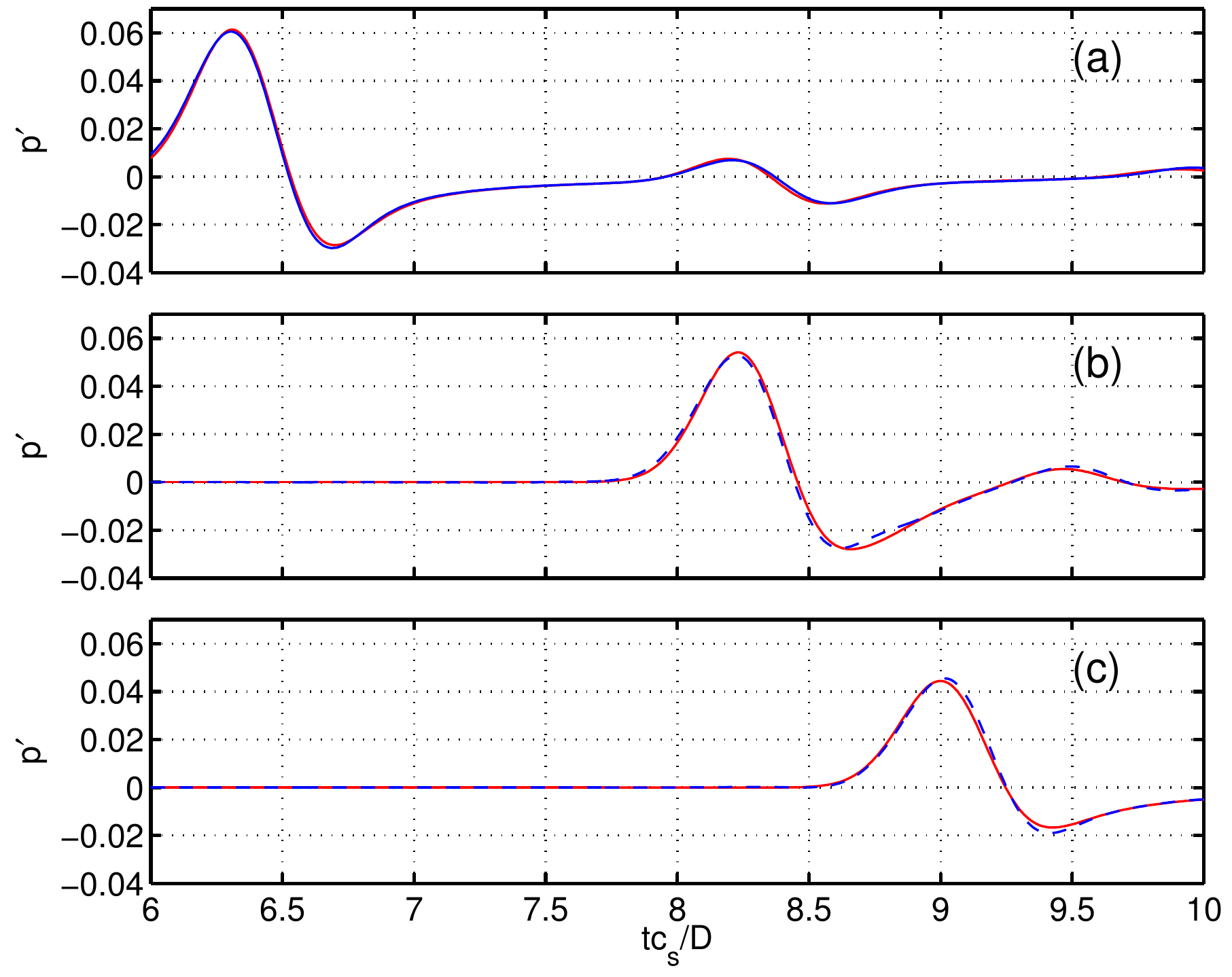}}
\end{center}
\caption{Comparison between computed (optimal MRT-LBM) and exact solutions of two-dimensional scattering problem at the three points . (a) The time series of $p^{\prime}$ at the point A. (b) The time series of $p^{\prime}$ at the point B. (c) The time series of $p^{\prime}$ at the point C. Line ({\color{red}---}): Exact solution. Dashed line ({\color{blue}- - -}): optimal MRT-LBM solutions.
}\label{scattering:optimalmrt}
\end{figure}

 \begin{figure}[!htbp]
\begin{center}
\scalebox{0.4}[0.4]{\includegraphics[angle=0]{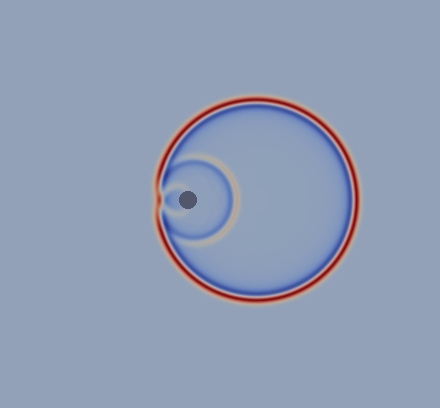}}\quad \scalebox{0.4}[0.4]{\includegraphics[angle=0]{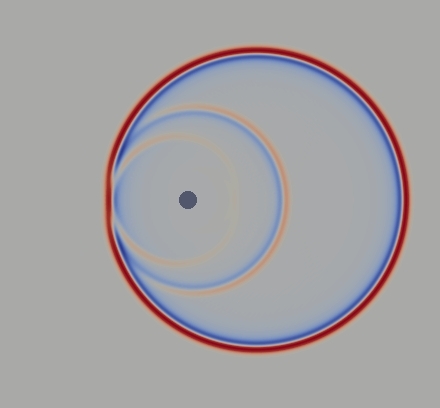}}
\end{center}
\caption{The acoustic wave patterns by the optimal MRT-LBM. (Left) $tc_s/D=5.77$; (Right) $tc_s/D=8.66$.}\label{scattering:propagation}
\end{figure}

\section{Conclusion}
In this paper, the researches are focused on the sensitivity analysis and the determination of the relaxation parameters in the MRT-LBM. By the investigations of the sensitivity, it is concluded that the numerical behaviors of the dispersion and dissipation are mainly dependent on the variation of the wave number ${\bf k}$. From the behaviors of the eigenvalues of $M^{\rm mrt}$ in the complex planes, it is observed that the distribution of the eigenvalues has the certain shapes if the wave number ${\bf k}$ is fixed and the velocity vector is regarded as a random variable. According to the sensitivity analysis, the integral spaces of the optimization strategies are simplified for 2D/3D problems. Then, by von Neumann analysis, the numerical performances of the optimization strategies are validated. Finally, by the benchmark acoustic problems and considering  the ignorable small viscosity in the LBS, the optimal MRT-LBM is compared with the BGK-LBM and the used high-order compact/DRP schemes . The numerical results demonstrate that the optimal MRT-LBM can be used to simulate the acoustic problems very well.  Meanwhile, the optimal MRT-LBM performs better than the BGK-LBM and the used high-order compact/DRP schemes numerically. Especially, some super-convergence properties are observed.

\section*{Acknowledgement}
This work was supported by the FUI project LaBS (Lattice Boltzmann Solver, \url{http://www.labs-project.org}).   Dr. Yan Zhang (Laboratoire Jacques-Louis Lions in Universit\'e Pierre et Marie Curie - Paris 6) is warmly acknowledged for useful discussions.


\appendix
\section{Transformation matrices and matrices $\Psi$ of MRT-LBM}\label{app:1}
\subsection{For two dimensional MRT-LBM with 9 discrete velocities}\label{App:d2q9}
The transformation matrice for MRT-LBM with 9 discrete velocities is described by
\begin{equation}
M=\left[ \begin {array}{rrrrrrrrr} 1&1&1&1&1&1&1&1&1\\\noalign{\medskip}0&1&0&-1&0&1&-1&-1&1\\\noalign{\medskip}0&0&1&0&-1&1&1&-1&-1\\\noalign{\medskip}-4&-1&-1&-1&-1&2&2&2&2\\\noalign{\medskip}4&-2&-2&-2&-2&1&1&1&1\\\noalign{\medskip}0&-2&0&2&0&1&-1&-1&1\\\noalign{\medskip}0&0&-2&0&2&1&1&-1&-1\\\noalign{\medskip}0&1&-1&1&-1&0&0&0&0\\\noalign{\medskip}0&0&0&0&0&1&-1&1&-1\end {array} \right] .
\end{equation}
For the weakly-compressible MRT-LBM, the matrix $\Psi$ is given by
\begin{equation}
\Psi=\left[ \begin {array}{rrrrrrrrr} 1&0&0&0&0&0&0&0&0\\\noalign{\medskip}0&1&0&0&0&0&0&0&0\\\noalign{\medskip}0&0&1&0&0&0&0&0&0\\\noalign{\medskip}{\it s_e}\, \left( -2-3\,{U}^{2}-3\,{V}^{2} \right) &6\,{\it s_e}\,U&6\,{\it s_e}\,V&1-{\it s_e}&0&0&0&0&0\\\noalign{\medskip}{\it s_\epsilon}\, \left( 1+3\,{U}^{2}+3\,{V}^{2} \right) &-6\,{\it s_\epsilon}\,U&-6\,{\it s_\epsilon}\,V&0&1-{\it s_\epsilon}&0&0&0&0\\\noalign{\medskip}0&-{\it s_q}&0&0&0&1-{\it s_q}&0&0&0\\\noalign{\medskip}0&0&-{\it s_q}&0&0&0&1-{\it s_q}&0&0\\\noalign{\medskip}{\it s_\nu}\, \left( -{U}^{2}+{V}^{2} \right) &2\,{\it s_\nu}\,U&-2\,{\it s_\nu}\,V&0&0&0&0&1-{\it s_\nu}&0\\\noalign{\medskip}-{\it s_\nu}\,UV&{\it s_\nu}\,V&{\it s_\nu}\,U&0&0&0&0&0&1-{\it s_\nu}\end {array} \right]  .
\end{equation}
 For the MRT-LBM similar to an incompressible LBS, the matrix $\Psi$ is given by
\begin{equation}
\Psi=\left[ \begin {array}{rrrrrrrrr} 1&0&0&0&0&0&0&0&0\\\noalign{\medskip}0&1&0&0&0&0&0&0&0\\\noalign{\medskip}0&0&1&0&0&0&0&0&0\\\noalign{\medskip}-2\,{\it s_e}&6\,{\it s_e}\,U&6\,{\it s_e}\,V&1-{\it s_e}&0&0&0&0&0\\\noalign{\medskip}{\it s_\epsilon}&-6\,{\it s_\epsilon}\,U&-6\,{\it s_\epsilon}\,V&0&1-{\it s_\epsilon}&0&0&0&0\\\noalign{\medskip}0&-{\it s_q}&0&0&0&1-{\it s_q}&0&0&0\\\noalign{\medskip}0&0&-{\it s_q}&0&0&0&1-{\it s_q}&0&0\\\noalign{\medskip}0&2\,{\it s_\nu}\,U&-2\,{\it s_\nu}\,V&0&0&0&0&1-{\it s_\nu}&0\\\noalign{\medskip}0&{\it s_\nu}\,V&{\it s_\nu}\,U&0&0&0&0&0&1-{\it s_\nu}\end {array} \right] 
\end{equation}
\subsection{For three dimensional MRT-LBM with 15 discrete velocities}\label{App:d3q15}
The transformation matrice for MRT-LBM with 15 discrete velocities is described by
\begin{equation}
M= \left[ \begin {array}{rrrrrrrrrrrrrrr} 1&1&1&1&1&1&1&1&1&1&1&1&1&1&1\\\noalign{\medskip}0&1&-1&0&0&0&0&1&-1&1&-1&1&-1&1&-1\\\noalign{\medskip}0&0&0&1&-1&0&0&1&1&-1&-1&1&1&-1&-1\\\noalign{\medskip}0&0&0&0&0&1&-1&1&1&1&1&-1&-1&-1&-1\\\noalign{\medskip}-2&-1&-1&-1&-1&-1&-1&1&1&1&1&1&1&1&1\\\noalign{\medskip}16&-4&-4&-4&-4&-4&-4&1&1&1&1&1&1&1&1\\\noalign{\medskip}0&-4&4&0&0&0&0&1&-1&1&-1&1&-1&1&-1\\\noalign{\medskip}0&0&0&-4&4&0&0&1&1&-1&-1&1&1&-1&-1\\\noalign{\medskip}0&0&0&0&0&-4&4&1&1&1&1&-1&-1&-1&-1\\\noalign{\medskip}0&2&2&-1&-1&-1&-1&0&0&0&0&0&0&0&0\\\noalign{\medskip}0&0&0&1&1&-1&-1&0&0&0&0&0&0&0&0\\\noalign{\medskip}0&0&0&0&0&0&0&1&-1&-1&1&1&-1&-1&1\\\noalign{\medskip}0&0&0&0&0&0&0&1&1&-1&-1&-1&-1&1&1\\\noalign{\medskip}0&0&0&0&0&0&0&1&-1&1&-1&-1&1&-1&1\\\noalign{\medskip}0&0&0&0&0&0&0&1&-1&-1&1&-1&1&1&-1\end {array} \right].
\end{equation}
\section{Coefficient Matrices}\label{App:coef}
In this section, the dissipation coefficient matrix will be divided into two parts: exact dissipation coefficient matrix ($\delta t\cdot D$); the part ($\delta t\cdot \widehat D$) of non-Galilean invariance related with mean flows. 
\subsection{Coefficient matrices of L-NSE with the first-order of $\delta t$ for 2D MRT-LBM}\label{App:firstorder2d}
The exact part $D$ of the dissipation coefficient matrix is given by
\begin{equation}
D= \left[ \begin {array}{rrr} 0&0&0\\\noalign{\medskip}0&-\frac{1}{3}{{k_x}
}^{2}\sigma_\nu -\frac{1}{3}{{k_y}}^{2}\sigma_\nu -\frac{1}{3}{{k_x}}^{2}
\sigma_e&-\frac{1}{3}{k_x}{k_y}\sigma_e\\\noalign{\medskip}0&-\frac{1}{3}
{k_x}{k_y}\sigma_e&-\frac{1}{3}{{k_x}}^{2}\sigma_\nu -\frac{1}{3}{{
k_y}}^{2}\sigma_\nu -\frac{1}{3}{{k_y}}^{2}\sigma_e\end {array}
 \right].
\end{equation}

The  part $\widehat{D}$ of  non-Galilean invariance is expressed by
\begin{equation}
\widehat{D}[1,1]=\widehat{D}[1,2]=\widehat{D}[1,3]=0
\end{equation}
\begin{equation}
\begin{array}{ll}
\widehat{D}[2,1]&=\frac{1}{3}{{k_x}}^{2}U\sigma_\nu +\frac{1}{3}{{k_y}}^{2}U\sigma_\nu +\frac{1}{3}{k_x}{k_y}V\sigma_e+\frac{1}{3}{{k_x}}^{2}U\sigma_e\\[2mm]
\widehat{D}[2,2]&={{k_x}}^{2}{V}^{2}\sigma_e
+2{{k_x}}^{2}{U}^{2}\sigma_e+{k_x}V{k_y}U\sigma_e-{{k_x}}^{2}{V}^{2}\sigma_\nu +2{{k_x}}^{2}{U}^{2}\sigma_\nu +4{k_x}V{k_y}U\sigma_\nu +{{k_y}}^{2}{V}^{2}\sigma_\nu\\[2mm]
\widehat{D}[2,3]&=2
{k_x}{k_y}{V}^{2}\sigma_e+{{k_x}}^{2}UV\sigma_e+{k_x}{U}^{2}{k_y}\sigma_e+2{k_x}{U}^{2}{k_y}\sigma_\nu -2{k_x}{k_y}{V}^{2}\sigma_\nu +3{{k_y}}^{2}VU\sigma_\nu -{{k_x}}^{2}UV\sigma_\nu \\[2mm]
\widehat{D}[3,1]&=\frac{1}{3}{{k_y}}^{2
}V\sigma_e+\frac{1}{3}{k_y}{k_x}U\sigma_e+\frac{1}{3}{{k_y}}^{2}V\sigma_\nu +\frac{1}{3}{{k_x}}^{2}V\sigma_\nu \\[2mm]
\widehat{D}[3,2]&={{k_y}}^{2}VU\sigma_e+{k_x}{k_y}{V}^{2}\sigma_e+2{k_x}{U}^{2}{k_y}\sigma_e+3{{k_x}}^{2}UV\sigma_\nu +2{k_x}{k_y}{V}^{2}
\sigma_\nu -2{k_x}{U}^{2}{k_y}\sigma_\nu -{{k_y}}^{2}VU\sigma_\nu\\[2mm]
\widehat{D}[3,3]&={{k_y}}^{2}{U}^{2}\sigma_e+{k_x}V{k_y}U\sigma_e+2{{k_y}}^{2}{V}^{2}\sigma_e-{{k_y}}^{2}{U}^{2}\sigma_\nu +4{k_x}V{k_y}U\sigma_\nu +2{{k_y}}^{2}{V}^{2}\sigma_\nu +{{k_x}}^{2}{U}^{2}\sigma_\nu
\end{array}
\end{equation}
\subsection{For three dimensional MRT-LBM with 19 discrete velocities}\label{App:d3q19}
\begin{equation}
 D= \left[ \begin {array}{rrrr} 0&0&0&0\\\noalign{\medskip}0&-\frac{1}{3}{ \sigma_\nu}|{\rm k}|^2-\frac{1}{9}{{ k_x}}
^{2}({ \sigma_\nu}+2{ \sigma_e})&-\frac{1}{9}{ k_y}{
 k_x} \left( { \sigma_\nu}+2{ \sigma_e} \right) &-\frac{1}{9}{ k_z}
{ k_x} \left( { \sigma_\nu}+2{ \sigma_e} \right) 
\\\noalign{\medskip}0&-\frac{1}{9}{ k_y}{ k_x} \left( { \sigma_\nu}+
2{ \sigma_e} \right) &-\frac{1}{3}{ \sigma_\nu}|{\rm k}|^2-\frac{1}{9}{{ k_y}}
^{2}({ \sigma_\nu}+2{ \sigma_e})&-\frac{1}{9}{ k_z}{ k_y} \left( { \sigma_\nu}+2{
 \sigma_e} \right) \\\noalign{\medskip}0&-\frac{1}{9}{ k_z}{ k_x}
 \left( { \sigma_\nu}+2{ \sigma_e} \right) &-\frac{1}{9}{ k_z}{ k_y
} \left( { \sigma_\nu}+2{ \sigma_e} \right) &-\frac{1}{3}{ \sigma_\nu}|{\rm k}|^2-\frac{1}{9}{{ k_y}}^{2}({ \sigma_\nu}+2{ \sigma_z})\end {array} \right] 
\end{equation}
where the wave-number vector ${\rm k}=(k_x,k_y,k_z)$ and $|{\rm k}|=\sqrt{k_x^2+k_y^2+k_z^2}$. The viscosity  is defined  by $\nu=\frac{1}{3}\sigma_\nu$ and the bulk viscosity is defined by $\eta=\frac{2}{9}\sigma_e$. Then, the effective bulk viscosity is given by $\eta+\frac{1}{3}\nu$.

The part $\widehat{D}$ of non-Galilean invariance is given by
\begin{equation}
\widehat{D}[1,1]=\widehat{D}[1,2]=\widehat{D}[1,3]=\widehat{D}[1,4]=0
\end{equation}
\begin{equation}
\begin{array}{ll}
\widehat{D}[2,1]=&\frac{1}{3}{{k_y}}^{2}U{\sigma_\nu}+\frac{4}{9}{{k_x}}^{2}U{\sigma_\nu}+\frac{1}{9}{k_x}V{k_y}{\sigma_\nu}+\frac{1}{9}{k_x}W{k_z}{
\sigma_\nu}+\frac{1}{3}{{k_z}}^{2}U{\sigma_\nu}+\frac{2}{9}{k_x}W{k_z}{
\sigma_e}+\frac{2}{9}{k_x}V{k_y}{\sigma_e}+\frac{2}{9}{{k_x}}^{2}
U{\sigma_e}\\[2mm]
\widehat{D}[2,2]=&\frac{13}{3}{k_x}W{k_z}U{\sigma_\nu}+2{k_z}V{k_y}W{
\sigma_\nu}+{{k_z}}^{2}{W}^{2}{\sigma_\nu}+\frac{13}{3}{k_x}V{
k_y}U{\sigma_\nu}+\frac{8}{3}{{k_x}}^{2}{U}^{2}{\sigma_\nu}-\frac{2}{3}{{
k_x}}^{2}{V}^{2}{\sigma_\nu}+{{k_y}}^{2}{V}^{2}{\sigma_\nu}-\\[2mm]&\frac{2}{
3}{{k_x}}^{2}{W}^{2}{\sigma_\nu}+\frac{2}{3}{k_x}W{k_z}U{
\sigma_e}+\frac{4}{3}{{k_x}}^{2}{U}^{2}{\sigma_e}+\frac{2}{3}{{k_x}}^{2}{W}^
{2}{\sigma_e}+\frac{2}{3}{k_x}V{k_y}U{\sigma_e}+\frac{2}{3}{{k_x}
}^{2}{V}^{2}{\sigma_e}\\[2mm]
\widehat{D}[2,3]=&\frac{7}{3}{k_y}{k_x}{U}^{2}{\sigma_\nu}+3{k_y}W{k_z}
U{\sigma_\nu}-\frac{4}{3}{k_x}{k_y}{V}^{2}{\sigma_\nu}-\frac{2}{3}{
k_x}{W}^{2}{k_y}{\sigma_\nu}-\frac{2}{3}{{k_x}}^{2}UV{
\sigma_\nu}+3{{k_y}}^{2}VU{\sigma_\nu}-\frac{2}{3}{k_x}W{k_z}V{
\sigma_\nu}+\\[2mm]&\frac{2}{3}{k_x}{U}^{2}{k_y}{\sigma_e}+\frac{2}{3}{k_x
}{W}^{2}{k_y}{\sigma_e}+\frac{2}{3}{k_x}W{k_z}V{
\sigma_e}+\frac{2}{3}{{k_x}}^{2}UV{\sigma_e}+\frac{4}{3}{k_x}{k_y}{V}
^{2}{\sigma_e}\\[2mm]
\widehat{D}[2,4]=&-\frac{2}{3}{{k_x}}^{2}UW{\sigma_\nu}+3{{k_z}}^{2}WU{\sigma_\nu}-2
/3{k_x}V{k_y}W{\sigma_\nu}-\frac{4}{3}{k_x}{k_z}{W}^{
2}{\sigma_\nu}+3{k_y}U{k_z}V{\sigma_\nu}+\frac{7}{3}{k_x}
{U}^{2}{k_z}{\sigma_\nu}-\frac{2}{3}{k_x}{V}^{2}{k_z}{
\sigma_\nu}+\\[2mm]&\frac{2}{3}{{k_x}}^{2}UW{\sigma_e}+\frac{2}{3}{k_x}{U}^{2}{
k_z}{\sigma_e}+\frac{2}{3}{k_x}V{k_y}W{\sigma_e}+\frac{4}{3}{k_x
}{k_z}{W}^{2}{\sigma_e}+\frac{2}{3}{k_x}{V}^{2}{k_z}{
\sigma_e}\\[2mm]
\widehat{D}[3,1]=&\frac{2}{9}{{k_y}}^{2}V{\sigma_e}+\frac{2}{9}{k_y}W{k_z}{\sigma_e
}+\frac{2}{9}{k_y}U{k_x}{\sigma_e}+\frac{1}{3}{{k_z}}^{2}V{
\sigma_\nu}+\frac{1}{9}{k_y}W{k_z}{\sigma_\nu}+\frac{1}{9}{k_y}U{
k_x}{\sigma_\nu}+\frac{4}{9}{{k_y}}^{2}V{\sigma_\nu}+\frac{1}{3}{{k_x}}^{
2}V{\sigma_\nu}\\[2mm]
\widehat{D}[3,2]=&\frac{4}{3}{k_x}{U}^{2}{k_y}{\sigma_e}+\frac{2}{3}{k_x}{W}^{2}{
k_y}{\sigma_e}+\frac{2}{3}{{k_y}}^{2}VU{\sigma_e}+\frac{2}{3}{k_y}
W{k_z}U{\sigma_e}+\frac{2}{3}{k_x}{k_y}{V}^{2}{\sigma_e
}-\frac{2}{3}{k_x}{W}^{2}{k_y}{\sigma_\nu}-\frac{4}{3}{k_y}{k_x
}{U}^{2}{\sigma_\nu}+\\[2mm]&3{{k_x}}^{2}UV{\sigma_\nu}+3{k_x}
W{k_z}V{\sigma_\nu}-\frac{2}{3}{{k_y}}^{2}VU{\sigma_\nu}+\frac{7}{3}{
k_x}{k_y}{V}^{2}{\sigma_\nu}-\frac{2}{3}{k_y}W{k_z}U{
\sigma_\nu}\\[2mm]
\widehat{D}[3,3]=&\frac{2}{3}{{k_y}}^{2}{W}^{2}{\sigma_e}+\frac{4}{3}{{k_y}}^{2}{V}^{2}{
\sigma_e}+\frac{2}{3}{k_x}V{k_y}U{\sigma_e}+\frac{2}{3}{k_y}W{
k_z}V{\sigma_e}+\frac{2}{3}{{k_y}}^{2}{U}^{2}{\sigma_e}+\frac{13}{3}{
k_x}V{k_y}U{\sigma_\nu}+2{k_x}W{k_z}U{\sigma_\nu}+\\[2mm]&
{{k_x}}^{2}{U}^{2}{\sigma_\nu}+{{k_z}}^{2}{W}^{2}{\sigma_\nu-U*t}-
\frac{2}{3}{{k_y}}^{2}{W}^{2}{\sigma_\nu}-\frac{2}{3}{{k_y}}^{2}{U}^{2}{
\sigma_\nu}+\frac{8}{3}{{k_y}}^{2}{V}^{2}{\sigma_\nu}+\frac{13}{3}{k_z}V
{k_y}W{\sigma_\nu}\\[2mm]
\widehat{D}[3,4]=&3{k_z}U{k_x}V{\sigma_\nu}-\frac{2}{3}{k_y}{U}^{2}{k_z}
{\sigma_\nu}-\frac{2}{3}{{k_y}}^{2}VW{\sigma_\nu}-\frac{4}{3}{k_y}{
k_z}{W}^{2}{\sigma_\nu}-\frac{2}{3}{k_y}U{k_x}W{\sigma_\nu}
+\frac{7}{3}{k_y}{V}^{2}{k_z}{\sigma_\nu}+3{{k_z}}^{2}WV{
\sigma_\nu}+\\[2mm]&\frac{2}{3}{k_y}{U}^{2}{k_z}{\sigma_e}+\frac{4}{3}{k_y
}{k_z}{W}^{2}{\sigma_e}+\frac{2}{3}{k_y}{V}^{2}{k_z}{
\sigma_e}+\frac{2}{3}{k_y}U{k_x}W{\sigma_e}+\frac{2}{3}{{k_y}}^{2}VW{
\sigma_e}\\[2mm]
\widehat{D}[4,1]=&\frac{2}{9}{{k_z}}^{2}W{\sigma_e}+\frac{2}{9}{k_z}V{k_y}{\sigma_e
}+\frac{2}{9}{k_z}U{k_x}{\sigma_e}+\frac{4}{9}{{k_z}}^{2}W{
\sigma_\nu}+\frac{1}{3}{{k_y}}^{2}W{\sigma_\nu}+\frac{1}{9}{k_z}U{k_x}{
\sigma_\nu}+\frac{1}{3}{{k_x}}^{2}W{\sigma_\nu}+\frac{1}{9}{k_z}V{k_y
}{\sigma_\nu}\\[2mm]
\widehat{D}[4,2]=&\frac{2}{3}{k_x}{k_z}{W}^{2}{\sigma_e}+\frac{2}{3}{k_z}V{k_y}
U{\sigma_e}+\frac{4}{3}{k_x}{U}^{2}{k_z}{\sigma_e}+\frac{2}{3}{
k_x}{V}^{2}{k_z}{\sigma_e}+\frac{2}{3}{{k_z}}^{2}WU{
\sigma_e}+3{{k_x}}^{2}UW{\sigma_\nu}-\frac{2}{3}{{k_z}}^{2}WU{
\sigma_\nu}+\\[2mm]&3{k_x}V{k_y}W{\sigma_\nu}+\frac{7}{3}{k_x}{k_z
}{W}^{2}{\sigma_\nu}-\frac{2}{3}{k_x}{V}^{2}{k_z}{\sigma_\nu}-
\frac{4}{3}{k_x}{U}^{2}{k_z}{\sigma_\nu}-\frac{2}{3}{k_y}U{k_z}
V{\sigma_\nu}\\[2mm]
\widehat{D}[4,3]=&\frac{2}{3}{k_z}U{k_x}V{\sigma_e}+\frac{2}{3}{{k_z}}^{2}WV{
\sigma_e}+\frac{4}{3}{k_y}{V}^{2}{k_z}{\sigma_e}+\frac{2}{3}{k_y}{
k_z}{W}^{2}{\sigma_e}+\frac{2}{3}{k_y}{U}^{2}{k_z}{
\sigma_e}-\frac{2}{3}{{k_z}}^{2}WV{\sigma_\nu}+3{{k_y}}^{2}VW{
\sigma_\nu}-\\[2mm]&\frac{2}{3}{k_y}{U}^{2}{k_z}{\sigma_\nu}+3{k_y}U{
k_x}W{\sigma_\nu}-\frac{4}{3}{k_y}{V}^{2}{k_z}{\sigma_\nu}+
\frac{7}{3}{k_y}{k_z}{W}^{2}{\sigma_\nu}-\frac{2}{3}{k_z}U{k_x}
V{\sigma_\nu}\\[2mm]
\widehat{D}[4,4]=&\frac{8}{3}{{k_z}}^{2}{W}^{2}{\sigma_\nu}+\frac{13}{3}{k_x}W{k_z}U{
\sigma_\nu}+{{k_y}}^{2}{V}^{2}{\sigma_\nu}+2{k_x}V{k_y}
U{\sigma_\nu}-\frac{2}{3}{{k_z}}^{2}{U}^{2}{\sigma_\nu}+{{k_x}}^{2
}{U}^{2}{\sigma_\nu}+\frac{13}{3}{k_z}V{k_y}W{\sigma_\nu}-\\[2mm]&\frac{2}{3}{
{k_z}}^{2}{V}^{2}{\sigma_\nu}+\frac{2}{3}{k_x}W{k_z}U{
\sigma_e}+\frac{2}{3}{{k_z}}^{2}{U}^{2}{\sigma_e}+\frac{4}{3}{{k_z}}^{2}{W}^
{2}{\sigma_e}+\frac{2}{3}{{k_z}}^{2}{V}^{2}{\sigma_e}+\frac{2}{3}{k_y}
W{k_z}V{\sigma_e}
\end{array}
\end{equation}



\bibliographystyle{elsarticle-num}
\bibliography{<your-bib-database>}

\begin{thebibliography}{00}

 \bibitem{chendoolen}{S. Chen, G. Doolen,}{ Lattice Boltzmann method for fluid flows, Annu. Rev. Fluid  Mech. 161 (1998) 329.}

\bibitem{buickgreatedcampbell}{J. M. Buick, C. A. Greated, D. M. Cmpbell,} {Lattice BGK simulation of sound waves, Eurohys. Lett. 43 (2) (1998) 235-240.}

\bibitem{simondenispierre}{S. Mari\'e, D. Ricot, P. Sagaut,}{ Comparison between lattice Boltzmann method and Navier-Stokes high order schemes for computational aeroacoustics, J. Comput. Phys. 228 (2009) 1056-1070.}

\bibitem{ricotmariesagautbailly}{D. Ricot, S. Mari\'e, P. Sagaut, C. Bailly,
}{Lattice Boltzmann method with selective viscosity filter,  J.
Comput. Phys. 228 (2009) 4478-4490. }

\bibitem{buickbuckleygreated}{J. M. Buick, C.Spectra and pseudospectra:
the behavior of nonnormal matrices and operators L. Buckley, C. A. Greated, }{Lattice Boltzmann BGK-simulation of non-linear sound waves: The development of a shock front,  J. Phys. A: Math. Gen. 33 (2000) 3917-3928.}




\bibitem{dhumieres2}{D. d'Humi\`eres, I. Ginzburg, M. Krafczyexact solution of k, P. Lallemand, L. S. Luo, }{Multiple-relaxation-time lattice Boltzmann models in three dimensions, Phil. Trans.  R. Soc. Lond. A, 360 (2002) 437-451}  

\bibitem{lallemandluo}{P. Lallemand, L. S. Luo, }{Theory of the lattice Boltzmann method: Dispersion, dissipation, isotropy,
Galilean invariance, and stability, Phys. Rev. E. 61(6) (2000)
6546-6562. }

\bibitem{xusagaut}{H. Xu, P. Sagaut, }{Optimal low-dispersion low-dissipation LBM schemes for computational aeroacoustics, J. Comput. Phys 230 (13): 5353-5382 (2011)}

\bibitem{Kinsler}{L. E. Kinsler, A. R. Frey, A. B. Coppens, J. V. Sanders, }{Fundamentals of Acoustics, John Wiley \& Sons, Inc, 2000.}

\bibitem{Lord}{L. Rayleigh, }{Theory of Sound, Macmillan and Co., London, 1929.}



\bibitem{Morse}{P. M. Morse, K. U. Ingard, }{Theoretical Acoustics, Princeton University Press, 1986.}


\bibitem{senguptadipankarsagaut}{T. Sengupta, A. Dipankar, P. Sagaut, }{Error dynamics: Beyond von Neumann analysis, J. Comput. Phys. 226 (2) (2007) 1211-1218.}

\bibitem{duboislallemand}{F. Dubois, P. Lallemand, }{Towards higher order lattice Boltzmann schemes, J. Stat. Mech. Theory E, (2009) P0600.6}

\bibitem{dubois}{F. Dubois, }{Third order equivalent equation of lattice Boltzmann scheme, Disc. \& Cont. Dyn. Syst. 23 (1/2) (2009) 221-248.}

\bibitem{landaulifshitz}{L.D Landau, E.M. Lifshitz, }{Fluid Mechanics,second ed., Oxford: Pergamon, 1987.}

\bibitem{junklarluo}{M. Junk, A. Klar, L. S. Luo,} {Asymptotic analysis of the lattice Boltzmann equation, J. Comput. Phys. 210 (2005) 676-704.}

\bibitem{stewartsun}{G. Stewart, J. G. Sun, }{Matrix Perturbation Theory, Boston: Academic Press, 1990.}

\bibitem{AndrewTan}{A. L. Andrew, R. C. E.  Tan, } {Computation of derivatives of repeated eigenvalues and the corresponding eigenvectors, SIAM J. Matrix Anal. Appl. 20(1)(1999): 78-100. }
\bibitem{aamorsche}{N.P. van der Aa, H.G. ter Morsche and R.R.M. Mattheij, }{  Computation of eigenvalue and eigenvector derivatives for a general complex-valued eigensystem, Electronic Journal of Linear Algebra, 16 (2007) 300-314}

\bibitem{palabos}{User's Guide, \url{http://www.lbmethod.org/palabos/}.}





\bibitem{tam}{C.K.W. Tam , J.C. Webb,}{ Dispersion-relation-preserving finite difference schemes for computational acoustics, J. Comput. Phys. 107 (1993) 262-281}


\bibitem{tamckw}{C. Bogey, C. Bailly, }{Three-dimensional non-reflective boundary conditions for acoustic simulations: far field formulation and validation test cases, ACTA ACUSTICA united with ACOUSTICA, 88 (2002) 463-471.}

\bibitem{tamaiaa}{C.K.W. Tam, F.Q. Hu,} {Second computational aeroacoustic (CAA) workshop on benchmark problems. No. NASA CP-3352, 1997.}

\bibitem{Daniel}{D.J. Bodony, }{Accuracy of the simultaneous-approximation term boundary condition for time dependent problems, J. Sci. Comput. 43 (2010) 118-133.}

\bibitem{tamaiaa2}{C.K.W. Tam, F.Q. Hu,} {An optimized multi-dimensional interpolation scheme for computational aeroacoustics applications using overset grids. Presented at the 10th AIAA/CEAS Aeroacoustics Conference, AIAA Paper 2004-2812, 2004.}

\bibitem{xinzhang}{G. Ashcroft, X. Zhang, }{Optimized prefactored compact schemes, J. Comput. Phys. 190 (2003) 459-477.}

\bibitem{Hu}{F.Q. Hu, M.Y. Hussaini, J. Manthey,} {Low-dissipation and -dispersion Runge-Kutta schemes for computational acoustics, J. Comput. Phys. 124 (1996) 177-191.}


\end{thebibliography}



\end{document}